\documentclass[onecolumn,preprintnumbers,amsmath,amssymb,nofootinbib,noshowpacs]{revtex4}

\usepackage{graphicx,amsmath,amssymb,txfonts,natbib}

\renewcommand{\citet}[1]{\citep{#1}}

\newcommand{\be}{\begin{equation}}
\newcommand{\ee}{\end{equation}}
\newcommand{\bdm}{\begin{displaymath}}
\newcommand{\edm}{\end{displaymath}}
\newcommand{\bea}{\begin{eqnarray}}
\newcommand{\eea}{\end{eqnarray}}
\newcommand{\beas}{\begin{eqnarray*}}
\newcommand{\eeas}{\end{eqnarray*}}

\newcommand{\av}[1]{\left< #1\right>}
\newcommand{\dk}{\mathrm{d}k}
\newcommand{\dcubedk}{\mathrm{d}^3\mathbf{k}}
\newcommand{\dy}{\mathrm{d}y}
\newcommand{\da}{\mathrm{d}a}
\newcommand{\Li}{\mathrm{Li}}
\newcommand{\tk}{\kappa}

\newcommand{\RR}{\tau\tau}
\newcommand{\RS}{\tau\textrm{\tiny{\emph{S}}}}
\renewcommand{\SS}{\textrm{\tiny{\emph{SS}}}}
\newcommand{\VV}{\textrm{\tiny{\emph{VV}}}}
\newcommand{\TT}{\textrm{\tiny{\emph{TT}}}}

\newcommand{\tcos}{\widetilde{\cos}^{-1}}
\newcommand{\erf}{\;\mathrm{erf}}

\newcommand{\PTrTr}{\mathcal{P}_{\tau\tau}}
\newcommand{\PTrTs}{\mathcal{P}_{\tau\textrm{\tiny{\emph{S}}}}}
\newcommand{\PTsTs}{\mathcal{P}_{\textrm{\tiny{\emph{SS}}}}}
\newcommand{\PTvTv}{\mathcal{P}_{\textrm{\tiny{\emph{VV}}}}}
\newcommand{\PTtTt}{\mathcal{P}_{\textrm{\tiny{\emph{TT}}}}}

\begin{document}

\title{Concerning the statistics of cosmic magnetism}

\author{Iain A. Brown}
\email{i.a.brown@astro.uio.no}
\affiliation{Institute of Theoretical Astrophysics, University of Oslo, 0315 Oslo, Norway}
\affiliation{Institut f\"ur Theoretische Physik, Universit\"at Heidelberg, Philosophenweg 16, 69120 Heidelberg, Germany}


\date{\today}

\begin{abstract}
Magnetic fields appear to be a generic feature of the early universe and are a natural source of secondary CMB non-Gaussianity. In recent years the statistical nature of the stresses of a primordial magnetic field has been well studied. In this paper we confirm and extend these studies at one- and two-point level, and present analytical results for a wide range of power-law spectra. We also consider two non-power law cases of interest: a blue spectrum with an extended damping tail on small scales, which could be generated by the non-linear mixing of density and vorticity; and a red spectrum with a damping tail on large scales. We then briefly consider the CMB impacts that result from such fields. While this paper focuses on the one- and two-point moments, the techniques we employ are designed to ease the analysis of the full bispectra induced by primordial magnetic fields.
\end{abstract}


\maketitle

\section{Introduction}
Large-scale magnetic fields appear to be inevitable. Fields have been observed on galactic and cluster scales and likely exist on supercluster scales \citep{Kronberg:1993vk,Grasso:2000wj,Giovannini:2003yn,Xu:2005rb}. There are a variety of potential generation mechanisms in both the early and late universe, many of which may well have been in operation, ranging from so-called Biermann batteries (e.g. \citep{Biermann:2003xh,Hanayama:2005hd}) to fields generated at reionisation \citep{Subramanian:1994,Gnedin:2000ax,Langer:2002tx,Langer:2005gw}. The number of well-motivated magnetogenesis mechanisms would appear to make the existence of magnetic fields across a large range of scales inevitable. In particular, plasma processes in the early, pre-recombination universe \citep{Gopal:2004ut,Matarrese:2004kq,Takahashi:2005nd,Ichiki:2006cd,Siegel:2006px,Kobayashi:2007wd,Maeda:2008dv} wherein linear scalar perturbations drive nonlinear vorticity, producing currents in the proton/electron plasmas, would suggest that at least some magnetic fields before recombination are inevitable even if the detailed predictions in the various models differ somewhat. The fields extent in the present universe need not necessarily be primordial; the fields observed on cluster scales can be accounted for in principle by non-linear processes before recombination, by production at reionisation or at recombination \citep{Hogan:2000gv,Berezhiani:2003ik}, and it seems probable that each of these processes have contributed. There also many exist potential mechanisms in the extremely early universe, such as magnetogenesis at a phase transition (for example \citet{Baym:1995fk,Martin:1995su,Hindmarsh:1997tj,Boyanovsky:2002wa,Kahniashvili:2009qi}).

However, all of these fields will have a distinctive, blue power spectrum \citep{Durrer:2003ja}, and the most recent bounds on fields present at or before recombination rather favour a red spectrum \citep{Yamazaki:2010nf,Paoletti:2010rx}. It seems reasonable, therefore, to also consider mechanisms that could produce such red fields. To achieve a near-scale invariant spectral index these will necessarily be produced during an inflationary epoch \citep{Durrer:2003ja}, and there is a large variety of potential genesis mechanisms that generate a field from the breaking of the conformal invariance of the electromagnetic field, with \citep{Turner:1987bw,Bassett:2000aw,Prokopec:2004au,Giovannini:2007rh,Campanelli:2007cg,Bamba:2008my} being a few examples.

Constraints on a field produced during the inflationary epoch have naturally focused on the cosmic microwave background (CMB) (\citet{Barrow:1997mj,Clarkson:2002dd,Giovannini:2009fu,Yamazaki:2010nf,Paoletti:2010rx}, for example) and have typically found that the field strength today is at most of the order of nanogauss. A stochastic background of primordial magnetic fields, such as would be reasonably expected from an inflationary process, impacts on the CMB in two chief ways. Firstly the magnetic field itself acts as an additional radiative fluid which, should there be enough energy in the field, impacts on the expansion rate of the universe. However, by far the most significant impact comes from the interactions of the magnetic field and the ionised plasmas extent before recombination, and the scattering between the magnetic fields and the geometry. There have been numerous studies into magnetised cosmological perturbation theory; for some interesting studies see for example \citep{Jedamzik:1996wp,Subramanian:1997gi,Tsagas:1999ft,Mack:2001gc,Giovannini:2003yn,Tsagas:2004kv,Banerjee:2004df,Giovannini:2005jw,Barrow:2006ch,Brown:2006wv,Giovannini:2007qn,Kojima:2009ms} and their references. Clearly the presence of the magnetic field will impact on the CMB angular power spectra. While the impact on the temperature auto-correlation and on the temperature/polarisation cross-correlation will be small except at larger multipoles, a magnetic field also excites $B$ mode polarisation. The $B$ mode angular power spectrum would therefore be of great interest for the study of primordial magnetic fields. See for example \citep{Subramanian:1998fn,Durrer:1998ya,Koh:2000qw,Kahniashvili:2000vm,Mack:2001gc,Lewis:2004ef,Kahniashvili:2006hy,Kahniashvili:2008hx,Yamazaki:2008gr,Finelli:2008xh,Paoletti:2008ck,Bonvin:2010nr} and their references for studies of the magnetised CMB angular power spectra.

The $B$ mode spectrum produced by a primordial magnetic field is, however, likely to be nearly degenerate with the gravitational wave spectrum produced by inflationary models \citep{Lewis:2004ef} and even with forthcoming results from for example the Planck and QUIET experiments it is possible that we will be unable to disentangle the impact of magnetic fields from other contributions and from foreground signals. An alternative is provided by the non-Gaussianity of the magnetic field, studied in for example \citep{Brown:2005kr,Brown:2006wv,Seshadri:2009sy,Caprini:2009vk}; since the stress-energy tensor and the Lorentz forces generated by a magnetic field are non-linear, the statistics imparted on the perturbations by the magnetic field are necessarily non-Gaussian, regardless of the statistics of the underlying magnetic field. Furthermore, we would expect the greatest impact to be on large scales, observations of which with the WMAP satellite are already cosmic-variance limited \citep{Komatsu:2010fb}. Study of the magnetised CMB bispectrum is, however, unfortunately a difficult issue. It requires knowledge of the full 3-point moment of the stress tensor of the magnetic field (the ``shape function'' or ``intrinsic bispectrum'') and to date only a few particular configurations have been studied. While these cases are probably sufficient for nearly scale-invariant magnetic fields, for which power appears to be concentrated on the so-called local configuration (with the bispectrum geometry set by three wavevectors $\mathbf{k}_1\approx -\mathbf{k}_2$ and $\mathbf{k}_3\approx 0$), it has not been possible to fully test this. Furthermore, for the blue cases expected to arise in non-inflationary magnetogenesis models neither the local, nor indeed any other, configuration will necessarily dominate and knowledge of the full intrinsic bispectrum will be required.

To consider the impact of a magnetic field on the CMB, then, we require a clear understanding of the statistical nature of the stress tensor (and, equivalently, the Lorentz forces). In \citep{Brown:2005kr} and \citep{Brown:2006wv} (hereafter BC05 and B06) we studied the 1-, 2- and a special case of the 3-point moments of the stress tensor of a Gaussian magnetic field with a power law power spectrum with index $n_B$, employing a mixture of realisations of the magnetic fields and numerical integration. The nature of the 2-point moments changes depending on whether $n_B>-3/2$ or $n_B<-3/2$ and we considered two cases in particular. For a field with a flat spectrum we found the 2-point moments with both methods, demonstrating good agreement and describing the gross features of such ``ultra-violet'' fields. For a field with a strongly red spectrum we only employed realisations and could only draw the broadest of conclusions. In B06 we further considered a ``damped causal'' power spectrum, resembling that produced by plasma processes in the pre-recombination universe, and the statistics of the Lorentz forces (as subsequently studied in, for example, \citep{Paoletti:2008ck}). More recent studies have improved the approach to numerical integration \citep{Yamazaki:2008gr}, and extended it to a wider range of spectral indices; and have even found analytical solutions for particular cases \citep{Finelli:2008xh,Paoletti:2008ck}. We argue in this paper that the most important spectral indices are at $n_B=2$, $n_B=0$, $n_B=-3/2$ and $n_B=-5/2$, all of which are included in their solutions.

However, while this means that we now have solutions from analysis, numerical integration and from statistical realisations work remains to be done. In principle the realisations are the most general approach, and can be applied to any field, possessing any statistical nature and any power spectrum. However, they are severely compromised by the mode coverage on large scales associated with the finite size of the grid. The extent of this issue was not considered before. Due to its reliance on Wick's theorem, numerical integration can at present only be applied to fields with Gaussian statistics, although arbitrary power spectra can be employed. Our approach in BC05 and B06 was restricted to blue power-law spectra with $n_B>-3/2$ which is overly restrictive. The approach of \citep{Yamazaki:2008gr} is more general but relatively unwieldy and has been applied only to power-law spectra. Finally, while \citep{Paoletti:2008ck} found analytic solutions for spectral indices of interest, they are missing in particular the case $n_B=-2$. As there are only two analytic solutions in the region $n_B\in(-3,-3/2)$, at $n_B=-5/2$ and $n_B=-2$, this is an important omission. The subsequent approach to CMB observables, given these spectra, has been rather piecemeal and model-dependant.

It is our intention in this paper to reconsider many of these issues, always focussing on retaining as much generality and flexibility as possible. In particular, we wish to explicitly decouple the consideration of the the intrinsic magnetic statistics from the evaluation of the CMB angular power spectra; in this manner, one may build the statistical nature of the magnetic stress tensor in any manner one wishes, and wrap it onto the CMB employing independently-derived analytical or numerical transfer functions.  Further, we wish to approach the issue of the intrinsic statistics in an extensible manner, unifying to some degree the approach to analysis and to numerical integration. In doing so we extend the set of useful analytic solutions, finding solutions for all $n_B$ between $n_B=3$ and $n_B=-5/2$. While we will not generalise beyond underlying Gaussian statistics for the magnetic field, we also consider two non-power law power spectra of interest. The first is a damped causal spectrum similar to that considered in B06, and the second resembles a strongly red field produced during an inflationary epoch with an extended decaying tail on large scales. We find analytical solutions for the former and rely on numerical integration for the latter. We then consider the CMB, employing a clearly-defined approximation based on the generic form of a magnetised transfer function. In this paper we consider for brevity only the 1- and 2-point moments but we emphasise throughout the paper that we have constructed our techniques to be immediately applicable to the study of magnetic bispectra, to which we will turn in a forthcoming study. 

We start in section \ref{MagneticFields} by introducing our toy model of primordial magnetic fields and briefly discussing the normalisation of their power spectrum. In section \ref{OnePoint} we briefly consider the one-point statistics of the stress tensor before considering in somewhat more detail the two-point moments in section \ref{2Points}. In section \ref{CMB-General} we present our general approach to the magnetised CMB angular power spectra, and in section \ref{CMB-Results} apply this to the CMB temperature auto-correlation generated by tensor perturbations. We briefly conclude in section \ref{Conclusions}. Appendix \ref{Appendix-SpectraSolutions} presents those analytical solutions not previously presented in \cite{Paoletti:2008ck}, appendix \ref{Appendix-LargeScale} presents the forms on the largest scales, and appendix \ref{Appendix-Yamazaki} briefly addresses a comment in \citep{Yamazaki:2008gr} concerning our previous work. Latin indices refer to spatial dimensions and are raised and lowered with the Minkowski metric $\eta_{ij}=\mathrm{diag}(1,1,1)$.

\section{Tangled magnetic fields}
\label{MagneticFields}
At the linear level a large-scale primordial magnetic field $b_a(\mathbf{k},\eta)$ decays as $a^2(\eta)$. We work with the scaled field $B_a(\mathbf{k})=a^2(\eta)b_a(\mathbf{k},\eta)$, which is constant over time \citep{Subramanian:1997gi,Jedamzik:1996wp}, and assume the electric fields to be negligible. Magnetic fields are damped on small scales, primarily from radiation viscosity, and we approximately model this effect with a time-dependent cut-off scale $k_c(\eta)$ above which they have no power. If the magnetic field is generated at some epoch $\eta_{\mathrm{in}}$ with some cut-off scale $k_c^{\mathrm{Prim}}$ and the viscous damping is denoted by $k_D(\eta)$, the effective damping scale of the magnetic field will then be
\be
  k_c(\eta)=\mathrm{min}\left(k_c^{\mathrm{Prim}},k_D(\eta)\right).
\ee
The magnetic fields are then static for a brief period before the viscous damping scale evolves through the primordial, and the field begins to damp on increasingly large scales. Note that there is then a time-dependence in the magnetic field $B_a(\mathbf{x})$ associated with the damping scale; in this section we will write this explicitly. The damping scale freezes at photon decoupling; hence $k_c(\eta_0)\approx k_c(\eta_\mathrm{rec})$.

Take as a toy model scaled magnetic fields -- not necessarily Gaussian -- with the power spectrum
\be
  \langle B_a(\mathbf{k},k_c(\eta))B^*_b(\mathbf{k}',k_c(\eta))\rangle=\mathcal{P}_B(k)P_{ab}(\mathbf{k})H(k_c(\eta)-k)(2\pi)^3\delta(\mathbf{k-k}')
\ee
where
\be
\label{MagneticSpectrum}
  P_{ab}(\mathbf{k})=\delta_{ab}-\frac{k_ak_b}{k^2}
\ee
is a projection tensor that projects an object onto a plane orthogonal to the wavevector $\mathbf{k}$, $\mathcal{P}_B(k)$ is the magnetic power spectrum and in the interests of simplicity we have neglected an antisymmetric, helical component. $H(x)$ is the Heaviside function, which we define to vanish for $x<0$ and to be unity for $x\geq 0$. For a magnetic field with a power-law spectrum
\be
\label{MagneticPowerSpectrum}
  \mathcal{P}_B(k)=A_Bk^{n_B} ,
\ee
we can immediately see that analyticity of $\mathcal{P}_BP_{ab}$ requires $n_B\geq 2$, corresponding to magnetic fields generated by causal processes. Fields produced by the generation of second-order vorticity from linear scalar perturbations, as considered with varying results in \citep{Matarrese:2004kq,Gopal:2004ut,Takahashi:2005nd,Ichiki:2006cd,Siegel:2006px,Kobayashi:2007wd,Maeda:2008dv}, will naturally be of this type on the largest scales, with a decaying tail on smaller scales. In this paper we model cases both with and without an extended damping tail.

It is common in the literature to normalise the magnetic spectrum to the average magnetic energy density at the present epoch and on a scale $\lambda$, typically of the order of a megaparsec and associated with cluster scales. Smoothing the field with a Gaussian filter $f_\lambda(k)=\exp(-\lambda^2k^2)$, one finds (e.g. \citet{Caprini:2009vk}) that
\be
\label{Blambda2}
  B_\lambda^2=\frac{1}{\pi^2}\int_0^{k_c(\eta_0)}\mathcal{P}_B(k)e^{-\lambda^2k^2}k^2\dk .
\ee
For the power-law spectrum (\ref{MagneticPowerSpectrum}) this becomes a standard integral and has the general solution
\be
  B_\lambda^2=\frac{A_B}{2\pi^2\lambda^{n_B+3}}\left(\Gamma\left(\frac{n_B+3}{2}\right)-\Gamma\left(\frac{n_B+3}{2},\lambda^2k_c^2(\eta_0)\right)\right)
\ee
where $\Gamma(a,x)$ is the incomplete Gamma function. (Note that this differs slightly from that presented in \citet{Finelli:2008xh} due a different definition of the smoothing function.) If $k_c(\eta_0)\gg k_\lambda$, or if $n_B\rightarrow -3$, this reduces to
\be
\label{ApproxMagneticNormalisation}
  B_\lambda^2\approx\frac{A_B}{2\pi^2}\frac{1}{\lambda^{n_B+3}}\Gamma\left(\frac{n_B+3}{2}\right) ,
\ee
in agreement with previous estimates in, for example, \citet{Mack:2001gc}. From this we can see that we require $n_B>-3$ to keep the integral finite. (See for example \citep{Caprini:2001nb,Durrer:2003ja,Caprini:2009vk,Bonvin:2010nr} for some further discussion of these issues.) Since the damping scale is typically rather small, we assume (\ref{ApproxMagneticNormalisation}) to hold for the power-law fields we consider. In cases where an analytic solution to (\ref{Blambda2}) does not exist the equation can in principle be solved numerically or approximately.

An alternative that has recently been employed \citep{Finelli:2008xh,Paoletti:2008ck,Caprini:2009vk,Bonvin:2010nr} is instead to normalise the magnetic spectrum with the mean square field itself,
\be
  \av{B^2}=\frac{1}{\pi^2}\int\mathcal{P}_B(k)k^2\dk .
\ee
The two normalisations are clearly closely related, and in this paper we employ the former.

In addition to the minimally causal case $n_B=2$, we can identify two further power-law spectra of particular interest. In BC05 and B06 we studied in detail the ``flat'' or white-noise field with $n_B=0$, which highlights the impact of each mode on the statistics. A field with $n_B=-5/2$ is also of special interest as it is a half-integer -- which will shortly prove extremely useful -- that lies in the centre of the current bounds on the magnetic spectral index \citep{Yamazaki:2010nf,Paoletti:2010rx} and could be generated during an inflationary epoch.  A field with $n_B=-2$ also satisfies the current bounds, but constraints are weakest for $n_B\rightarrow -3$ (e.g. \citet{Caprini:2001nb}). We typically refer to fields with $n_B\gtrsim -3$ as ``inflationary'' fields, although in principle such a field can take any $n_B>-3$.

The magnetic field contributes to the Euler and Einstein equations through the stress-energy tensor $\tau^\mu_\nu(\mathbf{k})$. As the Poynting vector vanishes to first order and the magnetic energy density is equivalent to the isotropic pressure it is sufficient to consider the stress tensor $\tau^a_b(\mathbf{k})$,
\be
  \tau^a_b(\mathbf{k})=\tilde{\tau}^i_i(\mathbf{k})\delta^a_b-\tilde{\tau}^a_b(\mathbf{k}),
\qquad \mathrm{where}
\quad
 \tilde{\tau}_{ab}(\mathbf{k})=\int B_a(\mathbf{k}')B_b({\mathbf{k-k}'})\dcubedk'
\ee
is the self-convolution of the magnetic field. The stress tensor can be separated into the isotropic pressure (scalar trace), anisotropic pressure (traceless scalar), vorticity and transverse-traceless (TT) tensor components, which we denote by $\tau$, $\tau_S$, $\tau_a^V$ and $\tau_{ab}^T$ respectively. For more details see B06. The statistics of the impact of the magnetic field can be characterised by studying the rotationally-invariant statistics of these components. We consider here the 1-point moments (the probability distribution functions, skewnesses and kurtoses of the scalar pressures) and on the 2-point moments (the auto-correlations and the cross-correlation $\av{\tau\tau_S^*}$). The magnetic field also contributes through the Lorentz force and this has been studied in B06 and in \citep{Paoletti:2008ck,Paoletti:2010rx}; however, here we focus on the statistics of the stresses themselves.

Assuming the fields to be Gaussian in nature allows us to both analytically and numerically integrate the expressions for the stress power spectra, as done numerically in BC05, B06 and \citet{Yamazaki:2008gr}, and analytically as in \citep{Finelli:2008xh,Paoletti:2008ck}, and we extend these approaches both to a wide range of spectral indices $n_B$ and to two non-power law spectra of interest. Some of our analytical results were previously known \citep{Paoletti:2008ck}, but others, including the damped causal spectrum, are original to this paper, as are the numerical results for the IR-controlled field.

\begin{figure}
\centering
\includegraphics[width=0.45\textwidth]{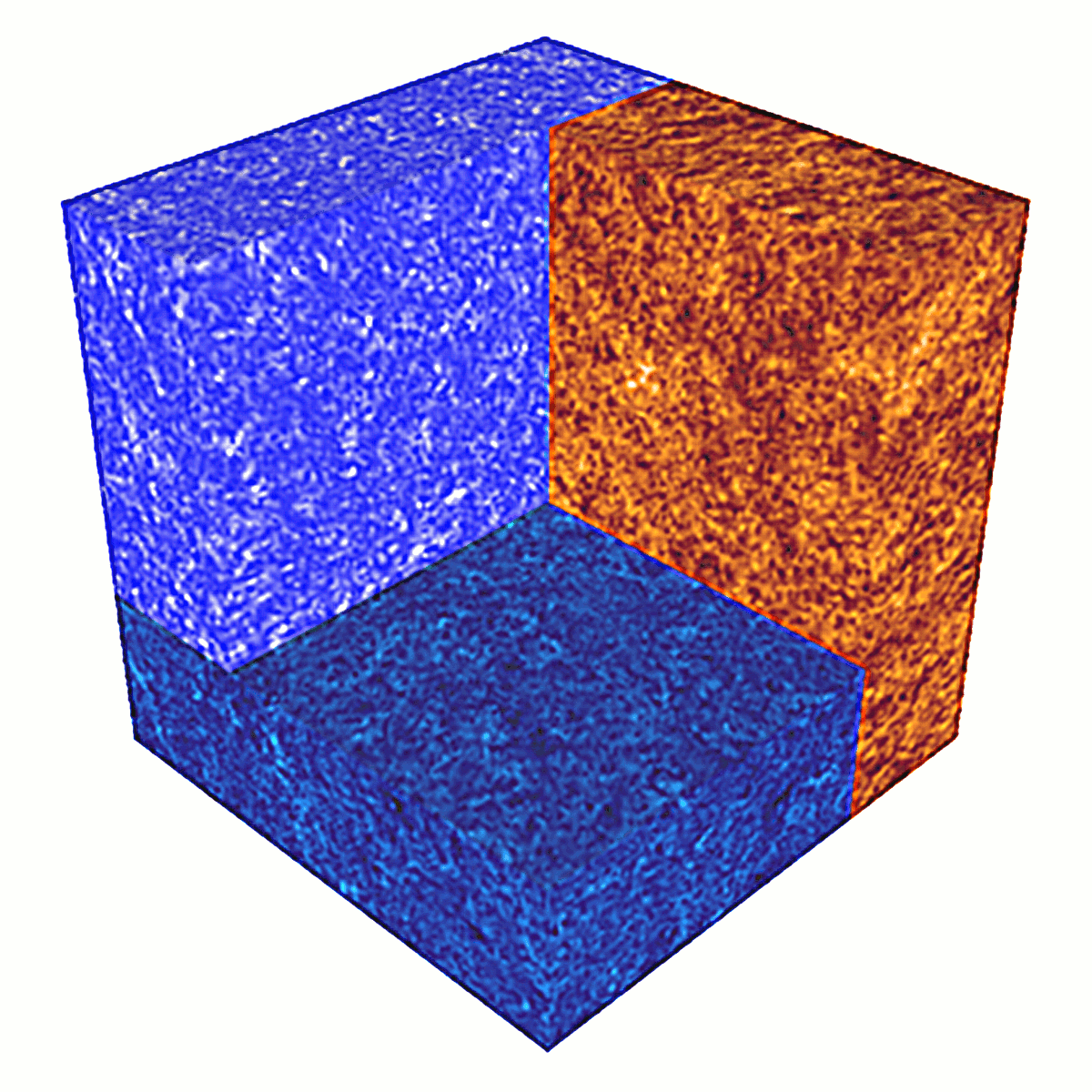}
\caption{Sample realisations of the magnetic field and scalar pressures for a spectrum $n_B=-5/2$ and side-length $l_\mathrm{dim}=256$. The magnetic field is on the back-right wall, the isotropic pressure on the back-left wall and the anisotropic pressure on the floor.}
\label{FieldSlices}
\end{figure}

To aid the analysis, particularly at the one-point level, we also employ extensions of the codes constructed in BC05 and B06, creating static realizations of $B_a(\mathbf{k})$ numerically. Details of the construction can be found in BC05 and B06, and the grid is characterised by its side-length $l_\mathrm{dim}$. In earlier studies we employed $l_\mathrm{dim}=192$, while in this paper we will instead employ $l_\mathrm{dim}=256$ or, in some cases $l_\mathrm{dim}=512$. The Nyquist frequency of the grid is $k_\mathrm{Nyquist}=l_\mathrm{dim}/2$; due to the quadratic nature of the stress tensor, we should then take the damping scale to be $k_c\leq l_\mathrm{dim}/4$ to avoid aliasing power from large to small scales. In Figure \ref{FieldSlices} we present a sample Gaussian realisation of a magnetic field with spectral index $n_B=-5/2$ along with its isotropic and anisotropic pressures, $\tau$ and $\tau_S$ respectively. All errors presented on quantities derived from statistical realisations are 1-$\sigma$ errors.

\section{One-Point Moments}
\label{OnePoint}
In BC05 and B06 we considered the probability distribution functions of the isotropic and anisotropic pressures of the magnetic field, and their skewnesses and kurtoses. The central moments of a distribution $P(x)$ with mean $\mu_1'$ are defined by $\mu_n=\av{(x-\mu'_1)^n}$. The second central moment is the variance, while the (normalised) skewness and kurtosis are
\be
  \gamma_1=\frac{\mu_3}{\mu_2^{3/2}}, \qquad \gamma_2=\frac{\mu_4}{\mu_2^2}-3
\ee
where the kurtosis is defined such that $\gamma_1=\gamma_2=0$ for a Gaussian distribution.

The isotropic pressure, $\tau\propto B_iB^i$ of a Gaussian magnetic field is expected to follow a $\chi^2$ distribution with three degrees of freedom. The skewness and kurtosis of a $\chi^2$ distribution with $p=3$ degrees of freedom are
\be
  \gamma_1=\sqrt{\frac{8}{p}}\approx 1.633, \qquad \gamma_2=\frac{12}{p}=4 .
\ee
In BC05 and B06 we confirmed the $\chi^2$ nature of the isotropic pressure employing twenty realisations at $l_\mathrm{dim}=192$, finding $\gamma_1^\tau=1.63\pm 0.01$ and $\gamma_2^\tau=3.99\pm 0.05$ for $n_B=0$. The anisotropic stress is harder to characterise but we argued that the distribution function should be composed of a mixture of $\chi^2$ and modified Bessel distributions and should therefore have a relatively small skewness. We found numerically that the statistics are dependent on the spectral index, with $\gamma_1^{\tau_S}=-0.24\pm 0.003$ and $\gamma_2^{\tau_S}=1.10\pm 0.01$ for $n_B=0$, and $\gamma_1^{\tau_S}=0.38\pm 0.01$ and $\gamma_2^{\tau_S}=0.86\pm 0.02$ for $n_B=-2.9$.

\begin{figure}
\centering
\includegraphics[width=0.45\textwidth]{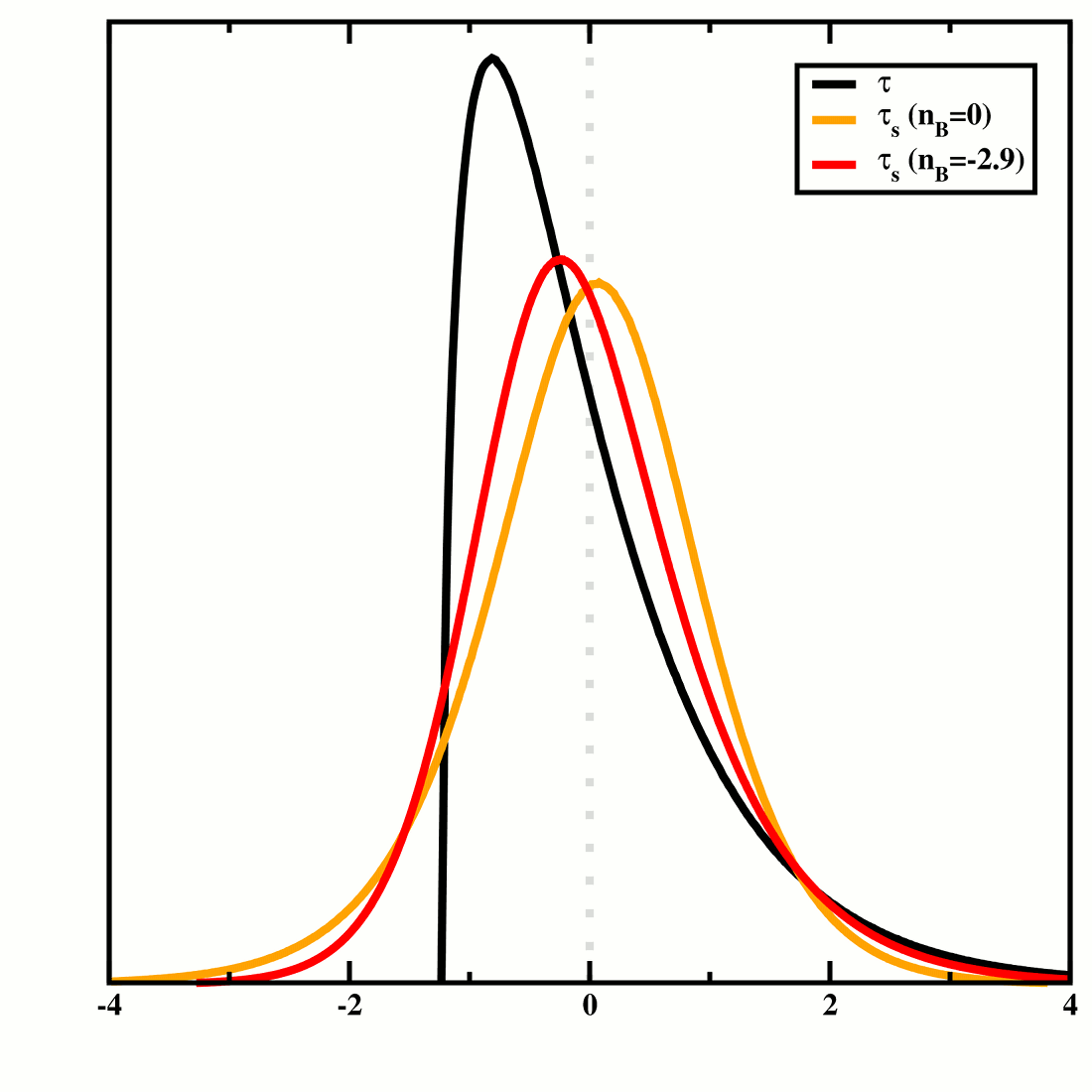}\includegraphics[width=0.45\textwidth]{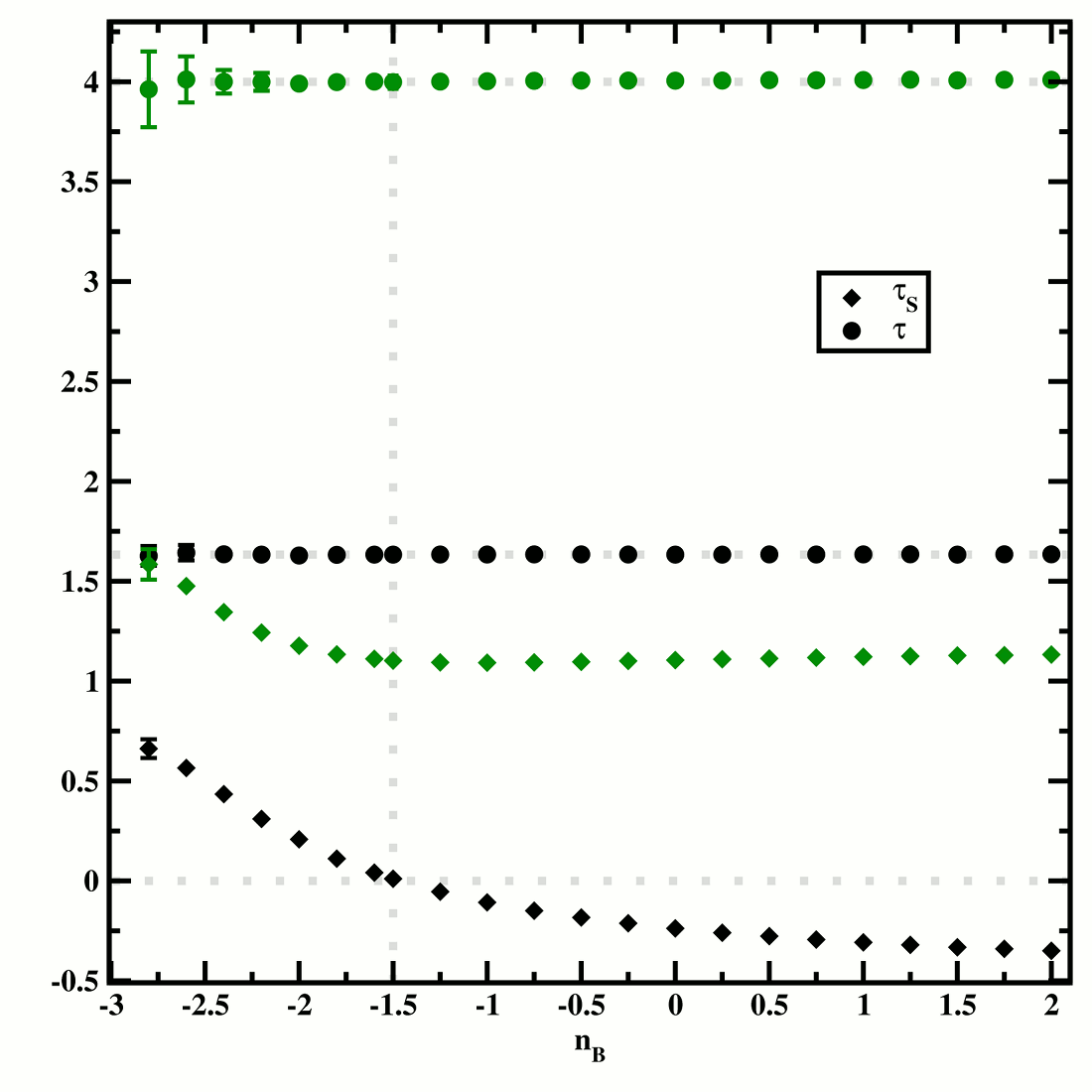}
\caption{Left: Probability distribution functions for the isotropic and anisotropic pressures. The $x$-axis is in units of the RMS amplitude of the field. Right: Skewnesses (black) and kurtoses (green) of the scalar pressures for $n_B\in (-3,2]$. In the interests of clarity, negligible error bars have been suppressed on both plots.}
\label{Figure-1Points}
\end{figure}

However, we did not probe this dependence in detail and we briefly consider it here. For $n_B\geq -3/2$ we employ twenty realisations at $l_\mathrm{dim}=256$ while for $n_B<-3/2$ we use twenty realisations at $l_\mathrm{dim}=512$. In the left panel of Figure \ref{Figure-1Points} we plot the distribution functions for $\tau$ and $\tau_S$. The isotropic pressure is invariant under changing $n_B$ to a good degree and we present it only for $n_B=0$, while we plot $\tau_S$ for both $n_B=0$ and $n_B=-2.9$. The $\chi^2$ nature of the isotropic pressure is very clear, as is the change in the skewness of the anisotropic pressure between the two cases.

In the right panel of Figure \ref{Figure-1Points} we plot for a wide range of $n_B$ the skewnesses (black) and kurtoses (green) of the scalar pressures. The statistics of the isotropic pressure are invariant across the range of $n_B$, with $\gamma_1^\tau\approx 1.633$ and $\gamma_2^\tau\approx 4$. The errors grow more significant and the mean of $\gamma_2^\tau$ reduces as $n_B\rightarrow -3$ due to the lack of modes with appreciable power in the most steeply-tilted cases. Explicitly, at $n_B=0$, $\gamma_1^\tau=1.634\pm 0.004$ and $\gamma_2^\tau=4.011\pm 0.031$, while at $n_B=-2.9$, $\gamma_1^\tau=1.620\pm 0.084$ and $\gamma_2^\tau=3.932\pm 0.284$.

The anisotropic pressure behaves quite differently. The kurtosis is invariant at $\gamma_2^{\tau_S}\approx 1.1$ for $n_B\in[-3/2,2]$, but after a transition at $n_B=-3/2$ the kurtosis climbs towards $\gamma_2^{\tau_S}\lesssim 2$ for $n_B\rightarrow -3$. The skewness depends even more strongly on $n_B$, rising monotonically for all $n_B$ as $n_B$ decreases. At $n_B=-3/2$ the skewness passes through zero. The magnetic field with $n_B=-3/2$ is therefore highlighted as a field of particular interest; for this field, the statistics of the anisotropic stress change their behaviour. Explicitly, at $n_B=0$, $\gamma_1^{\tau_S}=-0.238\pm 0.004$ and $\gamma_2^{\tau_S}=1.108\pm 0.011$, and at $n_B=-2.9$, $\gamma_1^{\tau_S}=0.697\pm 0.067$ and $\gamma_2^{\tau_S}=1.619\pm 0.105$.

Our findings for the isotropic pressure and the anisotropic pressure at $n_B=0$ are in full agreement with those in BC05 and B06. However, the statistics of the anisotropic pressure at $n_B=-2.9$ reveal a tension. This is an indication of a strong grid dependence as one tends $n_B$ towards scale-invariance. Having repeated the analysis with varying grid-sizes across a range of spectral indices we can determine that increasing the grid resolution tends to increase the skewness and kurtosis of the anisotropic pressure. While this is still the case for $l_\mathrm{dim}=512$, it is to be expected that the true results are close to, but slightly larger, than those presented in this paper.

Perhaps the most important result from this section then still holds -- that the field with $n_B=-3/2$ is selected as one of special interest and that at this point the skewness changes sign. This field also remains within observational bounds \citep{Yamazaki:2010nf,Paoletti:2010rx}. The other point of interest we can note is that, as $n_B\rightarrow -3$ as seems natural, the skewness of the anisotropic pressure tends towards $\gamma_1^{\tau_S}\lesssim 1$ and the kurtosis tends towards $\gamma_2^{\tau_S}\lesssim 2$. A firmer statement would require further analytical study.

\section{Intrinsic Power Spectra}
\label{2Points}
In BC05 and B06 we derived expressions for the correlations between components of the magnetic stress tensor, which take the form of an integral across two power spectra.\footnote{In contradiction to a statement in \citep{Yamazaki:2008gr}, these expressions were exact and did not neglect any terms. See Appendix \ref{Appendix-Yamazaki} for further details.} However, due to the presence of poles in the integrations for $n_B<-3/2$ we only integrated these equations numerically for $n_B=0$ and relied on statistical realisations for $n_B<-3/2$. These results were then compromised by limits on the dynamic range of the realisations and could neither be extended into the low-$k$ region nor to $k>k_c$. In this section we improve our analysis, rederiving and extending work by \citep{Yamazaki:2008gr,Paoletti:2008ck}.

In BC05 and B06 we pointed out that at the two-point level magnetic fields can be separated into ``ultra-violet'' and ``infra-red'' fields. Ultra-violet fields are those with $n_B\geq -3/2$; the stress spectra are dominated by the damping scale, and are white-noise on large scales. Infra-red fields have $n_B<-3/2$ and on large scales the stresses obey a power-law spectrum. The field with $n_B=-3/2$ is therefore also of interest for two-point moments. In \citep{Finelli:2008xh,Paoletti:2008ck} analytic solutions were found for $n_B\in\{3,2,1,0,-1,-3/2,-5/2\}$. These solutions encompass the fields we have highlighted as of most interest: the minimally causal case $n_B=2$, the white noise field $n_B=0$, the ``transitionary'' field $n_B=-3/2$, and an inflationary field with $n_B=-5/2$.

However, further analytic solutions can be found, and the integrations can be re-expressed in a form more explicitly amenable to numerical integration. Our approach to these allows us to consider a wide range of power spectra, not necessarily power-law; and, most importantly, the techniques presented in this section are readily adaptable to the study of magnetic bispectra, the integrands of which are significantly more complicated than those for the power spectra, and which are difficult to simulate statistically (see B06). In this section the amplitude of the power spectrum is assumed to be $A_B\equiv 1$ unless otherwise stated.

If $\tau_1$ and $\tau_2$ denote two components of the stress tensor (isotropic pressure, anisotropic pressure, vorticity or TT tensor), then the equal-time correlation between the two for a Gaussian magnetic field is \citep{Brown:2005kr,Brown:2006wv,Yamazaki:2008gr,Paoletti:2008ck,Caprini:2009vk}
\be
\label{RawSpectrum}
  \av{\tau_1(\mathbf{k})\tau_2^*(\mathbf{p})}=\mathcal{P}_{12}(k)\delta(\mathbf{k-p})=\delta(\mathbf{k-p})\int \mathcal{P}_B(k')\mathcal{P}_B\left({\left|\mathbf{k-k'}\right|}\right)\mathcal{F}_{12}(\mathbf{k},\mathbf{k'})\mathrm{d}^3\mathbf{k'} .
\ee
Three angles determine the relations between the wavevectors $\mathbf{k}$ and $\mathbf{k}'$,
\be
\label{RawSpectraAngles}
  \gamma=\hat{\mathbf{k}}\cdot\hat{\mathbf{k}'}, \qquad
  \mu=\hat{\mathbf{k}'}\cdot\widehat{\mathbf{k-k}'}, \qquad
  \beta=\hat{\mathbf{k}}\cdot\widehat{\mathbf{k-k}'}
\ee
and the angular terms in the integrations are given by
\bea
\centering
  \mathcal{F}_{\tau\tau}&=&\frac{1}{2}\left(1+\mu^2\right), \\
  \mathcal{F}_{\tau_S\tau_S}&=&\frac{1}{2}\left(4+\mu^2-3\left(\gamma^2+\beta^2\right)-6\gamma\mu\beta+9\gamma^2\beta^2\right), \\
  \mathcal{F}_{\tau\tau_S}&=&\frac{1}{2}\left(\mu^2+3\left(\gamma^2+\beta^2\right)-3\gamma\mu\beta-2\right), \\
  \mathcal{F}_{\tau^V\tau^V}&=&1-2\gamma^2\beta^2+\gamma\mu\beta, \\
  \mathcal{F}_{\tau^T\tau^T}&=&\left(1+\gamma^2\right)\left(1+\mu^2\right) .
\eea
Solving this integral is problematic due to the poles that can appear in $\mathcal{P}_B(k')$, $\mathcal{P}_B(|\mathbf{k-k}'|)$ and $\mathcal{F}_{12}$. To control this it is useful to express $k'$, $k$ and $p$ in units of the cut-off scale $k_c$, writing $k'=ak_c$ to ease the notation, and change variables such that we integrate across $y=\left|\mathbf{k-a}\right|^2$. This ensures that the poles in $\mathcal{P}(\left|\mathbf{k-a}\right|)$ can appear only at the lower limit of integration and are easier to control numerically.\footnote{This change of variables is similar to that employed in, for example, \citet{Ananda:2006af} where the focus was on the gravitational waves produced by first-order scalar perturbations. Such a change of variables might also be useful in studies of inflationary bispectra where again integrals across two or more power spectra must be taken.} Equation (\ref{RawSpectrum}) can then be recast as
\be
  \mathcal{P}_{12}(k)=\frac{\pi k_c^3}{k}\int_{a=0}^1a\mathcal{P}_B(a)\int_{y=(k-a)^2}^{(k+a)^2}\mathcal{P}_B(\sqrt{y})\mathcal{F}_{12}(k,a,y)\dy\da .
\ee
In terms of these units, the angles defined in (\ref{RawSpectraAngles}) are given by
\be
  \gamma=\frac{k^2+a^2-y}{2ak}, \qquad
  \mu=\frac{k\gamma-a}{\sqrt{y}}=\frac{k^2-a^2-y}{2a\sqrt{y}}, \qquad
  \beta=\frac{k-a\gamma}{\sqrt{y}}=\frac{k^2-a^2+y}{2k\sqrt{y}} .
\ee
For convenience, we define the scaled spectrum
\be
  Q_{12}(k)=\frac{k}{\pi k_c^3}\mathcal{P}_{12}(k) .
\ee
The power spectrum $\mathcal{P}_B(k)$ vanishes for arguments less than zero or greater than unity, which restricts the integration limits to particular values. These vary depending on the value of $k$:
\begin{itemize}
\item For $k\geq 2$, the lower limit on the integral across $y$ is always greater than unity, and the integral vanishes. Therefore $Q_{12}(k\geq 2)\equiv 0$, as has previously been found numerically (B05 and BC06, \citet{Yamazaki:2008gr}), and analytically \citep{Paoletti:2008ck,Caprini:2009vk} and is expected from the quadratic nature of the stress tensor.

\item For $k\in[1,2)$, the lower limit can be less than unity, and so there is a non-vanishing contribution to the power spectrum. This contribution decreases as $k\rightarrow 2$ and the range of integration vanishes. If $a>k-1$ then $(k-a)^2<1$; however, $(k+a)^2>1$ at all times. Therefore
\be
\label{Spectra1}
  Q_{12}(k\in[1,2))=\int_{k-1}^1a\mathcal{P}_B(a)\int_{(k-a)^2}^1\mathcal{P}_B(\sqrt{y})\mathcal{F}_{12}(k,a,y)\dy\da
\ee
and this tends smoothly to zero as $k$ tends to twice the damping scale.

\item For $k\in(0,1)$, the integral across $a$ can be separated into the regions $a\in[0,1-k]$ and $a\in[1-k,1]$. In the former region, $(k-a)^2$ and $(k+a)^2$ are both always less than unity, while in the latter $(k+a)^2\geq 1$. The integral is therefore
\be
\label{Spectra2}
Q_{12}(k\in(0,1))=\int_0^{1-k}a\mathcal{P}(a)\int_{(k-a)^2}^{(k+a)^2}\mathcal{P}(\sqrt{y})\mathcal{F}_{12}(k,a,y)\dy\da
   +\int_{1-k}^1a\mathcal{P}(a)\int_{(k-a)^2}^1\mathcal{P}(\sqrt{y})\mathcal{F}_{12}(k,a,y)\dy\da .
\ee
In the interests of numerical stability, this integral could be further separated depending on whether $a<k$ or $a>k$.

\item Finally, for $k=0$ the domain of the integration across $y$ vanishes, and $Q_{12}(k=0)\equiv 0$.
\end{itemize}

We can solve these integrals analytically for all integer and half-integer $n_B\in[-5/2,3]$, adding $n_B\in\{5/2,3/2,1/2,-1/2,-2\}$ to the solutions given in \citet{Paoletti:2008ck}, and verify these results with numerical integrations. Most importantly, this means that we now possess analytic solutions for both $n_B=-2$ and $n_B=-5/2$, which are infra-red and close to current observational bounds. Full analytic solutions not given in \cite{Paoletti:2008ck} are presented in Appendix \ref{Appendix-SpectraSolutions}. Numerical integration was performed with Monte-Carlo routines, using with 4,000$-$20,000 samples per integration, depending on $n_B$.\footnote{We employ the Numerical Recipes MISER routine. The complete set of solutions can be found in our notation in an online version of this paper found at the author's website {\tt http://folk.uio.no/ibrown}.}

\begin{figure}
\centering
\includegraphics[width=0.45\textwidth]{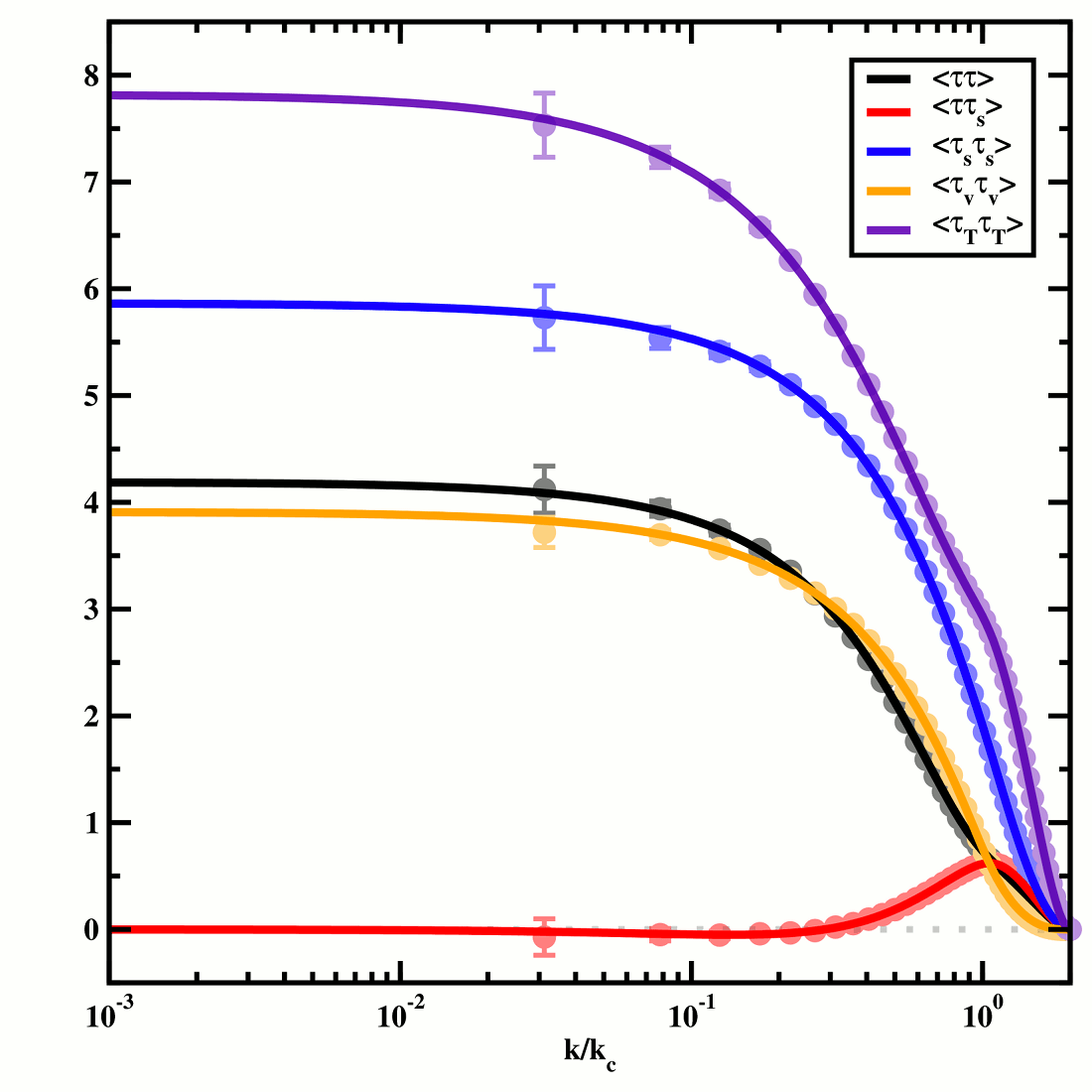}\includegraphics[width=0.45\textwidth]{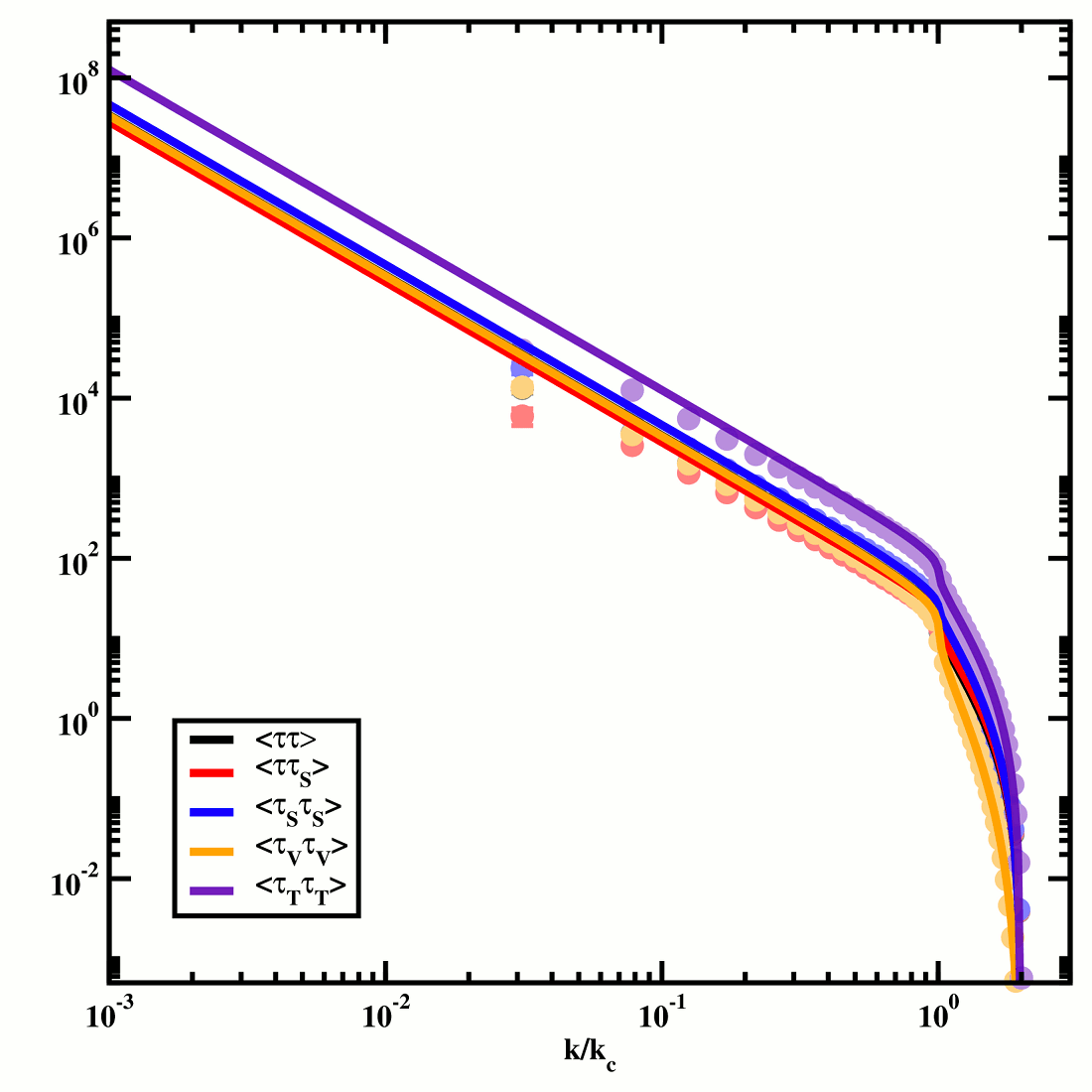}
\caption{The stress spectra for a white noise magnetic field with $n_B=0$ (left) and an inflationary field with $n_B=-5/2$ (right). The infra-red damping of the realisations of the inflationary field at $k<k_c$ is very apparent.}
\label{Figure-SpectraSolutions}
\end{figure}

In the left panel of Figure \ref{Figure-SpectraSolutions} we present the stress power spectra as found for $n_B=0$ from the analytic solution and by employing statistical realisations. The results from the numeric integration are indistinguishable from the analytical solutions. While in BC05 and B06 we employed twenty statistical realisations on a grid of size $l_\mathrm{dim}=192$, here we present instead the results from averaging one hundred realisations generated on a grid with $l_\mathrm{dim}=256$. The plots clearly demonstrate that the stress spectra are dominated by the damping scale -- at each $k$ both the gradient and the amplitude of the field are strongly dependant on it. Since this scale evolves in time, this implies that the statistics imprinted by a cosmological magnetic field vary across time. We refer to this as ``decoherence'' and it implies that studying the imprint of such fields on the CMB may require careful study, as for defect mechanisms (see for example \citet{Bevis:2006mj}). This behaviour holds for all $n_B>-3/2$. For $k<k_\mathrm{Coh}$ where $k_\mathrm{Coh}\approx k_c/100$, the stress spectra tend towards the white noise solutions discussed in the earlier literature (e.g. \citep{Caprini:2001nb,Mack:2001gc}). If the viscous damping is on a small enough scale these fields can be treated as white noise and the CMB signals recovered in the usual manner. However, this should be carefully verified to avoid wrongly estimating the impact on the CMB. In contrast to the auto-correlations, for $k<k_\mathrm{Coh}$ the cross correlation $\PTrTs$ tends to vanish and is only significant at $k\approx k_c$.

In the right panel we present the power spectra that are produced by an inflationary field with $n_B=-5/2$ -- this could be compared with Figure 4 of BC05. In BC05 and B06 we employed realisations at $l_\mathrm{dim}=192$. Furthermore, the damping scale was set at $k_c=l_\mathrm{dim}$ and we relied on the steeply tilted power spectrum to control the aliasing of power from large to small scales. This limited us to the study only of $k\ll k_c$. Both of these issues are addressed here; we use $l_\mathrm{dim}=256$ and set $k_c=l_\mathrm{dim}/4$ and employ the average of 100 realisations to reduce the errors further. Agreement with the analytic solutions for $k\geq k_c$ is extremely good. However, for $k<k_c$ we remain compromised by an infra-red damping arising from the infra-red cut-off associated with the finite grid size.

In contrast to the ultra-violet case, the gradients of the spectra produced by an infra-red field become independent of the damping scale at $k<k_\mathrm{Coh}$ where $k_\mathrm{Coh}=k_c$. They also scale as $\mathcal{P}_{12}(k)\propto k^{2n_B+3}$, in agreement with previous papers. The statistics imprinted by such a field on relatively large scales therefore do not evolve through time, and the CMB signals can be estimated with some confidence. This behaviour holds for all fields with $n_B<-3/2$ and we refer to these as ``coherent'' fields. However, even for these fields the stress statistics become strongly dependant on the damping scale for $k>k_\mathrm{Coh}$. Since the damping scale is in general significantly smaller than the scales relevant for CMB analysis we can safely assume that the infra-red fields are ``coherent'' across the entire range of $k$. Only if small scales must be taken into account must more care be taken. In comparison to the case with $n_B=0$, for this field the cross-correlation between the scalar components obeys the same behaviour as the auto-correlations.

To briefly summarise, then, we can consider the stresses of a white noise, decoherent field (and, by extension, fields with $n_B>-3/2$) to be white noise on scales $k\lesssim k_c/100$. The stresses of a coherent field with $n_B=-5/2$ (and, by extension, fields with $n_B<-3/2$), become a pure power-law for $k<k_c$, a far weaker condition that is easily satisfied for much of the history of the universe.

As we demonstrate in the next section, only the largest scales are required to predict the CMB signals. On such scales the analytic solutions are significantly simplified. We expect ultra-violet fields to possess a well-defined Taylor series, with the zeroth-order term producing the white-noise regime. In contrast, we expect the infra-red fields to admit Laurent series with the leading-order contribution $\propto k^{2n_B+3}$. In Appendix \ref{Appendix-LargeScale} we present the large-scale expansions for all integer and half-integer $n_B\in[-5/2,3]$. It is noteworthy that, for spectral indices $n_B\geq-3/2$, the correlations to zeroth order are related to one another by
\be
\label{CausalRatios}
  \PTrTs(k)=0, \qquad
  \frac{\PTsTs(k)}{\PTrTr(k)}
    =\frac{3}{2}\frac{\PTvTv(k)}{\PTrTr(k)}
    =\frac{3}{4}\frac{\PTtTt(k)}{\PTrTr(k)}
    =\frac{7}{5},
\ee
as predicted for ``causal'' defect mechanisms in \citet{Turok:1997gj}. (Note that the usages of the word ``causal'' differ between this paper and that work; here a ``causal'' field is one with $n_B\geq 2$ which can be produced by physical processes in a matter or radiation universe and retains the analyticity of the power spectrum.)

\begin{figure}
\centering
\includegraphics[width=0.45\textwidth]{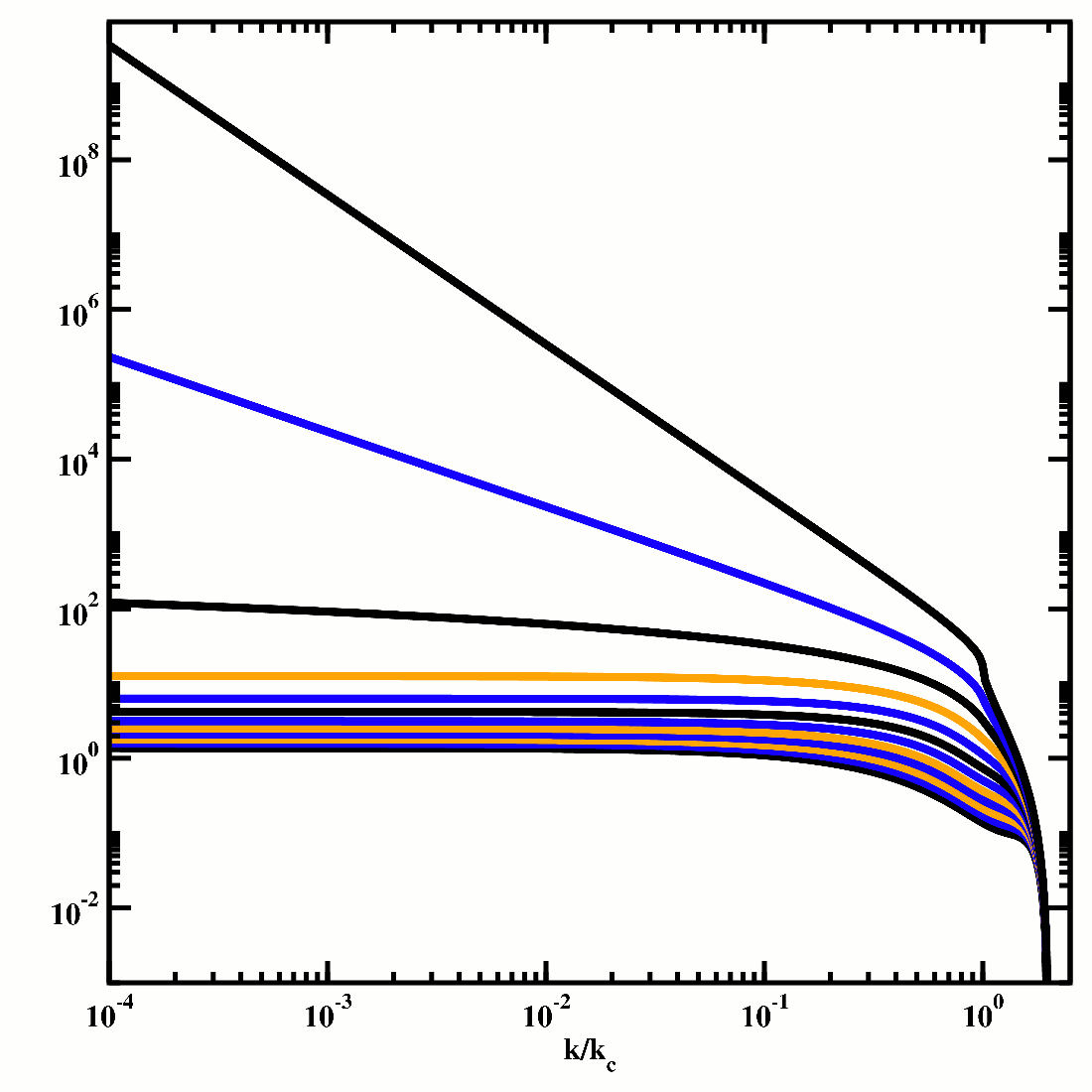}\includegraphics[width=0.45\textwidth]{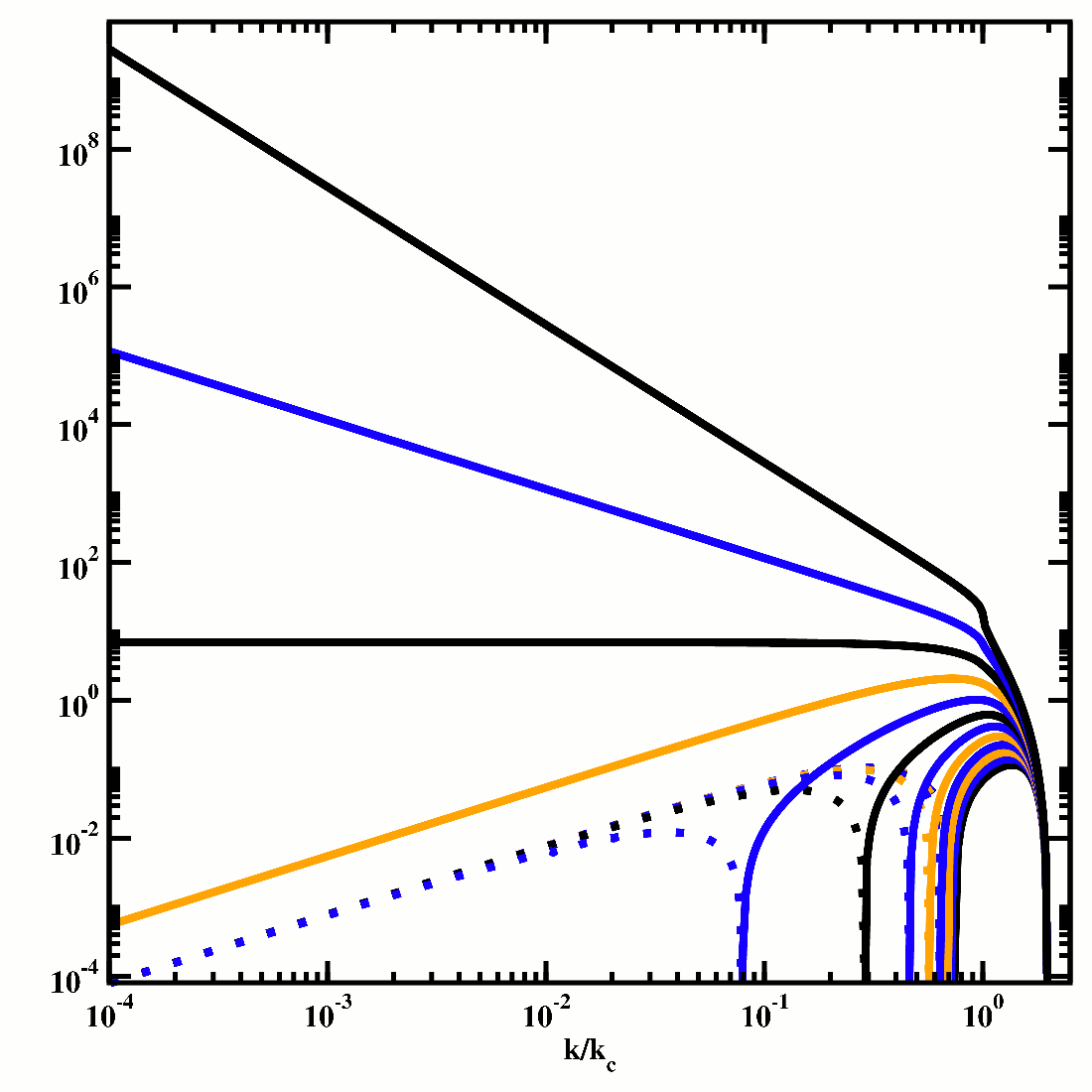}
\caption{$\PTrTr$ (left) and $\PTrTs$ (right) for integer and half-integer $n_B$ between $n_B=3$ (bottom, in black) and $n_B=-5/2$ (top, in black). Also highlighted in black are $n_B=0$ and $n_B=-3/2$.}
\label{Figure-TrTrTrTs}
\end{figure}

In Figure \ref{Figure-TrTrTrTs} we present $\PTrTr$ and $\PTrTs$ for $n_B\in[-5/2,3]$; the other auto-correlations follow the general behaviour of $\PTrTr$. Consider $\PTrTr$ first, plotted in the left panel. It is immediately noticeable that fields with $n_B>-3/2$ are decoherent, but below a ``coherence scale'' of $k_{\mathrm{Coh}}\approx k_c/100$ tend towards white noise. For $n_B<-3/2$, $k_\mathrm{Coh}=k_c$ and both cases for which we possess analytic solutions follow the expected scaling $\mathcal{P}_{AB}(k)\propto k^{2n_B+3}$. For $k>k_{\mathrm{Coh}}$ these fields are decoherent. At $n_B=-3/2$ itself the stresses diverge logarithmically with $k$.

The cross-correlation plotted in the right panel is of particular interest. In standard perturbation theory this cross-correlation is zero, and it is an interesting signal of a primordial magnetic field that it is in general non-vanishing. For $n_B>-3/2$ the leading-order term is the gradient; the signal is thus negligible on the largest scales compared to the auto-correlations. Interestingly, however, the leading-order contribution to the cross-correlation is invariant with respect to $n_B$ and is, moreover, an anti-correlation. This feature is clear in the figure; while the scale at which the signal changes from an anti-correlation to a correlation varies slowly with $n_B$ the behaviour on large scales remains
\be
  \PTrTs=-\frac{1}{4}A_B\pi k_c^3k .
\ee
This behaviour is broken at $n_B=-1$ where the cross-correlation becomes positive-definite and on large scales behaves as
\be
  \PTrTs=A_B\pi k_c^3k  .
\ee
The auto-correlations change their large-scale behaviour at $n_B=-3/2$, swapping from white-noise at $n_B>-3/2$ to a power law at $n_B<-3/2$, and diverge logarithmically at $n_B=-3/2$. In contrast, the cross-correlation is a power-law at both $n_B>-3/2$ and $n_B<-3/2$, while it is white-noise at $n_B=-3/2$. Furthermore, for $n_B<-3/2$, the cross-correlation rapidly becomes of equivalent magnitude to the auto-correlations. If $n_B\approx -5/2$, the existence of a non-vanishing scalar/traceless scalar correlation on the CMB is then an inevitable and characteristic consequence of a magnetic field.

We now turn to two toy models with the power spectra
\be
\label{DampedCausal}
  \mathcal{P}_B(k)=A_Bk^2\exp(-\xi k^2)H(1-k)
\ee
with $\xi\gtrsim 20$, corresponding to a ``causal'' magnetic field with a realistic damping tail -- this can be compared for example with the power spectrum presented in \citet{Matarrese:2004kq} -- and
\be
\label{IRControlled}
  \mathcal{P}_B(k)=\frac{2}{\pi}A_B\tan^{-1}\left(\xi k^6\right)k^{-5/2}H(1-k)
\ee
with $\xi\gtrsim 10^{30}$, which approximates a spectrum with $n_B=-5/2$ for much of the range of $k$ but is controlled on infra-red scales $k\lesssim 10^{-4}$. The damped causal spectrum will qualitatively resemble that for a magnetic field produced by physical processes in the post-inflationary universe -- whether produced by induced currents, at a phase transition or through some other process. While the IR-controlled spectrum could be motivated by a magnetogenesis model producing fields on an extremely small scale before or during an inflationary epoch, more concretely this spectrum models a realisation of an inflationary field generated on a grid, with the infra-red damping reproducing the grid cut-off.

Let us first consider the damped causal field, the stresses of which we expect to by similar to those of a power-law with $n_B=2$. Since $k_c$ remains a hard cut-off but the field is exponentially damped on a larger scale, we should see the gross features of the standard causal field move to smaller wavenumbers. We can solve equations (\ref{Spectra1}-\ref{Spectra2}) analytically and the general forms are presented in the appendices and plotted in the left panel of Figure \ref{Figure-SpectraAlternatives}. The Taylor series around $k=0$ gives us the behaviour on large scales:
\be
\label{DampedCausalStresses}
\begin{array}{c}
  \PTrTr=\pi A_Bk_c^3\left(T_0-\frac{1}{e^{2\xi}}k\right), \qquad
  \PTrTs=\pi A_Bk_c^3\left(-\frac{1}{4e^{2\xi}}k\right), \\
  \PTsTs=\pi A_Bk_c^3\left(\frac{7}{5}T_0-\frac{1}{e^{2\xi}}k\right), \qquad
  \PTvTv=\pi A_Bk_c^3\left(\frac{14}{15}T_0-\frac{5}{6e^{2\xi}}k\right), \\
  \PTtTt=\pi A_Bk_c^3\left(\frac{28}{15}T_0-\frac{7}{3e^{2\xi}}k\right)
\end{array} .
\ee
Here $\erf(x)$ is the error function and
\be
  T_0=\frac{1}{64}\frac{15\sqrt{2\pi}\erf(\sqrt{2\xi})\xi^3e^{2\xi}-60\xi^{7/2}-80\xi^{9/2}-64\xi^{11/2}}{\xi^{13/2}e^{2\xi}} .
\ee
The stress spectra for a damped causal field retain the ``causal'' relationships given in equation (\ref{CausalRatios}). Across the full range of $k$, the damped causal case qualitatively closely resembles the causal case; however, features seen in the causal case such as the flattening of the gradients at $k\approx k_c$ are suppressed and there is a distinct loss of power on large scales. The behaviour of the scalar cross-correlation is particularly interesting. On extremely large scales the cross-correlation is an anti-correlation growing linearly with $k$, as for the standard causal case. This linear term is, however, strongly suppressed with respect to the quadratic term for $k\gtrsim 10^{-13}$, above which the quadratic order dominates. Across almost the entire range of $k$, then, the cross-correlation is positive and growing quadratically. The level of suppression is dependant on $\xi$ and in principle the behaviour of the cross-correlation on large scales would allow us to clearly distinguish between the causal and damped causal cases.

The IR-controlled field is conceptually simpler than the damped causal field, since the modifications are on very large scales rather than on scales comparable to $k_c$. We would expect the stress power spectra for such a field to closely resemble the standard $n_B=-5/2$ case closely above a certain wavenumber $k_S$, but to decay on larger scales. Results from numerical integration are presented along with those for the standard case in the right panel of Figure \ref{Figure-SpectraAlternatives}; this spectrum is unfortunately not amenable to analytical integration. As expected, the small-scale results are almost exactly equivalent, although there is a consistent suppression of power for $k<k_c$. On scales larger than a transition scale $k_S$, however, the spectra flatten significantly compared to the undamped case. The most dramatic deviation from the previous behaviour is again in the scalar cross-correlation, which decays rapidly ($\PTrTs\sim k^{3}$) as $k\rightarrow 0$. In any magnetogenesis scenario that produces an infra-red cut-off, then, we can expect large changes in the behaviour of the scalar cross-correlation. The auto-correlations are strongly damped but remain non-negligible. For this particular spectrum, the transition wavenumber is $k_S\approx 10^{-3}$, but this will be strongly dependant on the form of the damping.

Comparison of the right panel of Figure \ref{Figure-SpectraAlternatives} with the realisations plotted in the right panel of Figure \ref{Figure-SpectraSolutions} shows a good qualitative agreement in the behaviour of the stresses. The damping effect is significantly more dramatic in the case of the realisations since the largest mode we can produce is only of order $\alpha/l_\mathrm{dim}\approx \alpha\cdot 10^{-3}$ where $\alpha>4$ to avoid aliasing; if required this could be modelled taking a smaller $\xi$ in equation (\ref{IRControlled}) or enforcing a hard cut-off at $k\approx 10^{-2}$.

We close this section by briefly discussing the advantages and disadvantages of the three approaches we have taken. We have principally relied on analytical solutions. These have the obvious advantages of being exact and, in particular, allowing us to recover the large-scale behaviour directly from a Laurent or Taylor series. In an ideal situation, one would always use analytic solutions and we do so whenever possible. However, analysis is extremely limited. In this and previous studies solutions have only been found for magnetic power law spectra with integer and half-integer exponent $n_B$, which lead to integer powers of $k$ on large scales; here we have added to this a spectrum composed of an exponential and a power law. For the most realistic case of a field with $n_B\approx -5/2$ this limits us to the two cases $n_B=-2$ and $n_B=-5/2$, which is not satisfactorary. Furthermore, since equations (\ref{Spectra1}-\ref{Spectra2}) have been derived employing Wick's theorem, the underlying magnetic fields must be Gaussian, which in most magnetogenesis scenarios is not to be guaranteed.

The other tool we have extensively employed is the numerical integration of equations (\ref{Spectra1}-\ref{Spectra2}). These are relatively rapid, and even more so since -- as we demonstrate in the next section -- for purposes of CMB analysis we generally need only recover the region where $k\ll k_c$. The approach that we have taken to the numerical integration helps control the various poles that can otherwise be problematic, and the agreement with the analytics is so good as to be indistinguishable. Furthermore, in a numerical integration we are not limited by the shape of the power spectrum, although for steeply-tilted spectra (see for example the case with $n_B=-2.9$ in the next section) we require an increasingly large number of samples in the integration to retain convergence. Numerical integration is therefore currently the most realistically flexible approach. Unfortunately, since it is based on equations (\ref{Spectra1}-\ref{Spectra2}) we are still limited to considering Gaussian magnetic fields.

The most general approach we have taken is to generate realisations of magnetic fields with a particular power spectrum and to construct and analyse the corresponding stress tensor. The advantages are clear: this approach is entirely general. Neither the form of the power spectrum nor (in principle, although we have focussed on Gaussian fields) the statistics are at all constrained. However, the disadvantages are also clear. This approach is slow, involving numerous operations spanning the entire grid and frequent Fourier transforms. More seriously, results can depend strongly on the grid resolution. For ultra-violet power-law fields we can acheive reasonable accuracy, although the minimum wavenumber we can consider is limited by the grid resolution. However, for a power spectrum with a red tilt -- with respect to the white noise case $\mathcal{P}_B(k)=\mathrm{const}.$ -- there is a strong infra-red damping for all $k<k_c$. This damping is lessened if we increase the resolution of the grid but this correspondingly slows the calculation further. It is possible to improve the results on large scales by employing a damping scale at $k_c\geq l_\mathrm{dim}$; due to the nature of the algorithms involved, power on smaller scales is strongly contaminated by aliasing from large scale modes, but the large modes themselves are less affected. Since it is precisely these scales that are most relevant for CMB studies, in cases where analytic or numerical solutions are lacking -- when the fields are non-Gaussian, for example -- this would be a crude method of finding the relevant behaviour of the spectra without running realisations on grids of size $l_\mathrm{dim}\geq 1,024$, but it should be only be employed with caution.

These issues only worsen when one considers instead the bispectra of the stress tensor. Since the wavenumbers must form a closed triangle, on large scales we have only a small set of modes on the grid contributing to each configuration. While for power spectra we can recover reasonable results with $\mathcal{O}(10)$ realisations, for the bispectra we are required to consider $\mathcal{O}(10^3)$ realisations, as in B06. Worse, as a bispectrum is effectively a two-dimensional structure, the computation is significantly slower than for the effectively one-dimensional power spectrum. The need to employ such a large number of realisations in an unavoidably lengthy algorithm makes generating bispectra from realisations an intensive task. See B06 for further details, where we considered the one-dimensional colinear case; even for this simple case we required $\sim 1,500$ realisations.

Ideally, we would always employ analytical solutions. However, where this is not possible the most realistic approach remains numerical integration of equations (\ref{Spectra1}-\ref{Spectra2}). Although this currently restricts us to Gaussian magnetic fields this is not necessarily problematic where a concrete magnetogenesis mechanism predicting a specific statistical nature is lacking. In the remainder of this paper we will neglect statistical realisations and focus where possible on analytical solutions, although we employ numerical solutions in cases where these are lacking.

\section{CMB Angular Power Spectra}
In this section we employ an approximation to the magnetised tensor transfer function to recover the CMB angular power spectrum induced by magnetised tensor perturbations in a general manner which will be directly applicable to the CMB angular bispectrum. This includes evaluations of the signals expected from the damped causal and the IR-controlled fields introduced in the previous section, but our focus is on the techniques and preperation for bispectrum evaluations rather than necessarily on the CMB angular power spectrum itself. However, the same approach could be employed using numerical transfer functions (as in \citep{Lewis:2004ef,Kojima:2009ms}) recovered from a magnetised Boltzmann code. While we consider only the temperature angular power spectrum arising from tensor perturbations, similar arguments apply to the polarisation and $\av{TE}$ angular power spectra, and to scalar and vector modes.

\subsection{General Considerations}
\label{CMB-General}
The CMB temperature angular power spectrum can be found by integrating across the transfer functions,
\be
\label{GeneralMagnetisedCMB}
  C_l=\frac{\pi}{4}\int_k\mathcal{P}(k)\left|\Delta_{Tl}(k,\eta_0)\right|^2k^2dk,
\ee
where in contrast to the normal definition we have taken the primordial power spectrum to be
\be
  \av{\xi(\mathbf{k})\xi^*(\mathbf{p})}=\mathcal{P}(k)(2\pi)^3\delta(\mathbf{k-p}) .
\ee
Here $\xi(\mathbf{k})$ is a random variable of unit variance that characterises the primordial statistics of the perturbations. The transfer function for tensor modes is given \citep{Seljak:1996is,Brown:2006wv} by
\be
  \Delta_{Tl}^{(T)}(k,\eta_0)=\sqrt{\frac{(l+2)!}{(l-2)!}}\int_{\eta=0}^{\eta_0}S_T^{(T)}(k,\eta)\frac{j_l(k(\eta_0-\eta))}{k^2(\eta_0-\eta)^2}\mathrm{d}\eta
\ee
where
\be
  S_T^{(T)}=-2\dot{h}^{(T)}+g\Phi^{(T)}
\ee
is a source term, $g$ is the visibility function and $\Phi^{(T)}$ depends on the lower moments of the tensor Boltzmann hierarchy -- see for example B06 for further details.

Neglecting the photon and neutrino anisotropic stresses, the tensor perturbation evolves as
\be
  \ddot{h}+2\frac{\dot{a}}{a}\dot{h}+k^2h=\frac{16\pi G}{a^2}\tau^{(T)}
\ee
where an overdot denotes a derivative with respect to conformal time. After matter/radiation equality the source then rapidly becomes negligible. It can then be shown \citep{Kahniashvili:2000vm,Mack:2001gc} that a good approximation for $\dot{h}$ in matter domination is
\be
  \dot{h}(k,\eta)\approx 32\pi Gk\eta_0^2z_\mathrm{eq}\ln\left(\frac{z_\mathrm{in}}{z_\mathrm{eq}}\right)\frac{j_l(k\eta)}{k\eta}
\ee
and then that, assuming instantaneous recombination, the tensor transfer function at the present day is approximately
\be
\label{ApproxDeltaTl}
  \Delta_{Tl}^{(T)}(k,\eta_0)=\mathcal{D}\sqrt{\frac{l(l+2)!}{(l-2)!}}\frac{J_{l+3}(k\eta_0)}{k^3\eta_0^3}
\ee
with $J_l(x)$ a Bessel function of the first kind and $\mathcal{D}$ a constant. Employing this expression assumes that we are working on large scales and it should therefore only be trusted for relatively low multipoles.

At the present epoch the transfer function is extremely tilted to low $k$ and we are restricted to $k\lesssim 10^{-3}-10^{-4}\mathrm{Mpc}^{-1}$. In contrast, the damping scale relevant for tensor modes is $k_c(\eta_\mathrm{eq})$ which is of the order of a few to a few hundred inverse megaparsecs. The transfer function then restricts us to extremely large scales: for inflationary fields we are well into the power-law regime, while for ultra-violet fields the sources are close to white-noise. The damped causal field also produces a white noise source on such scales. In general, we expect the majority of field configurations to produce source spectra that on large scales act as
\be
\label{CMBScaling}
 \mathcal{P}_{TT}(k)=P_\star\left(\frac{k}{k_\star}\right)^\alpha
\ee
for some amplitude $P_\star$ and pivot scale $k_\star$. For ultra-violet fields (including the damped causal case) we have $\alpha=0$, while for infra-red fields $\alpha=2n_B+3$. In the case of fields for which we possess an analytic solution we can further identify $P_\star/k_\star^\alpha$ from the Laurent series (see Appendix \ref{Appendix-LargeScale}). In a more general case, given sample points from a numerical integration or the results of an ensemble of realisations, we can recover the scaling $\alpha$ and amplitude $P_\star$ around a pivot wavenumber $k_\star$, which we can employ in the CMB integration. The structure of the power spectrum on small scales is practically irrelevant so long as equation (\ref{CMBScaling}) holds while the transfer function is non-negligible, which we expect to be true for all magnetic power spectra that do not possess features in the deep infra-red.

In principle the integral in equation (\ref{GeneralMagnetisedCMB}) runs only to $k/k_c(\eta_\mathrm{eq})=1$ since above this scale the stress spectra become decoherent, regardless of $n_B$. However, since the transfer function is so steeply-tilted, the error introduced by integrating to infinity is negligible for any realistic $\mathcal{P}_{TT}(k)$. Employing the approximation (\ref{CMBScaling}) then gives
\be
  C_l^{(T)}=\frac{\pi\mathcal{D}^2}{4}\frac{P_\star}{k_\star^\alpha}\frac{l(l+2)!}{(l-2)!}\int_0^\infty k^{\alpha-4}J_{l+3}^2(k\eta_0)\dk .
\ee
This is a standard integral and the solution is
\be
\label{MagnetisedCMB}
  C_l=\frac{\sqrt{\pi}}{8}\frac{\mathcal{D}^2P_\star}{\left(k_\star\eta_0\right)^\alpha}\frac{1}{\eta_0^3}\frac{\Gamma\left(\frac{4-\alpha}{2}\right)}{\Gamma\left(\frac{5-\alpha}{2}\right)}\frac{l(l+2)!}{(l-2)!}\frac{\Gamma\left(\frac{2l+3+\alpha}{2}\right)}{\Gamma\left(\frac{2l+11-\alpha}{2}\right)} .
\ee
This is true for all systems in which the approximate form for $\Delta_{Tl}$ holds, and for which $\mathcal{P}_{TT}(k)$ obeys a simple power-law while $\Delta_{Tl}^2$ is non-negligible. Since in evaluating the numerical results in section \ref{2Points} we set $A_B\equiv 1$ and worked with a wavenumber in units of $k_c$, then if we recover a value $\tilde{P}_\star$ at a pivot scale $k_\star=\tilde{k}_\star k_c$ we have that
\be
  P_\star=\tilde{P}_\star A_Bk_c^3, \qquad \PTtTt=\left(A_Bk_c^{3-\alpha}\tilde{P}_\star\right)\left(\frac{k}{\tilde{k}_\star}\right)^\alpha .
\ee

We should emphasise that equation (\ref{MagnetisedCMB}) has been derived based on the form of $\Delta_{Tl}$, which equally applies when considering the magnetic bispectra \citep{Shiraishi:2010sm}. The magnetic bispectrum is a 3D quantity dependent on wavenumbers $\{k,p,q\}$ which form a closed triangle. We can parameterise this with the variables $\{k,r,\phi\}$ where $r=p/k$ and $\phi$ is the angle between $\mathbf{k}$ and $\mathbf{p}$. It has been shown that with $r=1$, for $\phi=0$ (the so-called ``colinear'' case \citep{Brown:2005kr,Brown:2006wv,Caprini:2009vk}), and for both $\phi=2\pi/3$ (the equilateral case) and $\phi\rightarrow \pi$ (the degenerate, or squeezed case) \citep{Seshadri:2009sy,Caprini:2009vk} the bispectra for $n_B<-1$ follow scaling laws in $k$ for $k<k_\mathrm{Coh}$. In forthcoming work we further demonstrate that the coherence scale for $\phi=0$ is $k_{\mathrm{Coh}}\approx k_c/2$; for greater values of $\phi$ we expect $k_\mathrm{Coh}\lesssim k_c/2$. The nature of $\Delta_{Tl}$ then implies that we can employ the same reasoning as we have for the power spectra and focus only on extremely large scales, sampling only as many points as necessary to reconstruct the behaviour of the intrinsic bispectrum $\mathcal{B}_{TTT}(k)$. While the benefits of doing so for the angular power spectra are arguable, the saving in computational time and complexity for the bispectra will be extremely significant. We explore these issues thoroughly in a follow-up paper.

\subsection{Results}
\label{CMB-Results}
Consider first ultra-violet fields, with $n_B>-3/2$. In this case  the power spectra are white noise on the largest scales and
\be
 \alpha=0, \qquad P_\star=\pi A_Bk_c^3f(n_B)
\ee
where $f(n_B)$ is the scaled amplitude of the zeroth-order term in the Taylor series. The CMB angular power spectrum (\ref{MagnetisedCMB}) is therefore
\be
 C_l=\frac{16\pi^3}{3}\mathcal{D}^2B_\lambda^2\left(\frac{k_c\lambda}{\eta_0}\right)^3\frac{\lambda^{n_B}f(n_B)}{\Gamma\left(\frac{n_B+3}{2}\right)}\frac{l(l+2)!}{(l-2)!}\frac{(2l+1)!!}{(2l+9)!!} .
\ee
Here $a!!=a(a-2)(a-4)\ldots$ is the double factoral function. In the limit of large $l$ the angular power spectrum tends towards
\be
 l(l+1)C_l\propto l^3
\ee
in agreement with earlier estimates \citep{Mack:2001gc}. In particular, consider the causal case with $n_B=2$ and the flat case with $n_B=0$. For $n_B=2$, $f(n_B)=16/15$ and the CMB angular power spectrum is
\be
\label{CMBCausal}
  C_l=\frac{1,024\pi^{5/2}}{135}\mathcal{D}^2B_\lambda^2\left(\frac{k_c\lambda}{\eta_0}\right)^3\lambda^2\frac{l(l+2)!}{(l-2)!}\frac{(2l+1)!!}{(2l+9)!!} .
\ee
For $n_B=0$, $f(n_B)=112/45$ and so
\be
  C_l=\frac{3,584\pi^{5/2}}{135}\mathcal{D}^2B_\lambda^2\left(\frac{k_c\lambda}{\eta_0}\right)^3\frac{l(l+2)!}{(l-2)!}\frac{(2l+1)!!}{(2l+9)!!}=\frac{7}{2\lambda^2}\left(\frac{k_c(\eta_\mathrm{eq},n_B=0)}{k_c(\eta_\mathrm{eq},n_B=2)}\right)^3C_l(n_B=2) .
\ee
Since the damping scale for $n_B=0$ is significantly greater than that for $n_B=2$ \citep{Mack:2001gc,Caprini:2009vk}, all other parameters being equal the signal from a flat field will be larger than that for the causal field.

The situation is different for the infra-red fields as $\alpha$ becomes $n_B$-dependent. For $n_B<-3/2$,
\be
 \alpha=2n_B+3, \qquad P_\star k_\star^{-\alpha}=\pi A_Bk_c^{3-\alpha}f(n_B)
\ee
and ultimately
\be
 C_l=\frac{\pi^{7/2}}{4}\mathcal{D}^2B_\lambda^2\left(\frac{k_c\lambda}{\eta_0}\right)^3\frac{\lambda^{n_B}f(n_B)}{(k_c\eta_0)^{2n_B+3}\Gamma\left(\frac{n_B+3}{2}\right)}\frac{\Gamma\left(\frac{1}{2}-n_B\right)}{\Gamma\left(1-n_B\right)}\frac{l(l+2)!}{(l-2)!}\frac{\Gamma(l+3+n_B)}{\Gamma(l+4-n_B)} .
\ee
This holds for all power-law spectra with $n_B<-3/2$ but since $n_B$ is not necessarily an integer or half-integer the dependence on $l$ is not necessarily straightforward.

There are two fields in this regime for which we have exact solutions. If $n_B=-2$ then $f(n_B)=9\pi^2/4$ and the CMB angular power spectrum takes the simple form
\be
  C_l=\frac{27\pi^{11/2}}{128}\mathcal{D}^2B_\lambda^2\frac{\lambda k_c^4}{\eta_0^2}\frac{l^2(l-1)}{(l+5)(l+4)(l+3)}
\ee
while for $n_B=-5/2$, we have $f(n_B)=3,008/75$ and
\be
  C_l=\frac{770,048\pi^3}{1,125\Gamma(1/4)}\mathcal{D}^2B_\lambda^2\frac{k_c^5\lambda^{1/2}}{\eta_0}\frac{l(l+2)!}{(l-2)!}\frac{(2l-1)!!}{(2l+11)!!} .
\ee
In the limit of large-$l$ these become
\be
  l(l+1)C_l\propto\left\{\begin{array}{rcl} l^2&,&n_B=-2 \\ l&,&n_B=-5/2\end{array}\right. ,
\ee
again in agreement with previous results \citep{Mack:2001gc}. Note that these results have been found with relative ease from a flexible approach.

We also have an exact solution for the damped causal field. The amplitude of the spectrum is given by
\be
  B_\lambda^2=\frac{A_B}{8\pi^2(\xi+\lambda^2)^{5/2}}\left(3\sqrt{\pi}\erf\left(k_c(\xi+\lambda^2)^{1/2}\right)-\left(6+4k_c^2\left(\xi+\lambda^2\right)\right)k_c(\xi+\lambda^2)^{1/2}e^{-k_c^2(\xi+\lambda^2)}\right)
\ee
which for reasonable parameters ($k_c\lambda\gtrsim 5$) is rapidly dominated by the error function giving
\be
  A_B\approx \frac{8\pi^{3/2}\left(\xi+\lambda^2\right)^{5/2}}{3}B_\lambda^2 .
\ee
From the stress spectrum in equation (\ref{DampedCausalStresses}), on large scales we have $\alpha=0$ and
\be
  P_\star=\pi A_Bk_c^3\frac{7}{240}\frac{15\sqrt{2\pi}\xi^3e^{2\xi}\erf(\sqrt{2\xi})-4(15\xi^{7/2}+20\xi^{9/2}+16\xi^{11/2})}{\xi^{13/2}e^{2\xi}} .
\ee
The CMB angular power spectrum is then
\be
C_l\approx\frac{28\pi^{5/2}}{135}\mathcal{D}^2B_\lambda^2\left(\xi+\lambda^2\right)^{5/2}\left(\frac{k_c}{\eta_0}\right)^3\frac{l(l+2)!}{(l-2)!}\frac{(2l+1)!!}{(2l+9)!!}
  \left(\frac{15\sqrt{2\pi}\xi^3e^{2\xi}\erf\left(\sqrt{2\xi}\right)-4\left(15\xi^{7/2}+20\xi^{9/2}+16\xi^{11/2}\right)}{\xi^{13/2}e^{2\xi}}\right) .
\ee
In the limit $\xi\rightarrow 0$ this reduces to the angular power spectrum for an undamped causal field (\ref{CMBCausal}), while for large $\xi$ it becomes
\be
  C_l\approx \frac{28\pi^3}{9}\frac{\sqrt{2}}{\xi}\mathcal{D}^2B_\lambda^2\left(\frac{k_c}{\eta_0}\right)^3\frac{l(l+2)!}{(l-2)!}\frac{(2l+1)!!}{(2l+9)!!}
   \approx \frac{1}{\xi\lambda^5}\left(\frac{k_c(\mathrm{damped})}{k_c(\mathrm{undamped})}\right)^3C_l(\mathrm{undamped}) .
\ee
As should be expected since, on the scales where $\Delta_{Tl}$ is non-negligible the stress spectrum for the damped case is similar to that for the undamped case, the CMB signals are similar between the two cases.

Consider the ``transitionary'' case with $n_B=-3/2$. For this field, the stress spectrum on large scales diverges logarithmically and we cannot necessarily employ a simple scaling approximation for $\PTtTt$. The amplitude of the power spectrum is given by equation (\ref{ApproxMagneticNormalisation}). Employing the approximation in equation (\ref{LogTT}) and the general expression in equation (\ref{GeneralMagnetisedCMB}) gives, after some manipulation,
\be
C_l\approx\frac{1,792\pi^3}{135\Gamma(3/4)}\mathcal{D}^2B_\lambda^2\left(\frac{\lambda^{1/2}k_c}{\eta_0}\right)^3\left(\ln(k_c\eta_0)-\Psi\left(l+\frac{1}{2}\right)
 +\ln(2)-\frac{1}{6}\frac{f(l)}{2l+1}\frac{(2l+1)!!}{(2l+9)!!}\right)\frac{l(l+2)!}{(l-2)!}\frac{(2l+1)!!}{(2l+9)!!}
\ee
where $f(l)=160l^5+2,576l^4+15,152l^3+48,618l+20,529$. Of the terms summed in the brackets, the most significant contribution come from the first two; for sensible parameters we always expect have $k_c\eta_0\gg 1$, while the term dependent on the polynomial $f(l)$ decreases monotonically and is only marginally significant even at $l\approx 2$. We can therefore approximate
\be
  C_l\approx\frac{1,792\pi^3}{135\Gamma(3/4)}\mathcal{D}^2B_\lambda^2\left(\lambda^{1/2}\frac{k_c}{\eta_0}\right)^3\left(\ln(k_c\eta_0)-\Psi\left(l+\frac{1}{2}\right)\right)\frac{l(l+2)!}{(l-2)!}\frac{(2l+1)!!}{(2l+9)!!}
\ee
although in Figure \ref{Figure-CMB} we retain the full generality. In these expressions $\Psi(x)=\mathrm{d}(\ln(\Gamma(x)))/\mathrm{d}x$ is the digamma function, which is potentially significant for larger $l\gtrsim 1000$ but negligible for larger angular scales. Note also that on the scales on which $\Psi(x)$ might dominate the approximations for the transfer function will no longer be valid. The impact from this field then closely resembles that for ultra-violet fields with $n_B>-3/2$, although the similarity grows less for increasing $l$. This is to be expected since the divergence at large scales in $\mathcal{P}_{TT}(k)$ is only logarithmic and to some extent could be approximated by white noise. The modifications to the CMB angular power spectrum at large scales tend to decrease the $l$-dependance of the signal; this is consistent with the general flattening of $l(l+1)C_l$ as $n_B$ reduces and grows more negative.

The magnetic spectrum for an ``inflationary'' field controlled on large scales, equation (\ref{IRControlled}), was chosen to act as an ultra-violet field on large scales but to mimic a field with $n_B=-5/2$ on smaller scales. This induces a stress spectrum split into two regimes, characterised by a change in the gradient on large scales. If the gradient is $\alpha_1$ on scales $k<k_S$ and $\alpha_2$ on scales $k>k_S$, then assuming $k_\star<k_S$,
\be
  \PTtTt(k)\approx\frac{P_\star}{k_\star^{\alpha_1}}\left\{\begin{array}{rl}k^{\alpha_1}&,k\in(0,k_S] \\ k_S^{\alpha_1-\alpha_2}k^{\alpha_2}&,k\in(k_S,k_c]\end{array}\right. .
\ee
From the numerical results shown in Figure \ref{Figure-SpectraAlternatives}, for the IR-controlled field we can take $\alpha_1=0$ and $\alpha_2=-2$, the deep infra-red acting as white noise while smaller scales obey the same law $\alpha_2=2n_B+3$ as the infra-red fields. Substituting this into the general expression for the magnetised CMB in equation (\ref{GeneralMagnetisedCMB}) and solving the integral yields
\bea
  C_l&=&\frac{\pi}{4}\mathcal{D}^2\frac{P_\star}{\eta_0^3}x_s^2\frac{l(l+2)!}{(l-2)!}\left[\frac{512}{15\pi}\frac{(2l-1)!!}{(2l+11)!!}
   +\frac{1}{32[(l+3)!]^2}\left(\frac{x_s}{2}\right)^{2l+1}\left(\frac{1}{(2l+3)}\mathcal{F}^{l+\frac{3}{2},l+\frac{7}{2}}_{2l+7,l+4,l+\frac{5}{2}}(-x_s^2)
\right.\right. \nonumber \\ && \left.\left.\qquad\qquad
   -\frac{1}{(2l+1)(l-1)}\mathcal{F}^{l+\frac{7}{2},l+\frac{1}{2}}_{2l+7,l+4,l+\frac{3}{2}}(-x_s^2)\right)
 \right]
\eea
where $\mathcal{F}^{a_1,a_2,\ldots}_{b_1,b_2,\ldots}(x)$ is a hypergeometric function and $x_s=k_s\eta_0$. Corrections from the hypergeometric functions are restricted to very low $l$, and unless $k_s\eta_0\gtrsim 1$ are negligible. Since we expect $k_s\eta_0\ll 1$ by construction we have
\be
  C_l=\frac{128}{15}\mathcal{D}^2\frac{P_\star}{\eta_0^3}(k_s\eta_0)^2\frac{l(l+2)!}{(l-2)!}\frac{(2l-1)!!}{(2l+11)!!} .
\ee
Even though the stress spectrum decays on large scales, where the transfer function is most significant, this is indistinguishable in form from the imprint of the straight power law with $n_B=-5/2$. Writing the magnetic power spectrum as
\be
  \mathcal{P}_B(k)=A_Bk^{-5/2}\tan^{-1}\left(\xi k^6\right)H(k_c-k)
\ee
then the normalisation (\ref{Blambda2}) has an unwieldy exact solution dependant on a number of hypergeometric functions. However, we are assuming that $\xi$ is very large implying that the hypergeometric terms are negligible and the field can be normalised to
\be
  B_\lambda^2\approx\left(\frac{2}{3}\right)^{1/4}\frac{1}{32\pi^{3/2}}A_B\frac{\Gamma\left(\frac{13}{24}\right)}{\Gamma\left(\frac{19}{24}\right)}\sec\left(\frac{11\pi}{24}\right)\sec\left(\frac{7\pi}{24}\right)\sec\left(\frac{\pi}{8}\right)\csc\left(\frac{7\pi}{24}\right)\lambda^{-1/2}
   \approx\frac{\lambda^{-1/2}}{5.4}A_B .
\ee
From the numerical integration, the amplitude $\tilde{P}_\star$ evaluated at a pivot of $\tilde{k}_\star=k_\star/k_c\approx 7\times 10^{-6}$ is $P_\star=\tilde{P}_\star A_Bk_c^3\approx (4.5\times 10^7)A_Bk_c^3$. One can then find that the CMB angular power spectrum is
\be
  C_l\approx 46\tilde{P}_\star\tilde{k}^2_\star\mathcal{D}^2B_\lambda^2\frac{\lambda^{1/2}k_c^5}{\eta_0}\frac{l(l+2)!}{(l-2)!}\frac{(2l-1)!!}{(2l+11)!!}
   \approx 0.34\left(\frac{k_c(\mathrm{IR}\;\mathrm{controlled})}{k_c(n_B=-5/2)}\right)^5C_l(n_B=-5/2)
\ee
where in the second step we have used that $k_S\approx k_c/1000$. Since the damping scales are approximately of the same order of magnitude, the signal arising from such an infra-red controlled magnetic field is smaller than, but otherwise practically indistinguishable from, the undamped case.

The methods we have employed for the IR-controlled field demonstrate a general approach to the CMB angular power spectra applicable for any magnetic field for which we can numerically generate -- or produce with statistical realisations -- a stress spectrum. In the simplest cases we need only recover $\tilde{P}_\star$ at a pivot $\tilde{k}_\star$, find $\alpha$, and then employ these in equation (\ref{MagnetisedCMB}). Table \ref{CMBSpectra-Table} shows the values of $\tilde{P}_\star$ and $\alpha$ and the $1$-$\sigma$ errors on $\alpha$, for a range of spectra, with $k_\star=(5\times 10^{-4})k_c(\eta_\mathrm{eq})$,\footnote{Note that the \emph{physical} scale at which we are sampling is not constant; we are choosing a pivot point in units of $k_c=k_c(\eta_\mathrm{eq},n_B)$. For large values of $n_B$, we are therefore employing a pivot on much larger scales than we are for lower $n_B$.} We recover $\alpha$ from averaging the gradient of the numerical spectra across much of the coherent range -- typically from $\sim 20$ samples. Figure \ref{Figure-Reconstructions} shows the resulting approximations. Note that before recovering $\tilde{P}_\star$ and $\alpha$ we have smoothed the numerical integration for $n_B=-2.9$ to remove some residual noise arising from the limited sample volume. Since the scaling regime is clear, and only the accuracy with which $\alpha$ is recovered depends on the smoothing, there is no error introduced in doing so.
\begin{table}
\centering
\bdm
\begin{array}{c|c|c|c}
   n_B  & \tilde{P}_\star     &        \alpha & \alpha_\mathrm{expected} \\ \hline
  -2.9  & 9.233\times 10^{11} & -2.80\pm 0.04 & -2.80 \\
  -2.75 & 4.552\times 10^{10} & -2.50\pm 0.02 & -2.5 \\
  -2.5  & 5.024\times 10^8    & -2.00\pm 0.02 & -2 \\
    0   & 7.812               & (-2.47\pm 7.40)\times 10^{-4} &  0 \\
    2   & 3.345               & (-7.76\pm 8.38)\times 10^{-4} &  0 \\
  2\;(\mathrm{UV}\;\mathrm{controlled}) & 9.627\times 10^{-5} & (-0.49\pm 1.36)\times 10^{-3} & \sim 0
\end{array}
\edm
\caption{Stress spectra recovered for sample power-law magnetic spectra.}
\label{CMBSpectra-Table}
\end{table}
The saving in integration time when we need only consider a limited range in $k$ is significant even at the two-point level; when we consider the bispectrum in future study this will become even more important. 

By extension, we can model more complex spectra, as with the IR-controlled field, with an arbitrary number of distinct regimes, each characterised by a particular value for $\alpha$. However, in most cases we would expect to require only one or two regimes. (An alternative approach employing general fits for ranges of $n_B$ was recently taken in \citep{Paoletti:2010rx}.) Similar arguments will equally apply to the vector and scalar perturbations and to the polarisation power spectra, although the form of the transfer functions will be different. More directly, similar arguments apply for the tensor CMB angular bispectrum, which can be recovered employing the same transfer function.

To close this brief consideration of the magnetised CMB we consider two power-law fields, with $n_B=-2.9$ and $n_B=-2.75$. For these fields we can employ equation (\ref{MagnetisedCMB}) with the values of $\tilde{P}_\star$ and $\alpha$ from Table \ref{CMBSpectra-Table}. Employing the central value of $\alpha$ and neglecting the error we can wrap the infra-red power spectra onto the microwave background with equation (\ref{MagnetisedCMB}) to find
\bea
  C_l&=&\frac{256\pi^2}{15}\frac{\left(\tilde{P}_\star\tilde{k}_\star^2\right)}{\Gamma(1/4)}\mathcal{D}^2B_\lambda^2\frac{k_c^5\lambda^{1/2}}{\eta_0}\frac{l(l+2)!}{(l-2)!}\frac{(2l-1)!!}{(2l+1)!!} \nonumber \\
  &\approx& 5,831\mathcal{D}^2B_\lambda^2\frac{k_c^5\lambda^{1/2}}{\eta_0}\frac{l(l+2)!}{(l-2)!}\frac{(2l-1)!!}{(2l+1)!!}
    \approx 0.996C_l^{\mathrm{Analytic}}, \qquad n_B=-2.5,
\eea
\bea
  C_l&=&\frac{\pi^{5/2}}{4\Gamma(1/8)}\frac{\Gamma(13/4)}{\Gamma(15/4)}\left(\tilde{P}_\star\tilde{k}_\star^{5/2}\right)\mathcal{D}^2B_\lambda^2\left(\frac{\lambda k_c^{22}}{\eta_0^2}\right)^{1/4}\frac{l(l+2)!}{(l-2)!}\frac{\Gamma(l+1/4)}{\Gamma(l+27/4)} \nonumber \\
    &\approx& 85.10\mathcal{D}^2B_\lambda^2\left(\frac{\lambda k_c^{22}}{\eta_0^2}\right)^{1/4}\frac{l(l+2)!}{(l-2)!}\frac{\Gamma(l+1/4)}{\Gamma(l+27/4)}, \qquad n_B=-2.75,
\eea
\bea
 C_l&=&\frac{\pi^{5/2}}{4\Gamma(1/20)}\frac{\Gamma(34/10)}{\Gamma(39/10)}\left(\tilde{P}_\star\tilde{k}_\star^{2.8}\right)\mathcal{D}^2B_\lambda^2\left(\frac{\lambda^{1/2}k_c^{29}}{\eta_0}\right)^{2/10}\frac{\Gamma(l+1/10)}{\Gamma(l+69/10)} \nonumber \\
   &\approx&90.94\mathcal{D}^2B_\lambda^2\left(\frac{\lambda^{1/2}k_c^{29}}{\eta_0}\right)^{2/10}\frac{\Gamma(l+1/10)}{\Gamma(l+69/10)}, \qquad n_B=-2.9 .
\eea
The agreement between the numerical and analytic estimates for $n_B=-2.5$ is extremely good and confirms the approach, while the $l$-dependence for the other two cases is in agreement with previous estimates within their range of validity (for example \citep{Mack:2001gc,Paoletti:2008ck,Yamazaki:2008gr}).

In Figure \ref{Figure-CMB} we plot the large-scale CMB spectra for the magnetic fields we have considered in this paper normalised to $\left.l(l+1)C_l\right|_{l=2}=1$ to emphasise the $l$-dependences. In this plot we employ the full result for $n_B=-3/2$ field rather than neglecting the $\Psi(l+1/2)$ and $f(l)$ terms. We would emphasise that it is not our intention to derive exact CMB predictions -- that has been done in, for example, \citep{Lewis:2004ef,Yamazaki:2008gr,Paoletti:2008ck,Yamazaki:2010nf,Paoletti:2010rx} employing the results from magnetised Boltzmann codes. It has rather been our intention to demonstrate a rapid approach to the large-scale magnetised CMB that will prove extremely useful when considering the magnetised bispectra, as it allows us to focus purely on the $k\ll k_c$ region. However, we have here presented the impact on the microwave background of two interesting fields, one a causally-generated field damped in the ultra-violet across a wide range of $k$, and one an acausal field damped in the deep infra-red. In both cases, for the reasonable parameters we have chosen, the impacts are indistinguishable from a standard causal field and a standard inflationary field respectively.

\section{Conclusions}
\label{Conclusions}
In this paper we have considered in detail the statistics of the stress tensor of a tangled primordial magnetic field at the one- and two-point levels, employing a mixture of analytical techniques, numerical integration, and statistical realisations. Assuming Gaussian statistics and a power-law spectrum for the magnetic field, we are restricted to the region $n_B>-3$; a field with $n_B=-3$ would be exactly scale-invariant. Fields with $n_B\geq 2$ can be produced by causal processes. We have examined the one-point moments, extending the results of BC05 and B06 to a wide range of spectral indices and a much higher grid-resolution.  We have verified that the isotropic pressure of the magnetic field is a $\chi^2$ field to a very high degree, as has been argued before. We also considered the anisotropic pressure, which runs with spectral index. In particular, for indices $n_B\geq -3/2$ the statistics of the anisotropic pressure are constant, with a skewness $\gamma_1\approx -0.24$ and a kurtosis $\gamma_2\approx 1.1$. However, in the region $n_B<-3/2$ both the skewness and the kurtosis increase as $n_B\rightarrow -3$. There is a strong grid dependence for such steeply-tilted spectra which requires us to operate on grids of side-length $l_\mathrm{dim}=512$, which arises from the power being piled onto a small number of modes on the finite grid. This infra-red divergence is inevitable -- and re-emerges in the study of the 2-point moments -- but we can nevertheless predict that $\gamma_1\lesssim 1$ and $\gamma_2\lesssim 2$ at $n_B=-3$. Perhaps more importantly, we have demonstrated that at the one-point level the field with $n_B=-3/2$ separates the parameter space into two regions in which the anisotropic stress behaves quite differently.

Turning to the two-point moments, our chief focus has been to construct a formalism which can be equally applied to both analytic and to numeric evaluation of the stress spectra -- and which, more importantly, will be readily adaptable to the calculation of magnetic bispectra. We then applied this to the two-point moments of a power-law magnetic field, finding analytic solutions existing for all integer and half-integer $n_B\in[-5/2,3]$. While some of these solutions were previously found in \citep{Paoletti:2008ck}, others were not. As has been known for some time, the parameter space is split in two, again at $n_B=-3/2$. For $n_B>-3/2$ the integrals are readily solved, and across the range of $k$, the stress spectra are technically ``decoherent'', in that the amplitude and the gradient at each $k$ depend explicitly on the damping scale, which is time-dependent. On the very largest scales, for $k\lesssim k_\mathrm{Coh}\approx k_c/100$ the cross-correlation of the isotropic and anisotropic pressures vanishes (and is an anti-correlation on intermediate scales, except at $n_B=-1$), while the other correlations tend towards white noise. The ratio between the various correlations is constant across $n_B$, and is equivalent to the ``causal'' ratios found in \citep{Turok:1997gj}. Conversely, for $n_B<-3/2$, there is a large region in which the statistics are ``coherent'', for wavenumbers up to $k_\mathrm{Coh}\approx k_c$. On these scales the power spectra scale as $\mathcal{P}(k)\propto k^{2n_B+3}$ as has been appreciated for some time; it is worth noting that it is only for integer and half-integer $n_B$ that this results in integer powers of $k$. Interestingly, the cross-correlation between the scalar pressures is of the same order of magnitude as the auto-correlations. Above $k_\mathrm{Coh}$ the statistics remain entirely decoherent. In particular, we presented a solution for the case with $n_B=-2$. Since we cannot find a solution at $n_B=-3$, there are only two solutions in this ``coherent'' or ``infra-red'' region available, $n_B=-5/2$ and $n_B=-2$. Since these values remain in the centre of the allowed constraints, it is important that we possess the solutions. Finally, the field with $n_B=-3/2$ behaves quite differently. In this case the cross-correlation of the scalar pressures is a positive correlation, and white noise on large scales, while the auto-correlations all diverge logarithmically with $k$. Numerical integrations employing Monte-Carlo techniques confirmed the analytic solutions to an extremely good accuracy.

From our analysis, we then not only have an approach to analytical and numerical integrations that can be extended with ease to bispectra, but we have also clearly identified the important regions of the spectra solutions, broadly governed by a coherence scale $k_\mathrm{Coh}$.

We then considered two non-power law magnetic power spectra -- the first, a ``damped causal'' field has a spectral index $n_B=2$ on large scales but an extended damping tail on smaller scales, and resembles a field produced by a causal mechanism such as the nonlinear vorticity of the electron/proton plasmas; while the second, an ``IR-controlled'' field, has a spectral index $n_B=-5/2$ on relatively small scales but decays on large scales, and while it could in principle be produced by a carefully-chosen inflationary scenario is more concretely employed to model the infra-red cut-off of a sample realisation. Analytic solutions exist for the damped causal field we chose. The damped causal stress spectra resemble the standard causal case with the gross features moved to a smaller wavenumber corresponding to the peak of the magnetic power spectrum. The IR controlled spectra, in contrast, tend towards white-noise for $k<k_S$ where $k_S$ is related to the scale at which large-scale magnetic power begins to decay, but closely resemble the standard infra-red case on larger scales, albeit with a suppression of power for all $k<k_\mathrm{Coh}$. For both of these scenarios, though, the cross-correlation significantly differs from the undamped cases, which might in principle allow us to distinguish between them.

We also considered the power-law spectra employing statistical realisations, running on grids of side-length $l_\mathrm{dim}=256$. While the results in the ultra-violet regime with $n_B>-3/2$ are in extremely good agreement with the analytical solutions, the limitations of the approach are quickly apparent: we are limited to a minimum wavenumber $k/k_c\approx 4/l_\mathrm{dim}\approx 10^{-3}$-$10^{-2}$. While this is close to the white-noise regime, one would not necessarily wish to na\"ively rely on the output at such a large scale, particularly as the error at the minimum wavenumber is naturally rather large. When considering $n_B<-3/2$ the limitations of the realisations are even clearer; there is an extremely strong impact from the infra-red damping. While working on increasingly large grids eases the problem this rapidly becomes unwieldy and unrealistic. While realisations remain in principle the most flexible approach to the statistics of early universe magnetic fields, not least as they can easily take into account magnetic fields with non-Gaussian statistics, they should be treated with caution.

Finally, we outlined an approach to the CMB which will be of use in the study of the CMB angular bispectrum arising from magnetic perturbations. Working with the tensor perturbations sourced by a magnetic field we considered the temperature auto-correlation. Using that the transfer function is steeply tilted to large scales, we argued that generally only the regions $k<k_\mathrm{Coh}$ contribute to the CMB integrations. The stress spectrum can then with fair generality be written as a simple power law, where the amplitude $P_\star$ at a convenient pivot scale $k_\star$ can be recovered from an analytic solution, from a numerical integration, or from a statistical realisation. Assuming that the approximate form for the tensor transfer function is accurate, exact general solutions for the CMB signals can then be found, and specialised to particular power spectra with a suitable choice of $P_\star$ and $\alpha$. Where this is not true, an approach similar to that which we took for the IR-controlled field can be applied: the stress spectrum can be split into two or more regions, each characterised by a different $\alpha_i$. The solutions that result are unwieldy combinations of hypergeometric functions, but in most cases are likely to reduce to simpler forms for realistic parameters.

In addition to reproducing the previously-known forms for the CMB angular power spectrum from magnetised tensor modes and power-law magnetic fields, we considered the damped causal and the IR-controlled fields. For the parameters we chose in this paper, which seem reasonable, the signals turn out to be indistinguishable from those from the standard causal and inflationary cases respectively. However, we have presented the expressions for a wider variety of parameters. We have also evaluated the amplitudes of such fields when smoothed on a comoving scale $k_\lambda$.

This approach to the CMB entirely separates the form of the stress spectrum from the transfer functions in a transparent manner. Furthermore, the arguments we presented do not rely on the specific analytic forms of the stress spectra; rather, they rely on the form of the magnetised transfer functions. Naturally, these do not differ between the angular power spectra and the angular bispectra. The bispectrum can be described as with the coordinate system $\{k,r,\phi\}$ and foliated into planes of $\{k,r\}$; in each of the 1D lines of constant-$r$ through these planes it appears \citep{Brown:2005kr,Brown:2006wv,Caprini:2009vk,Seshadri:2009sy} that the bispectra on large scales behave as either white-noise in $k$ or with a power law in $k$. Our approach to the CMB angular power spectra, therefore, can equally be applied to the magnetised CMB angular bispectra, which are otherwise extremely challenging to consider. This will be the focus of future study.

\begin{acknowledgments}
The author wishes to thank Kishore Ananda, Christian Byrnes, Chiara Caprini, Sami Nurmi and particularly Robert Crittenden for extremely useful comments and discussions at various stages of this work and gratefully acknowledges support and hospitality from Christof Wetterich. Portions of this work employ codes developed in collaboration with Robert Crittenden and Richard W. Brown.
\end{acknowledgments}

\appendix
\section{Analytic Power Spectra}
\label{Appendix-SpectraSolutions}
Analytic solutions exist at least for all integer and half-integer $n_B>-3$. For integer $n_B$ they take the form of simple polynomials, while for half-integers they include logarithmic and inverse trigonometric terms. Here we present these for $n_B\in[-5/2,3]$ that have not been previously presented in \cite{Paoletti:2008ck}, covering the magnetic fields range of most interest. In the following equations, we define
\be
  \tk=\left|1-k\right|^{-1/2}=\left\{\begin{array}{rcl}(1-k)^{-1/2}&,&k<1\\(k-1)^{-1/2}&,&k>1\end{array}\right.
\ee
and
\be
  \tcos(k)=\left\{\begin{array}{rcl} \cosh^{-1}\left(1-\frac{2}{k}\right)-\frac{1}{2}\pi-2(1-k)^{-1/2}&,&k<1
      \\ \cos^{-1}\left(1-\frac{2}{k}\right)-\frac{1}{2}\pi-2(k-1)^{-1/2}&,&k>1 \end{array}\right. .
\ee
These functions are undefined at $k=1$ for which we state the solutions explicitly.

We can express correlations in the form
\be
  \PTrTr(k)=\pi A_Bk_c^3\left\{\begin{array}{rl}\sum_{m=-5}^8\mathcal{A}^{\RR}_m(k)k^m&,k\in(0,1) \\ \mathcal{B}^{\RR}&,k=1 \\ \sum_{m=-5}^8\mathcal{C}^{\RR}_m(k)k^m&,k\in(1,2) \\ 0&,\mathrm{otherwise}
      \end{array}\right.
\ee
with equivalent expressions for the other correlations. In the following subsections we tabulate the non-vanishing $\mathcal{A}_m(k)$, $\mathcal{B}(k)$ and $\mathcal{C}_m(k)$.

\subsection{$n_B=5/2$}
\be
  \begin{array}{rclcrcl}
    \mathcal{A}^{\RR}_{-1}&=&\frac{184}{1,755}\left(1-\tk^{-1/2}\right), && \mathcal{C}^{\RR}_{-1}&=&\frac{184}{1,755}\left(1+\kappa^{-1/2}\right), \\
    \mathcal{A}^{\RR}_0&=&\frac{1,939}{3,510}\tk, && \mathcal{C}^{\RR}_0&=&-\frac{1,939}{3,510}\tk^{-1/2}, \\
    \mathcal{A}^{\RR}_1&=&-\frac{4}{45}\left(1+\frac{8,059}{624}\tk^{-1/2}\right), && \mathcal{C}^{\RR}_1&=&-\frac{4}{45}\left(1-\frac{8,059}{624}\tk^{-1/2}\right), \\
    \mathcal{A}^{\RR}_2&=&\frac{32,281}{28,080}\tk^{-1/2}, && \mathcal{C}^{\RR}_2&=&-\frac{32,281}{28,080}\tk^{-1/2}, \\
    \mathcal{A}^{\RR}_3&=&\frac{1}{25}\left(1-\frac{144,827}{11,232}\tk^{-1/2}\right), && \mathcal{C}^{\RR}_3&=&\frac{1}{25}\left(1+\frac{144,827}{11,232}\tk^{-1/2}\right), \\
    \mathcal{A}^{\RR}_4&=&\frac{127,223}{2,246,400}\tk^{-1/2}, && \mathcal{C}^{\RR}_4&=&-\frac{127,223}{2,246,400}\tk^{-1/2}, \\
    \mathcal{A}^{\RR}_5&=&-\frac{1,793}{166,400}\tk^{-1/2}, && \mathcal{C}^{\RR}_5&=&\frac{1,793}{166,400}\tk^{-1/2}, \\
    \mathcal{A}^{\RR}_6&=&\frac{44,267}{1,996,800}\tk^{-1/2}, && \mathcal{C}^{\RR}_6&=&-\frac{44,627}{1,996,800}\tk^{-1/2}, \\
    \mathcal{A}^{\RR}_7&=&\frac{133}{184,320}\tk^{-1/2}, && \mathcal{C}^{\RR}_7&=&-\frac{133}{184,320}\tk^{-1/2}, \\
    \mathcal{A}^{\RR}_8&=&\frac{133}{122,880}\tcos(k), && \mathcal{C}^{\RR}_8&=&-\frac{133}{122,880}\tcos(k),
  \end{array}
\ee
\be
  \begin{array}{rclcrcl}
    \mathcal{A}^{\RS}_{-3}&=&\frac{32}{3,315}\left(1-\tk^{-1/2}\right), && \mathcal{C}^{\RS}_{-3}&=&\frac{32}{3,315}\left(1+\tk^{-1/2}\right), \\
    \mathcal{A}^{\RS}_{-2}&=&\frac{16}{3,315}\tk^{-1/2}, && \mathcal{C}^{\RS}_{-2}&=&-\frac{16}{3,315}\tk^{-1/2}, \\
    \mathcal{A}^{\RS}_{-1}&=&-\frac{124}{1,755}\left(1-\frac{536}{527}\tk^{-1/2}\right), && \mathcal{C}^{\RS}_{-1}&=&-\frac{124}{1,755}\left(1+\frac{536}{527}\tk^{-1/2}\right), \\
    \mathcal{A}^{\RS}_0&=&-\frac{1,036}{29,835}, && \mathcal{C}^{\RS}_0&=&\frac{1,036}{29,835}\tk^{-1/2}, \\
    \mathcal{A}^{\RS}_1&=&\frac{1}{9}\left(1-\frac{11,026}{3,315}\tk^{-1/2}\right), && \mathcal{C}^{\RS}_1&=&\frac{1}{9}\left(1+\frac{11,026}{3,315}\tk^{-1/2}\right), \\
    \mathcal{A}^{\RS}_2&=&\frac{44,339}{59,670}\tk^{-1/2}, && \mathcal{C}^{\RS}_2&=&-\frac{44,339}{59,670}\tk^{-1/2}, \\
    \mathcal{A}^{\RS}_3&=&-\frac{1}{50}\left(1+\frac{336,841}{11,934}\tk^{-1/2}\right), && \mathcal{C}^{\RS}_3&=&-\frac{1}{50}\left(1-\frac{336,841}{11,934}\tk^{-1/2}\right), \\
    \mathcal{A}^{\RS}_4&=&\frac{762,589}{4,773,600}\tk^{-1/2}, && \mathcal{C}^{\RS}_4&=&-\frac{762,589}{4,773,600}\tk^{-1/2}, \\
    \mathcal{A}^{\RS}_5&=&\frac{2,269}{3,182,400}\tk^{-1/2}, && \mathcal{C}^{\RS}_5&=&-\frac{2,269}{3,182,400}\tk^{-1/2}, \\
    \mathcal{A}^{\RS}_6&=&-\frac{14,837}{12,729,600}\tk^{-1/2}, && \mathcal{C}^{\RS}_6&=&\frac{14,837}{12,729,600}\tk^{-1/2}, \\
    \mathcal{A}^{\RS}_7&=&\frac{7}{23,040}, && \mathcal{C}^{\RS}_7&=&-\frac{7}{23,040}\tk^{-1/2}, \\
    \mathcal{A}^{\RS}_8&=&\frac{7}{15,360}\tcos(k), && \mathcal{C}^{\RS}_8&=&-\frac{7}{15,360}\tcos(k),
  \end{array}
\ee
\be
  \begin{array}{rclcrcl}
    \mathcal{A}^{\SS}_{-5}&=&\frac{512}{100,555}\left(1-\tk^{-1/2}\right), && \mathcal{C}^{\SS}_{-5}&=&\frac{512}{100,555}\left(1+\tk^{-1/2}\right), \\
    \mathcal{A}^{\SS}_{-4}&=&\frac{256}{100,555}\tk^{-1/2}, && \mathcal{C}^{\SS}_{-4}&=& -\frac{256}{100,555}\tk^{-1/2}, \\
    \mathcal{A}^{\SS}_{-3}&=&-\frac{128}{3,315}\left(1-\frac{185}{182}\tk^{-1/2}\right), &&\mathcal{C}^{\SS}_{-3}&=&-\frac{128}{3,315}\left(1+\frac{185}{182}\tk^{-1/2}\right), \\
    \mathcal{A}^{\SS}_{-2}&=&-\frac{5,728}{301,665}\tk^{-1/2}, && \mathcal{C}^{\SS}_{-2}&=&\frac{5,728}{301,665}\tk^{-1/2}, \\
    \mathcal{A}^{\SS}_{-1}&=&\frac{164}{585}\left(1-\frac{64,474}{63,427}\tk^{-1/2}\right), && \mathcal{C}^{\SS}_{-1}&=&\frac{164}{585}\left(1+\frac{64,474}{63,427}\tk^{-1/2}\right), \\
    \mathcal{A}^{\SS}_0&=&\frac{43,331}{51,714}\tk^{-1/2}, && \mathcal{C}^{\SS}_0&=&-\frac{43,331}{51,714}\tk^{-1/2}, \\
    \mathcal{A}^{\SS}_1&=&-\frac{4}{45}\left(1+\frac{634,775}{45,968}\tk^{-1/2}\right), && \mathcal{C}^{\SS}_1&=&-\frac{4}{45}\left(1-\frac{126,955}{103,428}\tk^{-1/2}\right), \\
    \mathcal{A}^{\SS}_2&=&\frac{5,223,497}{4,826,640}\tk^{-1/2}, && \mathcal{C}^{\SS}_2&=&-\frac{5,223,497}{4,826,640}\tk^{-1/2}, \\
    \mathcal{A}^{\SS}_3&=&\frac{1}{100}\left(1-\frac{673,341}{12,376}\tk^{-1/2}\right), && \mathcal{C}^{\SS}_3&=&\frac{1}{100}\left(1+\frac{673,341}{12,376}\tk^{-1/2}\right), \\
    \mathcal{A}^{\SS}_4&=&\frac{618,713}{5,241,600}\tk^{-1/2}, && \mathcal{C}^{\SS}_4&=&-\frac{618,713}{5,241,600}\tk^{-1/2}, \\
    \mathcal{A}^{\SS}_5&=&-\frac{169,157}{178,214,400}\tk^{-1/2}, && \mathcal{C}^{\SS}_5&=&\frac{169,157}{178,214,400}\tk^{-1/2}, \\
    \mathcal{A}^{\SS}_6&=&\frac{1,519,261}{712,857,600}\tk^{-1/2}, && \mathcal{C}^{\SS}_6&=&-\frac{1,519,261}{712,857,600}\tk^{-1/2}, \\
    \mathcal{A}^{\SS}_7&=&\frac{4}{3,465}\tk^{-1/2}, && \mathcal{C}^{\SS}_7&=&-\frac{217}{798,720}\tk^{-1/2}, \\
    \mathcal{A}^{\SS}_8&=&\frac{217}{532,480}\tcos(k), && \mathcal{C}^{\SS}_8&=&-\frac{217}{532,480}\tcos(k),
  \end{array}
\ee
\be
  \begin{array}{rclcrcl}
    \mathcal{A}^{\VV}_{-5}&=&-\frac{2,048}{904,995}\left(1-\tk^{-1/2}\right), && \mathcal{C}^{\VV}_{-5}&=&-\frac{2,048}{904,995}\left(1+\tk^{-1/2}\right), \\
    \mathcal{A}^{\VV}_{-4}&=&-\frac{1,024}{904,995}\tk^{-1/2}, && \mathcal{C}^{\VV}_{-4}&=&\frac{1,024}{904,995}\tk^{-1/2}, \\
    \mathcal{A}^{\VV}_{-3}&=&\frac{64}{9,945}\left(1-\frac{95}{91}\tk^{-1/2}\right)k^{-3}, &&\mathcal{C}^{\VV}_{-3}&=&\frac{64}{9,945}\left(1+\frac{95}{91}\tk^{-1/2}\right), \\
    \mathcal{A}^{\VV}_{-2}&=&\frac{928}{301,665}\tk^{-1/2}, && \mathcal{C}^{\VV}_{-2}&=&-\frac{928}{301,665}\tk^{-1/2}, \\
    \mathcal{A}^{\VV}_{-1}&=&\frac{184}{1,755}\left(1-\frac{35,338}{35,581}\tk^{-1/2}\right), && \mathcal{C}^{\VV}_{-1}&=&\frac{184}{1,755}\left(1+\frac{35,338}{35,581}\tk^{-1/2}\right), \\
    \mathcal{A}^{\VV}_0&=&\frac{201,463}{387,855}\tk^{-1/2}, && \mathcal{C}^{\VV}_0&=&-\frac{201,463}{387,855}\tk^{-1/2}, \\
    \mathcal{A}^{\VV}_1&=&-\frac{2}{45}\left(1+\frac{782,623}{34,476}\tk^{-1/2}\right), && \mathcal{C}^{\VV}_1&=&-\frac{2}{45}\left(1-\frac{782,623}{34,476}\tk^{-1/2}\right), \\
    \mathcal{A}^{\VV}_2&=&\frac{4,405,799}{4,343,976}\tk^{-1/2}, && \mathcal{C}^{\VV}_2&=&-\frac{4,405,799}{4,343,976}\tk^{-1/2}, \\
    \mathcal{A}^{\VV}_3&=&-\frac{142,333}{257,040}\tk^{-1/2}, && \mathcal{C}^{\VV}_3&=&\frac{142,333}{257,040}\tk^{-1/2}, \\
    \mathcal{A}^{\VV}_4&=&\frac{3,743,149}{26,732,160}\tk^{-1/2}, && \mathcal{C}^{\VV}_4&=&-\frac{3,743,149}{26,732,160}\tk^{-1/2}, \\
    \mathcal{A}^{\VV}_5&=&\frac{16,265}{3,564,288}\tk^{-1/2}, && \mathcal{C}^{\VV}_5&=&-\frac{16,265}{3,564,288}\tk^{-1/2}, \\
    \mathcal{A}^{\VV}_6&=&-\frac{131,905}{14,257,152}\tk^{-1/2}, && \mathcal{C}^{\VV}_6&=&\frac{131,905}{14,257,152}\tk^{-1/2}, \\
    \mathcal{A}^{\VV}_7&=&-\frac{35}{239,616}\tk^{-1/2}, && \mathcal{C}^{\VV}_7&=&\frac{35}{239,616}\tk^{-1/2}, \\
    \mathcal{A}^{\VV}_8&=&-\frac{35}{159,744}\tcos(k), && \mathcal{C}^{\VV}_8&=&\frac{35}{159,744}\tcos(k) ,
  \end{array}
\ee
\be
  \begin{array}{rclcrcl}
    \mathcal{A}^{\TT}_{-5}&=&\frac{1,024}{904,995}\left(1-\tk^{-1/2}\right), && \mathcal{C}^{\TT}_{-5}&=&\frac{1,024}{904,995}\left(1+\tk^{-1/2}\right), \\
    \mathcal{A}^{\TT}_{-4}&=&\frac{512}{904,995}\tk^{-1/2}, && \mathcal{C}^{\TT}_{-4}&=&-\frac{512}{904,995}\tk^{-12}, \\
    \mathcal{A}^{\TT}_{-3}&=&\frac{128}{9,945}\left(1-\frac{90}{91}\tk^{-1/2}\right), &&\mathcal{C}^{\TT}_{-3}&=&\frac{128}{9,945}\left(1+\frac{90}{91}\tk^{-1/2}\right), \\
    \mathcal{A}^{\TT}_{-2}&=&\frac{5,888}{904,995}\tk^{-1/2}, && \mathcal{C}^{\TT}_{-2}&=&-\frac{5,888}{904,995}\tk^{-1/2}, \\
    \mathcal{A}^{\TT}_{-1}&=&\frac{8}{117}\left(1-\frac{444}{455}\tk^{-1/2}\right), && \mathcal{C}^{\TT}_{-1}&=&\frac{8}{117}\left(1+\frac{444}{455}\tk^{-1/2}\right), \\
    \mathcal{A}^{\TT}_0&=&\frac{125,194}{129,285}\tk^{-1/2}, && \mathcal{C}^{\TT}_0&=&-\frac{125,194}{129,285}\tk^{-1/2}, \\
    \mathcal{A}^{\TT}_1&=&\frac{4}{45}\left(1-\frac{21,901}{676}\tk^{-1/2}\right), && \mathcal{C}^{\TT}_1&=&\frac{4}{45}\left(1+\frac{21,901}{676}\tk^{-1/2}\right), \\
    \mathcal{A}^{\TT}_2&=&\frac{13,321,673}{3,619,980}\tk^{-1/2}, && \mathcal{C}^{\TT}_2&=&-\frac{13,321,673}{3,619,980}\tk^{-1/2}, \\
    \mathcal{A}^{\TT}_3&=&\frac{1}{50}\left(1-\frac{2,021,309}{18,564}\tk^{-1/2}\right), && \mathcal{C}^{\TT}_3&=&-\frac{1}{50}\left(1+\frac{2,021,309}{18,564}\tk^{-1/2}\right), \\
    \mathcal{A}^{\TT}_4&=&\frac{10,455,203}{22,276,800}\tk^{-1/2}, && \mathcal{C}^{\TT}_4&=&-\frac{10,455,203}{22,276,800}\tk^{-1/2}, \\
    \mathcal{A}^{\TT}_5&=&-\frac{685,591}{44,553,600}\tk^{-1/2}, && \mathcal{C}^{\TT}_5&=&\frac{685,591}{44,553,600}\tk^{-1/2}, \\
    \mathcal{A}^{\TT}_6&=&\frac{5,768,543}{178,214,400}\tk^{-1/2}, && \mathcal{C}^{\TT}_6&=&-\frac{5,768,543}{178,214,400}\tk^{-1/2}, \\
    \mathcal{A}^{\TT}_7&=&\frac{371}{199,680}\tk^{-1/2}, && \mathcal{C}^{\TT}_7&=&-\frac{371}{199,680}\tk^{-1/2}, \\
    \mathcal{A}^{\TT}_8&=&\frac{371}{133,120}\tcos(k), && \mathcal{C}^{\TT}_8&=&-\frac{371}{133,120}\tcos(k),
  \end{array}
\ee
with
\be
\begin{array}{c}
   \mathcal{B}^{\RR}=\frac{491}{8,775}-\frac{133}{245760}\pi, \qquad
   \mathcal{B}^{\RS}=\frac{691}{22,950}-\frac{7}{30,720}\pi, \\
   \mathcal{B}^{\SS}=\frac{3,039,559}{18,099,900}-\frac{217}{1,064,960}\pi, \qquad
   \mathcal{B}^{\VV}=\frac{35,062}{542,997}+\frac{35}{319,488}\pi, \\
   \mathcal{B}^{\TT}=\frac{1,730,959}{9,049,950}-\frac{371}{266,240}\pi .
\end{array}
\ee

\subsection{A Damped Causal Field}
For the power spectrum defined in equation (\ref{DampedCausal}) analytic solutions can also be found. Here we define
\be
  A_1=e^{-2\xi}, \qquad
  A_2(k)=e^{-\xi k(k-2)}, \qquad
  E_f(k)=\erf\left(\frac{1}{2}\sqrt{2\xi}(k-2)\right) .
\ee
\be
  \begin{array}{ccccl}
    \mathcal{A}^{\RR}_{-1}&=&\mathcal{C}^{\RR}_{-1}&=&\frac{A_1}{8\xi^4}\left(4(1-A_2(k))\xi^2+8(1-A_2(k))\xi+5(1-A_2(k))\right), \\
    \mathcal{A}^{\RR}_0&=&\mathcal{C}^{\RR}_0&=&\frac{3A_1}{16\xi^3}\left((5+4\xi)A_2(k)-\frac{5}{4}\sqrt{\frac{2\pi}{\xi}}e^{\xi(2-k^2/2)}E_f(k)\right), \\
    \mathcal{A}^{\RR}_1&=&\mathcal{C}^{\RR}_1&=&-\frac{A_1}{32\xi^3}\left(4(2+A_2(k))\xi+8-3A_2(k)\right), \\
    \mathcal{A}^{\RR}_2&=&\mathcal{C}^{\RR}_2&=&\frac{A_1}{32\xi^{5/2}}e^{-\xi k^2}\left(\sqrt{2\pi}A_2(k)e^{\xi (k^2+4)/2}-2\frac{\sqrt{\xi}}{A_1}\right), \\
    \mathcal{A}^{\RR}_3&=&\mathcal{C}^{\RR}_3&=&\frac{A_1}{32\xi^2}\left(2-A_2(k)\right), \\
    \mathcal{A}^{\RR}_4&=&\mathcal{C}^{\RR}_4&=&-\frac{\sqrt{2\pi}}{64\xi^{3/2}}e^{-\xi k^2/2}A_2(k) ,
  \end{array}
\ee
\be
  \begin{array}{ccccl}
    \mathcal{A}^{\RS}_{-3}&=&\mathcal{C}^{\RS}_{-3}&=&\frac{3A_1}{8\xi^5}(2+\xi)\left(1-A_2(k)\right), \\
    \mathcal{A}^{\RS}_{-2}&=&\mathcal{C}^{\RS}_{-2}&=&\frac{3A_1}{4\xi^4}(2+\xi)A_2(k), \\
    \mathcal{A}^{\RS}_{-1}&=&\mathcal{C}^{\RS}_{-1}&=&-\frac{A_1}{16\xi^4}\left(\right), \\
    \mathcal{A}^{\RS}_0&=&\mathcal{C}^{\RS}_0&=&\frac{A_1}{8\xi^3}(1+6\xi)A_2(k), \\
    \mathcal{A}^{\RS}_1&=&\mathcal{C}^{\RS}_1&=&\frac{A_1}{16\xi^3}\left((5-2A_2(k))\xi+5\right), \\
    \mathcal{A}^{\RS}_2&=&\mathcal{C}^{\RS}_2&=&-\frac{A_1}{64\xi^{5/2}}e^{-\xi k^2}\left(4\frac{\sqrt{\xi}}{A_1}+\sqrt{2\pi}E_f(k)e^{\xi(k^2+4)/2}\right), \\
    \mathcal{A}^{\RS}_3&=&\mathcal{C}^{\RS}_3&=&-\frac{A_1}{32\xi^2}\left(1+A_2(k)\right), \\
    \mathcal{A}^{\RS}_4&=&\mathcal{C}^{\RS}_4&=&-\frac{\sqrt{2\pi}}{64\xi^{3/2}}e^{-\xi k^2/2}E_f(k),
  \end{array}
\ee
\be
  \begin{array}{ccccl}
    \mathcal{A}^{\SS}_{-5}&=&\mathcal{C}^{\SS}_{-5}&=&\frac{27A_1}{8\xi^6}\left(1-A_2(k)\right), \\
    \mathcal{A}^{\SS}_{-4}&=&\mathcal{C}^{\SS}_{-4}&=&\frac{27A_1}{4\xi^5}A_2(k), \\
    \mathcal{A}^{\SS}_{-3}&=&\mathcal{C}^{\SS}_{-3}&=&-\frac{3A_1}{8\xi^5}\left(2(2+7A_2(k))\xi+8+A_2(k)\right), \\
    \mathcal{A}^{\SS}_{-2}&=&\mathcal{C}^{\SS}_{-2}&=&\frac{3A_1}{4\xi^4}(1+2\xi)A_2(k), \\
    \mathcal{A}^{\SS}_{-1}&=&\mathcal{C}^{\SS}_{-1}&=&\frac{A_1}{16\xi^4}\left(4(5-2A_2(k))\xi^2+4(10-7A_2(k))\xi+31-10A_2(k)\right), \\
    \mathcal{A}^{\SS}_0&=&\mathcal{C}^{\SS}_0&=&-\frac{A_1}{16\xi^3}(1-12\xi)A_2(k)-\frac{21}{64}\sqrt{2\pi}e^{-\xi k^2/2}E_f(k), \\
    \mathcal{A}^{\SS}_1&=&\mathcal{C}^{\SS}_1&=&-\frac{A_1}{32\xi^3}\left(4(2+A_2(k))\xi+8+3A_2(k)\right), \\
    \mathcal{A}^{\SS}_2&=&\mathcal{C}^{\SS}_2&=&-\frac{A_1}{16\xi^{5/2}}e^{-\xi k^2}\left(\frac{\sqrt{\pi}}{A_1}+\sqrt{2\pi}e^{\xi(k^2+4)/2}E_f(k)\right), \\
    \mathcal{A}^{\SS}_3&=&\mathcal{C}^{\SS}_3&=&\frac{A_1}{64\xi^2}\left(1-2A_2(k)\right), \\
    \mathcal{A}^{\SS}_4&=&\mathcal{C}^{\SS}_4&=&-\frac{\sqrt{2\pi}}{64\xi^{3/2}}e^{-\xi k^2/2}E_f(k),
  \end{array}
\ee
\be
  \begin{array}{ccccl}
    \mathcal{A}^{\VV}_{-5}&=&\mathcal{C}^{\VV}_{-5}&=&-\frac{3A_1}{2\xi^6}\left(1-A_2(k)\right), \\
    \mathcal{A}^{\VV}_{-4}&=&\mathcal{C}^{\VV}_{-4}&=&-\frac{3A_1}{\xi^5}A_2(k), \\
    \mathcal{A}^{\VV}_{-3}&=&\mathcal{C}^{\VV}_{-3}&=&\frac{A_1}{4\xi^5}\left((1+11A_2(k))\xi+2(1+2A_2(k))\right), \\
    \mathcal{A}^{\VV}_{-2}&=&\mathcal{C}^{\VV}_{-2}&=&-\frac{A_1}{2\xi^4}(4+3\xi)A_2(k), \\
    \mathcal{A}^{\VV}_{-1}&=&\mathcal{C}^{\VV}_{-1}&=&\frac{A_1}{8\xi^4}\left(4\xi^2+2(4+3A_2(k))\xi+5-3A_2(k)\right), \\
    \mathcal{A}^{\VV}_0&=&\mathcal{C}^{\VV}_0&=&-\frac{A_1}{8\xi^3}A_2(k)-\frac{7\sqrt{2\pi}}{32\xi^{7/2}}e^{-\xi k^2/2}E_f(k), \\
    \mathcal{A}^{\VV}_1&=&\mathcal{C}^{\VV}_1&=&-\frac{A_1}{16\xi^3}\left(2\xi+2+A_2(k)\right), \\
    \mathcal{A}^{\VV}_2&=&\mathcal{C}^{\VV}_2&=&-\frac{\sqrt{2\pi}}{32\xi^{5/2}}e^{-\xi k^2/2}E_f(k),
  \end{array}
\ee
\be
  \begin{array}{ccccl}
    \mathcal{A}^{\TT}_{-5}&=&\mathcal{C}^{\TT}_{-5}&=&\frac{3A_1}{4\xi^6}\left(1-A_2(k)\right), \\
    \mathcal{A}^{\TT}_{-4}&=&\mathcal{C}^{\TT}_{-4}&=&\frac{3A_1}{2\xi^5}A_2(k), \\
    \mathcal{A}^{\TT}_{-3}&=&\mathcal{C}^{\TT}_{-3}&=&\frac{A_1}{4\xi^5}\left(2(1-4A_2(k))\xi+4-7A_2(k)\right), \\
    \mathcal{A}^{\TT}_{-2}&=&\mathcal{C}^{\TT}_{-2}&=&\frac{A_1}{2\xi^4}(7+4\xi)A_2(k), \\
    \mathcal{A}^{\TT}_{-1}&=&\mathcal{C}^{\TT}_{-1}&=&\frac{A_1}{8\xi^4}\left(4(1-4A_2(k))\xi^2+8(1-5A_2(k))\xi-(1+10A_2(k))\right), \\
    \mathcal{A}^{\TT}_0&=&\mathcal{C}^{\TT}_0&=&\frac{3}{4\xi^3}(1+4\xi)-\frac{7\sqrt{2\pi}}{16\xi^{7/2}}e^{-\xi k^2/2}E_f(k), \\
    \mathcal{A}^{\TT}_1&=&\mathcal{C}^{\TT}_1&=&\frac{A_1}{8\xi^3}\left(2(1-2A_2(k))\xi+2+A_2(k)\right), \\
    \mathcal{A}^{\TT}_2&=&\mathcal{C}^{\TT}_2&=&-\frac{A_1}{4\xi^2}A_2(k), \\
    \mathcal{A}^{\TT}_3&=&\mathcal{C}^{\TT}_3&=&\frac{A_1}{32\xi^2}\left(1-4A_2(k)\right), \\
    \mathcal{A}^{\TT}_4&=&\mathcal{C}^{\TT}_4&=&-\frac{\sqrt{2\pi}}{16\xi^{3/2}}e^{-\xi k^2/2}E_f(k) .
  \end{array}
\ee

\subsection{$n_B=3/2$}
\be
\begin{array}{c}
  \begin{array}{ccccl}
    \mathcal{A}^{\RR}_{-1}&=&\mathcal{C}^{\RR}_{-1}&=&\frac{296}{1,617}\left(1-\tk^{-1/2}\right), \\
    \mathcal{A}^{\RR}_0&=&\mathcal{C}^{\RR}_0&=&\frac{1,226}{1,617}\tk^{-1/2}, \\
    \mathcal{A}^{\RR}_1&=&\mathcal{C}^{\RR}_1&=&-\frac{4}{21}\left(1+\frac{1,811}{308}\tk^{-1/2}\right), \\
    \mathcal{A}^{\RR}_2&=&\mathcal{C}^{\RR}_2&=&\frac{2,693}{4,312}\tk^{-1/2}, \\
    \mathcal{A}^{\RR}_3&=&\mathcal{C}^{\RR}_3&=&\frac{1}{9}\left(1-\frac{3,449}{8,624}\tk^{-1/2}\right), \\
    \mathcal{A}^{\RR}_4&=&\mathcal{C}^{\RR}_4&=&\frac{1,927}{44,352}\tk^{-1/2}, \\
    \mathcal{A}^{\RR}_5&=&\mathcal{C}^{\RR}_5&=&-\frac{8,797}{88,704}\tk^{-1/2},
  \end{array}
  \\
  \mathcal{A}^{\RR}_6=-\frac{55}{5,376}\left(\tcos(k)+\pi\right),
  \\
  \mathcal{C}^{\RR}_6=-\frac{55}{5,376}\tcos(k),
\end{array}
\qquad
\begin{array}{c}
  \begin{array}{ccccl}
    \mathcal{A}^{\RS}_{-3}&=&\mathcal{C}^{\RS}_{-3}&=&\frac{32}{1,155}\left(1-\tk^{-1/2}\right), \\
    \mathcal{A}^{\RS}_{-2}&=&\mathcal{C}^{\RS}_{-2}&=&\frac{16}{1,155}\tk^{-1/2}, \\
    \mathcal{A}^{\RS}_{-1}&=&\mathcal{C}^{\RS}_{-1}&=&-\frac{244}{1,617}\left(1-\frac{312}{305}\tk^{-1/2}\right), \\
    \mathcal{A}^{\RS}_0&=&\mathcal{C}^{\RS}_0&=&-\frac{596}{8,085}, \\
    \mathcal{A}^{\RS}_1&=&\mathcal{C}^{\RS}_1&=&\frac{5}{21}\left(1-\frac{818}{385}\tk^{-1/2}\right), \\
    \mathcal{A}^{\RS}_2&=&\mathcal{C}^{\RS}_2&=&\frac{737}{980}\tk^{-1/2}, \\
    \mathcal{A}^{\RS}_3&=&\mathcal{C}^{\RS}_3&=&-\frac{1}{18}\left(1+\frac{62,903}{10,780}\tk^{-1/2}\right), \\
    \mathcal{A}^{\RS}_4&=&\mathcal{C}^{\RS}_4&=&-\frac{1,079}{110,880}\tk^{-1/2}, \\
    \mathcal{A}^{\RS}_5&=&\mathcal{C}^{\RS}_5&=&\frac{3,821}{221,760}\tk^{-1/2},
  \end{array}
  \\
  \mathcal{A}^{\RS}_6=-\frac{5}{2,688}\left(\tcos(k)+\pi\right),
  \\
  \mathcal{C}^{\RS}_6=-\frac{5}{2,688}\tcos(k), \\
\end{array}
\ee
\be
\begin{array}{c}
  \begin{array}{ccccl}
    \mathcal{A}^{\SS}_{-5}&=&\mathcal{C}^{\SS}_{-5}&=&\frac{1,536}{80,465}\left(1-\tk^{-1/2}\right), \\
    \mathcal{A}^{\SS}_{-4}&=&\mathcal{C}^{\SS}_{-4}&=&\frac{768}{80,465}\tk^{-1/2}, \\
    \mathcal{A}^{\SS}_{-3}&=&\mathcal{C}^{\SS}_{-3}&=&-\frac{128}{1,155}\left(1-\frac{427}{418}\tk^{-1/2}\right), \\
    \mathcal{A}^{\SS}_{-2}&=&\mathcal{C}^{\SS}_{-2}&=&-\frac{13,088}{241,395}\tk^{-2}, \\
    \mathcal{A}^{\SS}_{-1}&=&\mathcal{C}^{\SS}_{-1}&=&\frac{76}{147}\left(1-\frac{223,942}{218,405}\tk^{-1/2}\right), \\
    \mathcal{A}^{\SS}_0&=&\mathcal{C}^{\SS}_0&=&\frac{2,003,102}{1,689,765}\tk^{-1/2}, \\
    \mathcal{A}^{\SS}_1&=&\mathcal{C}^{\SS}_1&=&-\frac{4}{21}\left(1+\frac{2,053,913}{321,860}\tk^{-1/2}\right), \\
    \mathcal{A}^{\SS}_2&=&\mathcal{C}^{\SS}_2&=&\frac{275,949}{409,640}\tk^{-1/2}, \\
    \mathcal{A}^{\SS}_3&=&\mathcal{C}^{\SS}_3&=&\frac{1}{36}\left(1-\frac{1,182,337}{204,820}\tk^{-1/2}\right), \\
    \mathcal{A}^{\SS}_4&=&\mathcal{C}^{\SS}_4&=&\frac{19,583}{4,213,440}\tk^{-1/2}, \\
    \mathcal{A}^{\SS}_5&=&\mathcal{C}^{\SS}_5&=&-\frac{104,837}{8,426,880}\tk^{-1/2},
  \end{array}
  \\
  \mathcal{A}^{\SS}_6=-\frac{155}{2,688}\left(\tcos(k)+\pi\right) \\ \mathcal{C}^{\SS}_6=-\frac{155}{59,136}\tcos(k),
  \end{array}
\begin{array}{c}
  \begin{array}{rclcrcl}
    \mathcal{A}^{\VV}_{-5}&=&\mathcal{C}^{\VV}_{-5}&=&-\frac{2,048}{241,395}\left(1-\tk^{-1/2}\right), \\
    \mathcal{A}^{\VV}_{-4}&=&\mathcal{C}^{\VV}_{-4}&=&-\frac{1,024}{241,395}\tk^{-1/2}, \\
    \mathcal{A}^{\VV}_{-3}&=&\mathcal{C}^{\VV}_{-3}&=&\frac{64}{3,465}\left(1-\frac{221}{209}\tk^{-1/2}\right), \\
    \mathcal{A}^{\VV}_{-2}&=&\mathcal{C}^{\VV}_{-2}&=&\frac{6,304}{724,185}\tk^{-1/2}, \\
    \mathcal{A}^{\VV}_{-1}&=&\mathcal{C}^{\VV}_{-1}&=&\frac{296}{1,617}\left(1-\frac{114,742}{115,995}\tk^{-1/2}\right), \\
    \mathcal{A}^{\VV}_0&=&\mathcal{C}^{\VV}_0&=&\frac{1,207,628}{1,689,765}\tk^{-1/2}, \\
    \mathcal{A}^{\VV}_1&=&\mathcal{C}^{\VV}_1&=&-\frac{2}{21}\left(1+\frac{866,661}{80,465}\tk^{-1/2}\right), \\
    \mathcal{A}^{\VV}_2&=&\mathcal{C}^{\VV}_2&=&\frac{38,609}{55,860}\tk^{-1/2}, \\
    \mathcal{A}^{\VV}_3&=&\mathcal{C}^{\VV}_3&=&-\frac{249,103}{1,228,920}\tk^{-1/2}, \\
    \mathcal{A}^{\VV}_4&=&\mathcal{C}^{\VV}_4&=&-\frac{403}{33,440}\tk^{-1/2}, \\
    \mathcal{A}^{\VV}_5&=&\mathcal{C}^{\VV}_5&=&\frac{12,519}{468,160}\tk^{-1/2},
  \end{array}
  \\
  \mathcal{A}^{\VV}_6=\frac{65}{29,568}\left(\tcos(k)+\pi\right),
  \\
  \mathcal{C}^{\VV}_6=\frac{65}{29,568}\tcos(k),
\end{array}
\ee

\be
\begin{array}{c}
  \begin{array}{ccccl}
    \mathcal{A}^{\TT}_{-5}&=&\mathcal{C}^{\TT}_{-5}&=&\frac{1,024}{241,395}\left(1-\tk^{-1/2}\right), \\
    \mathcal{A}^{\TT}_{-4}&=&\mathcal{C}^{\TT}_{-4}&=&\frac{512}{241,395}\tk^{-1/2}, \\
    \mathcal{A}^{\TT}_{-3}&=&\mathcal{C}^{\TT}_{-3}&=&\frac{128}{3,465}\left(1-\frac{206}{209}\tk^{-1/2}\right), \\
    \mathcal{A}^{\TT}_{-2}&=&\mathcal{C}^{\TT}_{-2}&=&\frac{13,568}{724,185}\tk^{-1/2}, \\
    \mathcal{A}^{\TT}_{-1}&=&\mathcal{C}^{\TT}_{-1}&=&\frac{104}{1,617}\left(1-\frac{37,724}{40,755}\tk^{-1/2}\right), \\
    \mathcal{A}^{\TT}_0&=&\mathcal{C}^{\TT}_0&=&\frac{2,161,256}{1,689,765}\tk^{-1/2}, \\
    \mathcal{A}^{\TT}_1&=&\mathcal{C}^{\TT}_1&=&\frac{4}{21}\left(1-\frac{1,324,971}{80,465}\tk^{-1/2}\right), \\
    \mathcal{A}^{\TT}_2&=&\mathcal{C}^{\TT}_2&=&\frac{857,909}{307,230}\tk^{-1/2}, \\
    \mathcal{A}^{\TT}_3&=&\mathcal{C}^{\TT}_3&=&\frac{1}{18}\left(1-\frac{1,521,059}{102,410}\tk^{-1/2}\right), \\
    \mathcal{A}^{\TT}_4&=&\mathcal{C}^{\TT}_4&=&\frac{49,501}{1,053,360}\tk^{-1/2}, \\
    \mathcal{A}^{\TT}_5&=&\mathcal{C}^{\TT}_5&=&-\frac{35,737}{300,960}\tk^{-1/2},
  \end{array}
  \\
  \mathcal{A}^{\TT}_6=-\frac{305}{14,784}\left(\tcos(k)+\pi\right), \\ \mathcal{C}^{\TT}_6=-\frac{305}{14,784}\tcos(k),
\end{array}
\ee
with
\be
\begin{array}{c}
   \mathcal{B}^{\RR}=\frac{503}{4,851}-\frac{55}{10,752}\pi, \qquad
   \mathcal{B}^{\RS}=\frac{2,879}{48,510}-\frac{5}{5,376}\pi,\\
   \mathcal{B}^{\SS}=\frac{5,324,279}{20,277,180}-\frac{155}{118,272}\pi, \qquad
   \mathcal{B}^{\VV}=\frac{495,794}{5,069,295}+\frac{65}{59,136}\pi, \\
   \mathcal{B}^{\TT}=\frac{3,564,031}{10,138,590}-\frac{305}{29,568}\pi .
\end{array}
\ee

\subsection{$n_B=1/2$}
\be
  \begin{array}{rclcrcl}
    \mathcal{A}^{\RR}_{-1}&=&\frac{104}{225}\left(1-\tk^{1/2}\right), && \mathcal{C}^{\RR}_{-1}&=&\frac{104}{225}\left(1+\tk^{-1/2}\right)k^{-1}, \\
    \mathcal{A}^{\RR}_0&=&\frac{173}{225}\tk^{1/2}, && \mathcal{C}^{\RR}_0&=&-\frac{277}{225}\tk^{-1/2}, \\
    \mathcal{A}^{\RR}_1&=&-\frac{4}{5}\left(1-\frac{19}{120}\tk^{1/2}\right), && \mathcal{C}^{\RR}_1&=&-\frac{4}{5}\left(1-\frac{289}{360}\tk^{-1/2}\right), \\
    \mathcal{A}^{\RR}_2&=&-\frac{103}{360}\tk^{1/2}, && \mathcal{C}^{\RR}_2&=&\frac{743}{1,800}\tk^{-1/2}, \\
    \mathcal{A}^{\RR}_3&=&\left(1-\frac{113}{144}\tk^{1/2}\right), && \mathcal{C}^{\RR}_3&=&\left(1+\frac{359}{720}\tk^{-1/2}\right), \\
    \mathcal{A}^{\RR}_4&=&\frac{51}{160}\left(\tcos(k)+2\tk^{1/2}\right), && \mathcal{C}^{\RR}_4&=&-\frac{51}{160}\left(\tcos(k)+\frac{2,048}{459}\tk^{-1/2}\right),
  \end{array}
\ee
\be
  \begin{array}{rclcrcl}
    \mathcal{A}^{\RS}_{-3}&=&\frac{32}{195}\left(1-\tk^{-1/2}\right), && \mathcal{C}^{\RS}_{-3}&=&\frac{32}{195}\left(1+\tk^{-1/2}\right), \\
    \mathcal{A}^{\RS}_{-2}&=&\frac{16}{195}\tk^{-1/2}, && \mathcal{C}^{\RS}_{-2}&=&-\frac{16}{195}\tk^{-1/2}, \\
    \mathcal{A}^{\RS}_{-1}&=&-\frac{148}{225}\left(1-\frac{496}{481}\tk^{-1/2}\right), && \mathcal{C}^{\RS}_{-1}&=&-\frac{148}{225}\left(1+\frac{496}{481}\tk^{-1/2}\right), \\
    \mathcal{A}^{\RS}_0&=&-\frac{932}{2,925}\tk^{-1/2}, && \mathcal{C}^{\RS}_0&=&\frac{932}{2,925}\tk^{-1/2}, \\
    \mathcal{A}^{\RS}_1&=&\left(1-\frac{3,878}{2,925}\tk^{-1/2}\right), && \mathcal{C}^{\RS}_1&=&\left(1+\frac{3,878}{2,925}\tk^{-1/2}\right), \\
    \mathcal{A}^{\RS}_2&=&\frac{7,147}{5,850}\tk^{-1/2}, && \mathcal{C}^{\RS}_2&=&-\frac{7,147}{5,850}\tk^{-1/2}, \\
    \mathcal{A}^{\RS}_3&=&-\frac{1}{2}\left(1-\frac{523}{1,170}\tk^{-1/2}\right), && \mathcal{C}^{\RS}_3&=&-\frac{1}{2}\left(1+\frac{523}{1,170}\tk^{-1/2}\right), \\
    \mathcal{A}^{\RS}_4&=&-\frac{3}{40}\left(\tcos(k)+\frac{2,560}{351}\tk^{-1/2}\right)k^4, && \mathcal{C}^{\RS}_4&=&\frac{3}{40}\left(\tcos(k)+\frac{2,560}{351}\tk^{-1/2}\right),
  \end{array}
\ee
\be
  \begin{array}{rclcrcl}
    \mathcal{A}^{\SS}_{-5}&=&\frac{512}{3,315}\left(1-\tk^{-1/2}\right)k^{-5}, && \mathcal{C}^{\SS}_{-5}&=&\frac{512}{3,315}\left(1+\tk^{-1/2}\right), \\
    \mathcal{A}^{\SS}_{-4}&=&\frac{256}{3,315}\tk^{-1/2}, && \mathcal{C}^{\SS}_{-4}&=&-\frac{256}{3,315}\tk^{-1/2}, \\
    \mathcal{A}^{\SS}_{-3}&=&-\frac{128}{195}\left(1-\frac{35}{34}\tk^{-1/2}\right), && \mathcal{C}^{\SS}_{-3}&=&-\frac{128}{195}\left(1+\frac{35}{34}\tk^{-1/2}\right), \\
    \mathcal{A}^{\SS}_{-2}&=&-\frac{352}{1,105}\tk^{-1/2}, && \mathcal{C}^{\SS}_{-2}&=&\frac{352}{1,105}\tk^{-1/2}, \\
    \mathcal{A}^{\SS}_{-1}&=&\frac{356}{225}\left(1-\frac{20,614}{19,669}\tk^{-1/2}\right), && \mathcal{C}^{\SS}_{-1}&=&\frac{356}{225}\left(1+\frac{20,614}{19,669}\tk^{-1/2}\right), \\
    \mathcal{A}^{\SS}_0&=&\frac{107,123}{49,725}\tk^{-1/2}, && \mathcal{C}^{\SS}_0&=&-\frac{107,123}{49,725}\tk^{-1/2}, \\
    \mathcal{A}^{\SS}_1&=&-\frac{4}{5}\left(1+\frac{72,071}{79,560}\tk^{-1/2}\right), && \mathcal{C}^{\SS}_1&=&-\frac{4}{5}\left(1-\frac{72,071}{79,560}\tk^{-1/2}\right), \\
    \mathcal{A}^{\SS}_2&=&-\frac{43,057}{397,800}\tk^{-1/2}, && \mathcal{C}^{\SS}_2&=&\frac{43,057}{397,800}\tk^{-1/2}, \\
    \mathcal{A}^{\SS}_3&=&\frac{1}{4}\left(1-\frac{3,005}{7,956}\tk^{-1/2}\right), && \mathcal{C}^{\SS}_3&=&\frac{1}{4}\left(1+\frac{3,005}{7,956}\tk^{-1/2}\right), \\
    \mathcal{A}^{\SS}_4&=&\frac{9}{160}\left(\tcos(k)+\frac{83,668}{17,901}\tk^{-1/2}\right), && \mathcal{C}^{\SS}_4&=&-\frac{9}{160}\left(\tcos(k)+\frac{83,968}{17,901}\tk^{-1/2}\right),
  \end{array}
\ee
\be
  \begin{array}{rclcrcl}
    \mathcal{A}^{\VV}_{-5}&=&-\frac{2,048}{29,835}\left(1-\tk^{-1/2}\right), && \mathcal{C}^{\VV}_{-5}&=&-\frac{2,048}{29,835}\left(1+\tk^{-1/2}\right), \\
    \mathcal{A}^{\VV}_{-4}&=&-\frac{1,024}{29,835}\tk^{-1/2}, && \mathcal{C}^{\VV}_{-4}&=&\frac{1,024}{29,835}\tk^{-1/2}, \\
    \mathcal{A}^{\VV}_{-3}&=&\frac{64}{585}\left(1-\frac{55}{51}\tk^{-1/2}\right), &&\mathcal{C}^{\VV}_{-3}&=&\frac{64}{585}\left(1+\frac{55}{51}\tk^{-1/2}\right), \\
    \mathcal{A}^{\VV}_{-2}&=&\frac{1,504}{29,835}\tk^{-1/2}, && \mathcal{C}^{\VV}_{-2}&=&-\frac{1,504}{29,835}\tk^{-2}, \\
    \mathcal{A}^{\VV}_{-1}&=&\frac{104}{225}\left(1-\frac{8,414}{8,619}\tk^{-1/2}\right), && \mathcal{C}^{\VV}_{-1}&=&\frac{104}{225}\left(1+\frac{8,414}{8,619}\tk^{-1/2}\right), \\
    \mathcal{A}^{\VV}_0&=&\frac{174,446}{149,175}\tk^{-1/2}, && \mathcal{C}^{\VV}_0&=&-\frac{174,446}{149,175}\tk^{-1/2}, \\
    \mathcal{A}^{\VV}_1&=&-\frac{2}{5}\left(1+\frac{41,737}{19,890}\tk^{-1/2}\right), && \mathcal{C}^{\VV}_1&=&-\frac{2}{5}\left(1-\frac{41,737}{19,890}\tk^{-1/2}\right), \\
    \mathcal{A}^{\VV}_2&=&\frac{32,641}{198,900}\tk^{-1/2}, && \mathcal{C}^{\VV}_2&=&-\frac{32,641}{198,900}\tk^{-1/2}, \\
    \mathcal{A}^{\VV}_3&=&\frac{1,067}{26,520}\tk^{-1/2}, && \mathcal{C}^{\VV}_3&=&-\frac{1,067}{26,520}\tk^{-1/2}, \\
    \mathcal{A}^{\VV}_4&=&-\frac{11}{240}\left(\tcos(k)+\frac{2,048}{663}\tk^{-12}\right), && \mathcal{C}^{\VV}_4&=&\frac{11}{240}\left(\tcos(k)+\frac{2,048}{663}\tk^{-1/2}\right),
  \end{array}
\ee
\be
  \begin{array}{rclcrcl}
    \mathcal{A}^{\TT}_{-5}&=&\frac{1,024}{29,835}\left(1-\tk^{-1/2}\right), && \mathcal{C}^{\TT}_{-5}&=&\frac{1,024}{29,835}\left(1+\tk^{-1/2}\right), \\
    \mathcal{A}^{\TT}_{-4}&=&\frac{512}{29,835}\tk^{-1/2}, && \mathcal{C}^{\TT}_{-4}&=&-\frac{512}{29,835}\tk^{-1/2}, \\
    \mathcal{A}^{\TT}_{-3}&=&\frac{128}{585}\left(1-\frac{50}{51}\tk^{-1/2}\right), &&\mathcal{C}^{\TT}_{-3}&=&\frac{128}{585}\left(1+\frac{50}{51}\tk^{-1/2}\right), \\
    \mathcal{A}^{\TT}_{-2}&=&\frac{256}{2,295}\tk^{-2}, && \mathcal{C}^{\TT}_{-2}&=&-\frac{256}{2,295}\tk^{-1/2}, \\
    \mathcal{A}^{\TT}_{-1}&=&-\frac{88}{225}\left(1-\frac{7,828}{7,293}\tk^{-1/2}\right), && \mathcal{C}^{\TT}_{-1}&=&-\frac{88}{225}\left(1+\frac{7,828}{7,293}\tk^{-1/2}\right), \\
    \mathcal{A}^{\TT}_0&=&\frac{251,468}{149,175}\tk^{-1/2}, && \mathcal{C}^{\TT}_0&=&-\frac{251,468}{149,175}\tk^{-1/2}, \\
    \mathcal{A}^{\TT}_1&=&\frac{4}{5}\left(1-\frac{34,031}{6,630}\tk^{-1/2}\right), && \mathcal{C}^{\TT}_1&=&\frac{4}{5}\left(1+\frac{34,031}{6,630}\tk^{-1/2}\right), \\
    \mathcal{A}^{\TT}_2&=&\frac{15,673}{7,650}\tk^{-1/2}, && \mathcal{C}^{\TT}_2&=&-\frac{15,673}{7,650}\tk^{-1/2}, \\
    \mathcal{A}^{\TT}_3&=&\frac{1}{2}\left(1-\frac{3,977}{6,630}\tk^{-1/2}\right), && \mathcal{C}^{\TT}_3&=&\frac{1}{2}\left(1+\frac{3,977}{6,630}\tk^{-1/2}\right), \\
    \mathcal{A}^{\TT}_4&=&\frac{41}{120}\left(\tcos(k)+\frac{2,048}{663}\tk^{-1/2}\right), && \mathcal{C}^{\TT}_4&=&-\frac{41}{120}\left(\tcos(k)+\frac{2,048}{663}\tk^{-1/2}\right),
  \end{array}
\ee
with
\be
\begin{array}{c}
   \mathcal{B}^{\RR}=\frac{149}{225}-\frac{51}{320}\pi ,\qquad
   \mathcal{B}^{\RS}=\frac{37}{5,850}+\frac{3}{80}\pi ,\\
   \mathcal{B}^{\SS}=\frac{8,113}{15,300}-\frac{9}{320}\pi ,\qquad
   \mathcal{B}^{\VV}=\frac{15,362}{149,175}+\frac{11}{480}\pi , \\
   \mathcal{B}^{\TT}=\frac{346,687}{298,350}-\frac{41}{240}\pi .
\end{array}
\ee

\subsection{$n_B=-1/2$}
\be
\begin{array}{l}
  \begin{array}{ccccl}
    \mathcal{A}^{\RR}_{-1}&=&\mathcal{C}^{\RR}_{-1}&=&\frac{8}{21}\left(1-\tk^{-1/2}\right), \\
    \mathcal{A}^{\RR}_0&=&\mathcal{C}^{\RR}_0&=&\frac{46}{21}\tk^{-1/2}, \\
    \mathcal{A}^{\RR}_1&=&\mathcal{C}^{\RR}_1&=&\frac{4}{3}\left(1-\frac{69}{28}\tk^{-1/2}\right) ,
  \end{array}
  \\
  \mathcal{A}^{\RR}_2=-\frac{3}{2}\left(\tcos(k)+\pi-\frac{53}{21}\tk^{-1/2}\right),
  \\
  \mathcal{C}^{\RR}_2=-\frac{3}{2}\left(\tcos(k)-\frac{53}{21}\tk^{-1/2}\right),
  \\
  \begin{array}{ccccl}
    \mathcal{A}^{\RR}_3&=&\mathcal{C}^{\RR}_3&=&1-\frac{64}{21}\tk^{-1/2} ,
  \end{array}
\end{array}
\qquad
\begin{array}{l}
  \begin{array}{ccccl}
    \mathcal{A}^{\RS}_{-3}&=&\mathcal{C}^{\RS}_{-3}&=&-\frac{32}{77}\left(1-\tk^{-1/2}\right), \\
    \mathcal{A}^{\RS}_{-2}&=&\mathcal{C}^{\RS}_{-2}&=&-\frac{16}{77}\tk^{-1/2}, \\
    \mathcal{A}^{\RS}_{-1}&=&\mathcal{C}^{\RS}_{-1}&=&-\frac{4}{3}\left(1-\frac{80}{77}\tk^{-1/2}\right), \\
    \mathcal{A}^{\RS}_0&=&\mathcal{C}^{\RS}_0&=&\frac{148}{231}\tk^{-1/2}, \\
    \mathcal{A}^{\RS}_1&=&\mathcal{C}^{\RS}_1&=&-\frac{5}{3}\left(1-\frac{362}{385}\tk^{-1/2}\right),
  \end{array}
  \\
  \mathcal{A}^{\RS}_2=\tcos(k)+\pi-\frac{32}{33}\tk^{-1/2} ,
  \\
  \mathcal{C}^{\RS}_2=\tcos(k)-\frac{32}{33}\tk^{-1/2},
  \\
  \begin{array}{ccccl}
    \mathcal{A}^{\RS}_3&=&\mathcal{C}^{\RS}_3&=&-\frac{1}{2}\left(1-\frac{128}{33}\tk^{-1/2}\right) ,
  \end{array}
\end{array}
\ee
\be
\begin{array}{l}
  \begin{array}{ccccl}
    \mathcal{A}^{\SS}_{-5}&=&\mathcal{C}^{\SS}_{-5}&=&-\frac{1,536}{2,695}\left(1-\tk^{-1/2}\right), \\
    \mathcal{A}^{\SS}_{-4}&=&\mathcal{C}^{\SS}_{-4}&=&-\frac{768}{2,695}\tk^{-1/2}, \\
    \mathcal{A}^{\SS}_{-3}&=&\mathcal{C}^{\SS}_{-3}&=&\frac{128}{77}\left(1-\frac{73}{70}\tk^{-1/2}\right), \\
    \mathcal{A}^{\SS}_{-2}&=&\mathcal{C}^{\SS}_{-2}&=&\frac{2,144}{2,695}\tk^{-1/2}, \\
    \mathcal{A}^{\SS}_{-1}&=&\mathcal{C}^{\SS}_{-1}&=&-\frac{4}{7}\left(1-\frac{102}{77}\tk^{-1/2}\right), \\
    \mathcal{A}^{\SS}_0&=&\mathcal{C}^{\SS}_0&=&\frac{1,002}{385}\tk^{-1/2}, \\
    \mathcal{A}^{\SS}_1&=&\mathcal{C}^{\SS}_1&=&\frac{4}{3}\left(1-\frac{619}{220}\tk^{-1/2}\right) , \\
  \end{array}
  \\
  \mathcal{A}^{\SS}_2=-\frac{11}{14}\left(\tcos(k)+\pi-\frac{1,216}{1,815}\tk^{-1/2}\right) , \\
  \mathcal{C}^{\SS}_2=-\frac{11}{14}\left(\tcos(k)-\frac{1,216}{1,815}\tk^{-1/2}\right), \\
  \begin{array}{ccccl}
    \mathcal{A}^{\SS}_3&=&\mathcal{C}^{\SS}_3&=&\frac{1}{4}\left(1-\frac{4,864}{1,155}\tk^{-1/2}\right) ,
  \end{array}
\end{array}
\qquad
\begin{array}{l}
  \begin{array}{ccccl}
    \mathcal{A}^{\VV}_{-5}&=&\mathcal{C}^{\VV}_{-5}&=&\frac{2,048}{8,085}\left(1-\tk^{-1/2}\right), \\
    \mathcal{A}^{\VV}_{-4}&=&\mathcal{C}^{\VV}_{-4}&=&\frac{1,024}{8,085}\tk^{-1/2}, \\
    \mathcal{A}^{\VV}_{-3}&=&\mathcal{C}^{\VV}_{-3}&=&-\frac{64}{231}\left(1-\frac{39}{35}\tk^{-1/2}\right), \\
    \mathcal{A}^{\VV}_{-2}&=&\mathcal{C}^{\VV}_{-2}&=&-\frac{992}{8,085}, \\
    \mathcal{A}^{\VV}_{-1}&=&\mathcal{C}^{\VV}_{-1}&=&\frac{8}{21}\left(1-\frac{82}{77}\tk^{-1/2}\right), \\
    \mathcal{A}^{\VV}_0&=&\mathcal{C}^{\VV}_0&=&\frac{788}{385}\tk^{-1/2}, \\
    \mathcal{A}^{\VV}_1&=&\mathcal{C}^{\VV}_1&=&\frac{2}{3}\left(1-\frac{1,381}{385}\tk^{-1/2}\right),
  \end{array}
  \\
  \mathcal{A}^{\VV}_2=-\frac{3}{7}\left(\tcos(k)+\pi-\frac{64}{165}\tk^{-1/2}\right),
  \\
  \mathcal{C}^{\VV}_2=-\frac{3}{7}\left(\tcos(k)-\frac{64}{165}\tk^{-1/2}\right),
  \\
  \begin{array}{ccccl}
    \mathcal{A}^{\VV}_3&=&\mathcal{C}^{\VV}_3&=&-\frac{128}{385}\tk^{-1/2},
  \end{array}
\end{array}
\ee
\be
\begin{array}{l}
  \begin{array}{ccccl}
    \mathcal{A}^{\TT}_{-5}&=&\mathcal{C}^{\TT}_{-5}&=&-\frac{1,024}{8,085}\left(1-\tk^{-1/2}\right), \\
    \mathcal{A}^{\TT}_{-4}&=&\mathcal{C}^{\TT}_{-4}&=&-\frac{512}{8,085}\tk^{-1/2}, \\
    \mathcal{A}^{\TT}_{-3}&=&\mathcal{C}^{\TT}_{-3}&=&-\frac{128}{231}\left(1-\frac{34}{35}\tk^{-1/2}\right), \\
    \mathcal{A}^{\TT}_{-2}&=&\mathcal{C}^{\TT}_{-2}&=&-\frac{768}{2,695}\tk^{-1/2}, \\
    \mathcal{A}^{\TT}_{-1}&=&\mathcal{C}^{\TT}_{-1}&=&\frac{24}{7}\left(1-\frac{236}{231}\tk^{-1/2}\right), \\
    \mathcal{A}^{\TT}_0&=&\mathcal{C}^{\TT}_0&=&\frac{568}{105}\tk^{-1/2}, \\
    \mathcal{A}^{\TT}_1&=&\mathcal{C}^{\TT}_1&=&-\frac{4}{3}\left(1+\frac{711}{385}\tk^{-1/2}\right),
  \end{array}
  \\
  \mathcal{A}^{\TT}_2=-\frac{2}{7}\left(\tcos(k)+\pi-\frac{64}{65}\tk^{-1/2}\right) ,
  \\
  \mathcal{C}^{\TT}_2=-\frac{2}{7}\left(\tcos(k)-\frac{64}{65}\tk^{-1/2}\right) ,
  \\
  \begin{array}{ccccl}
    \mathcal{A}^{\TT}_3&=&\mathcal{C}^{\TT}_3&=&\frac{1}{2}\left(1-\frac{512}{385}\tk^{-1/2}\right) ,
  \end{array}
\end{array}
\ee
with
\be
\begin{array}{c}
   \mathcal{B}^{\RR}=\frac{19}{17}-\frac{3}{4}\pi ,\qquad
   \mathcal{B}^{\RS}=\frac{1}{2}\pi-\frac{577}{462} ,\qquad
   \mathcal{B}^{\SS}=\frac{68,053}{32,340}-\frac{11}{28}\pi , \\
   \mathcal{B}^{\VV}=\frac{8,278}{8,085}-\frac{3}{14}\pi , \qquad
   \mathcal{B}^{\TT}=\frac{19}{7}-\frac{3}{4}\pi .
\end{array}
\ee

\subsection{$n_B=-2$}
The infra-red fields are dominated by a term proportional to $k^{2n_B+3}$; in this case, this is $k^{-1}$. To ease the presentation, for this field we tabulate these coefficients separately from the others.
\be
  \begin{array}{rclcl}
    \mathcal{A}^{\RR}_0&=&\mathcal{C}^{\RR}_0&=&-\frac{1}{2}, \\
    \mathcal{A}^{\RR}_1&=&\mathcal{C}^{\RR}_1&=&-\frac{1}{2}\ln(\tk), \\
    \mathcal{A}^{\RR}_3&=&\mathcal{C}^{\RR}_3&=&\frac{1}{16},
  \end{array}
\qquad
  \begin{array}{rclcl}
    \mathcal{A}^{\RS}_{-3}&=&\mathcal{C}^{\RS}_{-3}&=&\frac{3}{8}\ln(\tk), \\
    \mathcal{A}^{\RS}_{-2}&=&\mathcal{C}^{\RS}_{-2}&=&\frac{3}{8}, \\
    \mathcal{A}^{\RS}_0&=&\mathcal{C}^{\RS}_0&=&\frac{5}{8}, \\
    \mathcal{A}^{\RS}_1&=&\mathcal{C}^{\RS}_1&=&-\frac{9}{32}\left(1-\frac{20}{9}\ln(\tk)\right), \\
    \mathcal{A}^{\RS}_3&=&\mathcal{C}^{\RS}_3&=&-\frac{1}{32},
  \end{array}
\ee
\be
  \begin{array}{rclcl}
    \mathcal{A}^{\SS}_{-5}&=&\mathcal{C}^{\SS}_{-5}&=&\frac{9}{8}\ln(\tk), \\
    \mathcal{A}^{\SS}_{-4}&=&\mathcal{C}^{\SS}_{-4}&=&\frac{9}{8}, \\
    \mathcal{A}^{\SS}_{-3}&=&\mathcal{C}^{\SS}_{-3}&=&\frac{9}{16}\left(1-\frac{8}{3}\ln(\tk)\right), \\
    \mathcal{A}^{\SS}_{-2}&=&\mathcal{C}^{\SS}_{-2}&=&-\frac{9}{8}, \\
    \mathcal{A}^{\SS}_0&=&\mathcal{C}^{\SS}_0&=&-\frac{1}{2}, \\
    \mathcal{A}^{\SS}_1&=&\mathcal{C}^{\SS}_1&=&\frac{3}{8}\left(1-\frac{4}{3}\ln(\tk)\right)k, \\
    \mathcal{A}^{\SS}_3&=&\mathcal{C}^{\SS}_3&=&\frac{1}{64},
  \end{array}
\qquad
  \begin{array}{rclcl}
    \mathcal{A}^{\VV}_{-5}&=&\mathcal{C}^{\VV}_{-5}&=&-\frac{1}{2}\ln(\tk), \\
    \mathcal{A}^{\VV}_{-4}&=&\mathcal{C}^{\VV}_{-4}&=&-\frac{1}{2}, \\
    \mathcal{A}^{\VV}_{-3}&=&\mathcal{C}^{\VV}_{-3}&=&-\frac{1}{4}\left(1-\ln(\tk)\right), \\
    \mathcal{A}^{\VV}_{-2}&=&\mathcal{C}^{\VV}_{-2}&=&\frac{1}{12}, \\
    \mathcal{A}^{\VV}_0&=&\mathcal{C}^{\VV}_0&=&-\frac{1}{4}, \\
    \mathcal{A}^{\VV}_1&=&\mathcal{C}^{\VV}_1&=&\frac{7}{48}\left(1-\frac{12}{7}\ln(\tk)\right),
  \end{array}
\ee
\be
  \begin{array}{rclcl}
    \mathcal{A}^{\TT}_{-5}&=&\mathcal{C}^{\TT}_{-5}&=&\frac{1}{4}\ln(\tk), \\
    \mathcal{A}^{\TT}_{-4}&=&\mathcal{C}^{\TT}_{-4}&=&\frac{1}{4}, \\
    \mathcal{A}^{\TT}_{-3}&=&\mathcal{C}^{\TT}_{-3}&=&\frac{1}{8}\left(1+4\ln(\tk)\right), \\
    \mathcal{A}^{\TT}_{-2}&=&\mathcal{C}^{\TT}_{-2}&=&\frac{7}{12}, \\
    \mathcal{A}^{\TT}_0&=&\mathcal{C}^{\TT}_0&=&\frac{1}{2}, \\
    \mathcal{A}^{\TT}_1&=&\mathcal{C}^{\TT}_1&=&-\frac{13}{24}\left(1-\frac{12}{13}\ln(\tk)\right), \\
    \mathcal{A}^{\TT}_3&=&\mathcal{C}^{\TT}_3&=&\frac{1}{32},
  \end{array}
\ee
with
\be
\begin{array}{c}
   \mathcal{B}^{\RR}=\frac{1}{4}\pi^2-\frac{7}{16} ,\qquad
   \mathcal{B}^{\RS}=\frac{1}{8}\pi^2+\frac{7}{8} ,\qquad
   \mathcal{B}^{\SS}=\frac{3}{8}\pi^2-\frac{1}{64} , \\
   \mathcal{B}^{\VV}=\frac{1}{4}\pi^2-\frac{37}{48} , \qquad
   \mathcal{B}^{\TT}=\frac{3}{4}\pi^2+\frac{121}{96} ,
\end{array}
\ee
and
\be
\begin{array}{ccc}
  & \mathcal{A}^{\RR}_{-1}=3\left(\Li_2(k)+\ln(k)\ln(\tk)+\frac{1}{6}\ln(\tk)+\frac{1}{12}\pi^2\right), & \\ & \qquad
    \mathcal{C}^{\RR}_{-1}=\frac{3}{2}\left(\Li_2(k)-\Li_2(\tk)+\ln(k)\ln(\tk)+\frac{1}{3}\ln(\tk)\right), &
  \\ && \\
  & \mathcal{A}^{\RS}_{-1}=\frac{3}{2}\left(\Li_2(k)+\ln(k)\ln(\tk)-\frac{2}{3}\ln(\tk)+\frac{1}{8}+\frac{1}{12}\pi^2\right), & \\ & \qquad
    \mathcal{C}^{\RS}_{-1}=\frac{3}{4}\left(\Li_2(k)-\Li_2(\tk)+\ln(k)\ln(\tk)-\frac{4}{3}\ln(\tk)+\frac{1}{4}\right), &
  \\ && \\
  & \mathcal{A}^{\SS}_{-1}=\frac{9}{2}\left(\Li_2(k)+\ln(k)\ln(\tk)+\frac{7}{36}\ln(\tk)+\frac{1}{12}\pi^2-\frac{5}{48}\right), & \\ & \qquad
    \mathcal{C}^{\SS}_{-1}=\frac{9}{4}\left(\Li_2(k)-\Li_2(\tk)+\ln(k)\ln(\tk)+\frac{7}{18}\ln(\tk)-\frac{5}{24}\right), &
  \\ && \\
  & \mathcal{A}^{\VV}_{-1}=3\left(\Li_2(k)+\ln(k)\ln(\tk)+\frac{1}{6}\ln(\tk)+\frac{1}{12}\pi^2\right), &\\ & \qquad
    \mathcal{C}^{\VV}_{-1}=\frac{3}{2}\left(\Li_2(k)-\Li_2(\tk)+\ln(k)\ln(\tk)+\frac{1}{3}\ln(\tk)\right), &
  \\ && \\
  & \mathcal{A}^{\TT}_{-1}=9\left(\Li_2(k)+\ln(k)\ln(\tk)-\frac{5}{36}\ln(\tk)+\frac{5}{144}+\frac{1}{12}\pi^2\right), &\\&\qquad
    \mathcal{C}^{\TT}_{-1}=\frac{9}{2}\left(\Li_2(k)-\Li_2(\tk)+\ln(k)\ln(\tk)-\frac{5}{18}\ln(\tk)+\frac{5}{72}\right) .
\end{array}
\ee

\section{Analytic Power Spectra on Large Scales}
\label{Appendix-LargeScale}
Only the largest scales are required for work with the CMB. On such scales the analytic solutions are significantly simplified; generating their Laurent series and keeping up to quadratic order in $k$, the analytic solutions for all $n_B\in[-5/2,3]$ reduce to the following.

\subsection{$n_B=3$}
\be
\begin{array}{c}
 \begin{array}{rclrcl}
   \PTrTr(k)&\approx&\pi A_Bk_c^3\left(\frac{4}{9}-k+\frac{20}{21}k^2\right), &
   \PTrTs(k)&\approx&\pi A_Bk_c^3\left(-\frac{1}{4}k+\frac{64}{105}k^2\right), \\
   \PTsTs(k)&\approx&\pi A_Bk_c^3\left(\frac{28}{45}-k+\frac{628}{735}k^2\right), &
   \PTvTv(k)&\approx&\pi A_Bk_c^3\left(\frac{56}{135}-\frac{5}{6}k+\frac{592}{735}k^2\right),
 \end{array}
  \\
 \PTtTt(k)\approx\pi A_Bk_c^3\left(\frac{112}{135}-\frac{7}{3}k+\frac{2224}{735}k^2\right) .
\end{array}
\ee

\subsection{$n_B=5/2$}
\be
\begin{array}{c}
 \begin{array}{rclrcl}
   \PTrTr(k)&\approx&\pi A_Bk_c^3\left(\frac{1}{2}-k+\frac{3}{4}k^2\right), &
   \PTrTs(k)&\approx&\pi A_Bk_c^3\left(-\frac{1}{4}k+\frac{17}{30}k^2\right), \\
   \PTsTs(k)&\approx&\pi A_Bk_c^3\left(\frac{7}{10}-k+\frac{293}{420}k^2\right), &
   \PTvTv(k)&\approx&\pi A_Bk_c^3\left(\frac{7}{15}-\frac{5}{6}k+\frac{47}{70}k^2\right),
 \end{array}
  \\
 \PTtTt(k)\approx\pi A_Bk_c^3\left(\frac{14}{15}-\frac{7}{3}k+\frac{271}{105}k^2\right) .
\end{array}
\ee

\subsection{$n_B=2$}
\be
\begin{array}{c}
  \begin{array}{rclrcl}
    \PTrTr(k)&\approx&\pi A_Bk_c^3\left(\frac{4}{7}-k+\frac{8}{15}k^2\right), &
    \PTrTs(k)&\approx&\pi A_Bk_c^3\left(-\frac{1}{4}k+\frac{8}{15}k^2\right), \\
    \PTsTs(k)&\approx&\pi A_Bk_c^3\left(\frac{4}{5}-k+\frac{8}{15}k^2\right), &
    \PTvTv(k)&\approx&\pi A_Bk_c^3\left(\frac{8}{15}-\frac{5}{6}k+\frac{8}{15}k^2\right), \\
  \end{array}
  \\
  \PTtTt(k)\approx\pi A_Bk_c^3\left(\frac{16}{15}-\frac{7}{3}k+\frac{32}{15}k^2\right) .
\end{array}
\ee

\subsection{A damped causal field}
\be
\begin{array}{rcl}
  \PTrTr&=&\pi A_Bk_c^3\left(T_0-\frac{1}{e^{2\xi}}k-\frac{1}{384}\left(\frac{57\sqrt{2\pi}\erf(\sqrt{2\xi})e^{2\xi}-228\xi^{1/2}-304\xi^{3/2}-448\xi^{5/2}+256\xi^{7/2}}{\xi^{5/2}e^{2\xi}}\right)k^2\right), \\
  \PTrTs&=&\pi A_Bk_c^3\left(-\frac{1}{4e^{2\xi}}k+\frac{1}{960}\left(\frac{15\sqrt{2\pi}\erf(\sqrt{2\xi})e^{2\xi}-60\xi^{1/2}-80\xi^{3/2}+448\xi^{5/2}-256\xi^{7/2}}{\xi^{5/2}e^{2\xi}}\right)k^2\right), \\
  \PTsTs&=&\pi A_Bk_c^3\left(\frac{7}{5}T_0-\frac{1}{e^{2\xi}}k-\frac{1}{13,440}\left(\frac{1,365\sqrt{2\pi}\erf(\sqrt{2\xi})e^{2\xi}-5,460\xi^{1/2}-7,285\xi^{3/2}-12,992\xi^{5/2}+7,424\xi^{7/2}}{\xi^{5/2}e^{2\xi}}\right)k^2\right), \\
  \PTvTv&=&\pi A_Bk_c^3\left(\frac{14}{15}T_0-\frac{5}{6e^{2\xi}}k-\frac{1}{6,720}\left(\frac{525\sqrt{2\pi}\erf(\sqrt{2\xi})e^{2\xi}-2,100\xi^{1/2}-2,800\xi^{3/2}-5,824\xi^{5/2}+3,328\xi^{7/2}}{\xi^{5/2}e^{2\xi}}\right)k^2\right), \\
  \PTtTt&=&\pi A_Bk_c^3\left(\frac{28}{15}T_0-\frac{7}{3e^{2\xi}}k-\frac{1}{3,360}\left(\frac{735\sqrt{2\pi}\erf(\sqrt{2\xi})e^{2\xi}-2,940\xi^{1/2}-3,920\xi^{3/2}-10,304\xi^{5/2}+5,888\xi^{7/2}}{\xi^{5/2}e^{2\xi}}\right)k^2\right)
\end{array}
\ee
where
\be
  T_0=\frac{1}{64}\frac{15\sqrt{2\pi}\erf(\sqrt{2\xi})\xi^3e^{2\xi}-60\xi^{7/2}-80\xi^{9/2}-64\xi^{11/2}}{\xi^{13/2}e^{2\xi}} .
\ee

\subsection{$n_B=3/2$}
\be
\begin{array}{c}
  \begin{array}{rclrcl}
    \PTrTr(k)&\approx&\pi A_Bk_c^3\left(\frac{2}{3}-k+\frac{7}{24}k^2\right), &
    \PTrTs(k)&\approx&\pi A_Bk_c^3\left(-\frac{1}{4}k+\frac{31}{60}k^2\right), \\
    \PTsTs(k)&\approx&\pi A_Bk_c^3\left(\frac{14}{15}-k+\frac{299}{840}k^2\right), &
    \PTvTv(k)&\approx&\pi A_Bk_c^3\left(\frac{28}{45}-\frac{5}{6}k+\frac{163}{420}k^2\right),
  \end{array}
  \\
  \PTtTt(k)\approx\pi A_Bk_c^3\left(\frac{56}{45}-\frac{7}{3}k+\frac{353}{210}k^2\right) .
\end{array}
\ee

\subsection{$n_B=1$}
\be
\begin{array}{c}
  \begin{array}{rclrcl}
    \PTrTr(k)&\approx&\pi A_Bk_c^3\left(\frac{4}{5}-k\right), &
    \PTrTs(k)&\approx&\pi A_Bk_c^3\left(-\frac{1}{4}k+\frac{8}{15}k^2\right), \\
    \PTsTs(k)&\approx&\pi A_Bk_c^3\left(\frac{28}{25}-k+\frac{16}{105}k^2\right), &
    \PTvTv(k)&\approx&\pi A_Bk_c^3\left(\frac{56}{75}-\frac{5}{6}k+\frac{8}{35}k^2\right),
  \end{array}
  \\
  \PTtTt(k)\approx\pi A_Bk_c^3\left(\frac{112}{75}-\frac{7}{3}k+\frac{128}{105}k^2\right)
\end{array}
\ee

\subsection{$n_B=1/2$}
\be
\begin{array}{c}
  \begin{array}{rclrcl}
    \PTrTr(k)&\approx&\pi A_Bk_c^3\left(1-k-\frac{5}{12}k^2\right), &
    \PTrTs(k)&\approx&\pi A_Bk_c^3\left(-\frac{1}{4}k+\frac{19}{30}k^2\right), \\
    \PTsTs(k)&\approx&\pi A_Bk_c^3\left(\frac{7}{5}-k-\frac{7}{60}k^2\right), &
    \PTvTv(k)&\approx&\pi A_Bk_c^3\left(\frac{14}{15}-\frac{5}{6}k+\frac{1}{30}k^2\right),
  \end{array}
  \\
  \PTtTt(k)\approx\pi A_Bk_c^3\left(\frac{28}{15}-\frac{7}{3}k+\frac{11}{15}k^2\right)
\end{array}
\ee

\subsection{$n_B=0$}
\be
\begin{array}{c}
  \begin{array}{rclrcl}
    \PTrTr(k)&\approx&\pi A_Bk_c^3\left(\frac{4}{3}-k-\frac{4}{3}k^2\right), &
    \PTrTs(k)&\approx&\pi A_Bk_c^3\left(-\frac{1}{4}k+\frac{16}{15}k^2\right), \\
    \PTsTs(k)&\approx&\pi A_Bk_c^3\left(\frac{28}{15}-k+\frac{68}{105}k^2\right), &
    \PTvTv(k)&\approx&\pi A_Bk_c^3\left(\frac{56}{45}-\frac{5}{6}k-\frac{32}{105}k^2\right),
  \end{array}
  \\
  \PTtTt(k)\approx\pi A_Bk_c^3\left(\frac{112}{45}-\frac{7}{3}k+\frac{16}{105}k^2\right)
\end{array}
\ee

\subsection{$n_B=-1/2$}
\be
\begin{array}{rcl}
  \PTrTr(k)&\approx&\pi A_Bk_c^3\left(2-k+\frac{3}{2}\left(\ln(\frac{1}{4}k)-\frac{1}{2}\pi+\frac{43}{18}\right)k^2\right), \\
  \PTrTs(k)&\approx&\pi A_Bk_c^3\left(-\frac{1}{4}k-\left(\ln(\frac{1}{4}k)-\frac{1}{2}\pi+\frac{7}{3}\right)k^2\right), \\
  \PTsTs(k)&\approx&\pi A_Bk_c^3\left(\frac{14}{5}-k+\frac{11}{14}\left(\ln(\frac{1}{4}k)-\frac{1}{2}\pi+\frac{251}{54}\right)k^2\right), \\
  \PTvTv(k)&\approx&\pi A_Bk_c^3\left(\frac{28}{15}-\frac{5}{6}k+\frac{3}{7}\left(\ln(\frac{1}{4}k)-\frac{1}{2}\pi+\frac{143}{126}\right)k^2\right), \\
  \PTtTt(k)&\approx&\pi A_Bk_c^3\left(\frac{56}{15}-\frac{7}{3}k+\frac{2}{7}\left(\ln(\frac{k}{4})-\frac{1}{2}\pi+\frac{33}{14}\right)k^2\right) .
\end{array}
\ee

\subsection{$n_B=-1$}
\be
\begin{array}{c}
  \begin{array}{rclrcl}
    \PTrTr(k)&\approx&\pi A_Bk_c^3\left(4-5k+\frac{4}{3}k^2\right), &
    \PTrTs(k)&\approx&\pi A_Bk_c^3\left(\frac{7}{4}k-\frac{16}{15}k^2\right), \\
    \PTsTs(k)&\approx&\pi A_Bk_c^3\left(\frac{28}{5}-5k+\frac{68}{105}k^2\right), &
    \PTvTv(k)&\approx&\pi A_Bk_c^3\left(\frac{56}{15}-\frac{7}{2}k+\frac{32}{105}k^2\right),
  \end{array}
  \\
  \PTtTt(k)\approx\pi A_Bk_c^3\left(\frac{112}{15}-5k-\frac{16}{105}k^2\right) .
\end{array}
\ee

\subsection{$n_B=-3/2$}
There is a logarithmic divergence for this field at low $k$, as part of the transition between the white noise nature of the stresses for $n_B>-3/2$ and the scaling behaviour for $n_B<-3/2$. We retain this logarithmic divergence and expand the remaining terms up to quadratic order in $k$; the dominant contribution as $k\rightarrow 0$ comes from the logarithmic term.
\be
\label{LogTT}
\begin{array}{c}
  \begin{array}{rclrcl}
    \PTrTr(k)&\approx&\pi A_Bk_c^3\left(-4\ln(\frac{1}{4}k)+\frac{20}{9}-2\pi-k+\frac{5}{12}k^2\right), \\
    \PTrTs(k)&\approx&\pi A_Bk_c^3\left(\frac{20}{9}-\frac{1}{4}k-\frac{19}{30}k^2\right), \\
    \PTsTs(k)&\approx&\pi A_Bk_c^3\left(-\frac{28}{5}\ln(\frac{1}{4}k)+\frac{932}{225}-\frac{14}{5}\pi-k+\frac{7}{60}k^2\right), \\
    \PTvTv(k)&\approx&\pi A_Bk_c^3\left(-\frac{56}{15}\ln(\frac{1}{4}k)+\frac{608}{225}-\frac{28}{15}\pi-\frac{5}{6}k-\frac{1}{30}k^2\right),
  \end{array}
  \\
  \PTtTt(k)\approx\pi A_Bk_c^3\left(-\frac{112}{15}\ln(\frac{1}{4}k)+\frac{2,096}{225}-\frac{56}{15}\pi-\frac{7}{3}k-\frac{11}{15}k^2\right) .
\end{array}
\ee
On large scales, the ratio between the scalar cross-correlation and trace auto-correlation is
\bdm
  \frac{\PTrTs}{\PTrTr}\approx-\frac{5}{9\ln(k/4)}\rightarrow 0 .
\edm
The divergences in the other ratios cancel, and even though the white noise behaviour is broken, this transitionary field remains ``causal''.

\subsection{$n_B=-2$}
\be
\begin{array}{c}
  \begin{array}{rclrcl}
    \PTrTr(k)&\approx&\pi A_Bk_c^3\left(\frac{3}{4}\pi^2k^{-1}-4-k\right), &
    \PTrTs(k)&\approx&\pi A_Bk_c^3\left(\frac{3}{8}\pi^2k^{-1}-\frac{1}{4}k\right), \\
    \PTsTs(k)&\approx&\pi A_Bk_c^3\left(\frac{9}{8}\pi^2k^{-1}-\frac{28}{5}-k\right), &
    \PTvTv(k)&\approx&\pi A_Bk_c^3\left(\frac{3}{4}\pi^2k^{-1}-\frac{56}{15}-\frac{5}{6}k\right),
  \end{array}
  \\
  \PTtTt(k)\approx\pi A_Bk_c^3\left(\frac{9}{4}\pi^2k^{-1}-\frac{112}{15}-\frac{7}{3}k\right) .
\end{array}
\ee
The ratios between the spectra on the largest scales are therefore
\bdm
  3\frac{\PTrTs}{\PTrTr}=\frac{\PTsTs}{\PTrTr}=\frac{3}{2}\frac{\PTvTv}{\PTrTr}=\frac{1}{2}\frac{\PTtTt}{\PTrTr}=\frac{3}{2} .
\edm

\subsection{$n_B=-5/2$}
\be
\begin{array}{c}
  \begin{array}{rclrcl}
    \PTrTr(k)&\approx&\pi A_Bk_c^3\left(\frac{272}{25}k^{-2}-2-k\right), &
    \PTrTs(k)&\approx&\pi A_Bk_c^3\left(\frac{224}{25}k^{-2}-\frac{1}{4}k\right), \\
    \PTsTs(k)&\approx&\pi A_Bk_c^3\left(\frac{368}{25}k^{-2}-\frac{14}{5}-k\right), &
    \PTvTv(k)&\approx&\pi A_Bk_c^3\left(\frac{32}{3}k^{-2}-\frac{28}{15}-\frac{5}{6}k\right),
  \end{array}
  \\
  \PTtTt(k)\approx\pi A_Bk_c^3\left(\frac{3,008}{75}k^{-2}-\frac{56}{15}-\frac{7}{3}k\right) .
\end{array}
\ee
The ratios between the spectra on the largest scales are therefore
\bdm
  \frac{23}{4}\frac{\PTrTs}{\PTrTr}=\frac{\PTsTs}{\PTrTr}=\frac{69}{50}\frac{\PTvTv}{\PTrTr}=\frac{69}{188}\frac{\PTtTt}{\PTrTr}=\frac{23}{17} .
\edm

\section{Discrepancy with Yamazaki \emph{et. al.}}
\label{Appendix-Yamazaki}
In Yamazaki \emph{et. al.} \cite{Yamazaki:2008gr} it was claimed [page 4] that ``In almost all previous work the sum of the terms in brackets which include $\mathcal{C}$ in the $k$ integral of Eq. (13) have been set to unity'' and, in particular [page 7], that ``These results are almost the same as in Brown and Crittenden [46] [61]. The constant ratios of all modes for $n_B>-1.5$, however, are not unity because the terms including cosine factors [e.g. sum of terms within the bracket of Eq. (15)], is not unity as was assumed in the previous approximation.''

This is a misconception which we would like here to correct; in neither BC05, B06 nor the present work have we neglected any cosine factors. We briefly demonstrate this for the scalar trace auto-correlation (their equation (13)). The scalar trace auto-correlation, equation (\ref{RawSpectrum}) with $\mathcal{F}_{\tau\tau}=\frac{1}{2}\left(1+\mu^2\right)$ can be rewritten as an integral across $\gamma$, which is equal to Yamazaki \emph{et. al.}'s $\mathcal{C}$. Doing so, and assuming a power law spectrum, yields
\be
  \av{\tau^2}\propto \int_{k'=0}^{k_c}k'^{n_B+2}\int_{\gamma=-1}^1\left|\mathbf{k-k}'\right|^{n_B}\left(1+\mu^2\right)\mathrm{d}\gamma\dk' .
\ee
If we expand out $\mu$ then this can be explicitly rewritten as
\be
  \av{\tau^2}\propto \int_{k'=0}^{k_c}k'^{n_B+2}\int_{\gamma=-1}^1\left|\mathbf{k-k}'\right|^{n_B-2}\left((1+\gamma^2)k^2-4kk'\gamma+2k'^2\right)\mathrm{d}\gamma\dk' .
\ee
This is exactly Yamazaki \emph{et. al.}'s equation (13); neither we nor they have neglected any terms. The ``constant ratios'' to which they refer are presumably the causal ratios presented in the previous Appendix, which are valid to the zeroth order in a Laurent expansion around $k=0$. They are valid for all $n_B\geq -3/2$ on sufficiently large scales; naturally, including the linear order in $k$ will disrupt this.

However, their statement (in their footnote [61]) that we did not fully consider the case with $n_B<-3/2$ is, of course, entirely valid, as we merely considered these with realisations, an approach that is severely limited by an infra-red cut-off and by grid resolution, and did not attempt to integrate the expressions numerically, let alone analytically.

\bibliography{Statistics}

\newpage

\begin{figure}
\centering
\includegraphics[width=0.45\textwidth]{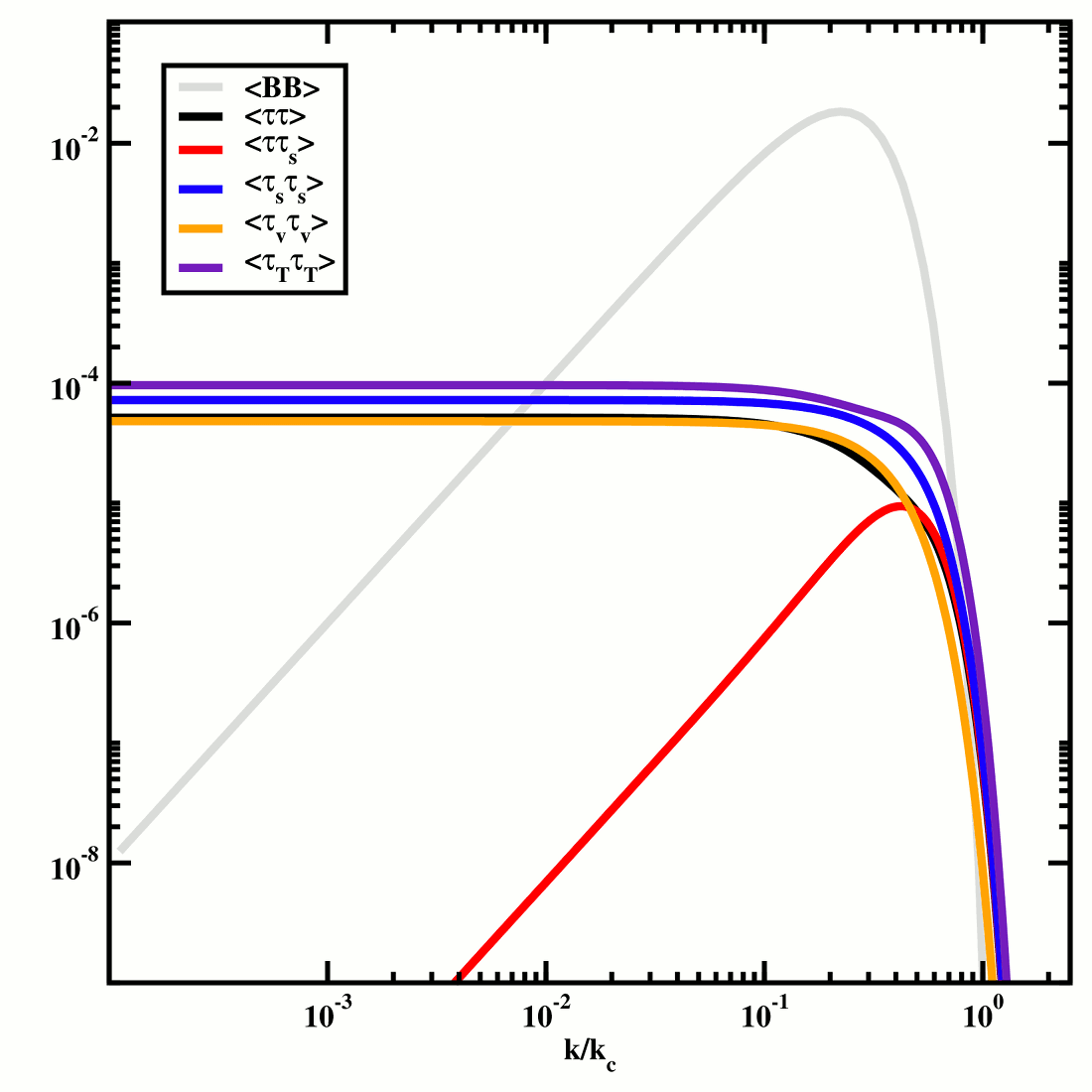}\includegraphics[width=0.45\textwidth]{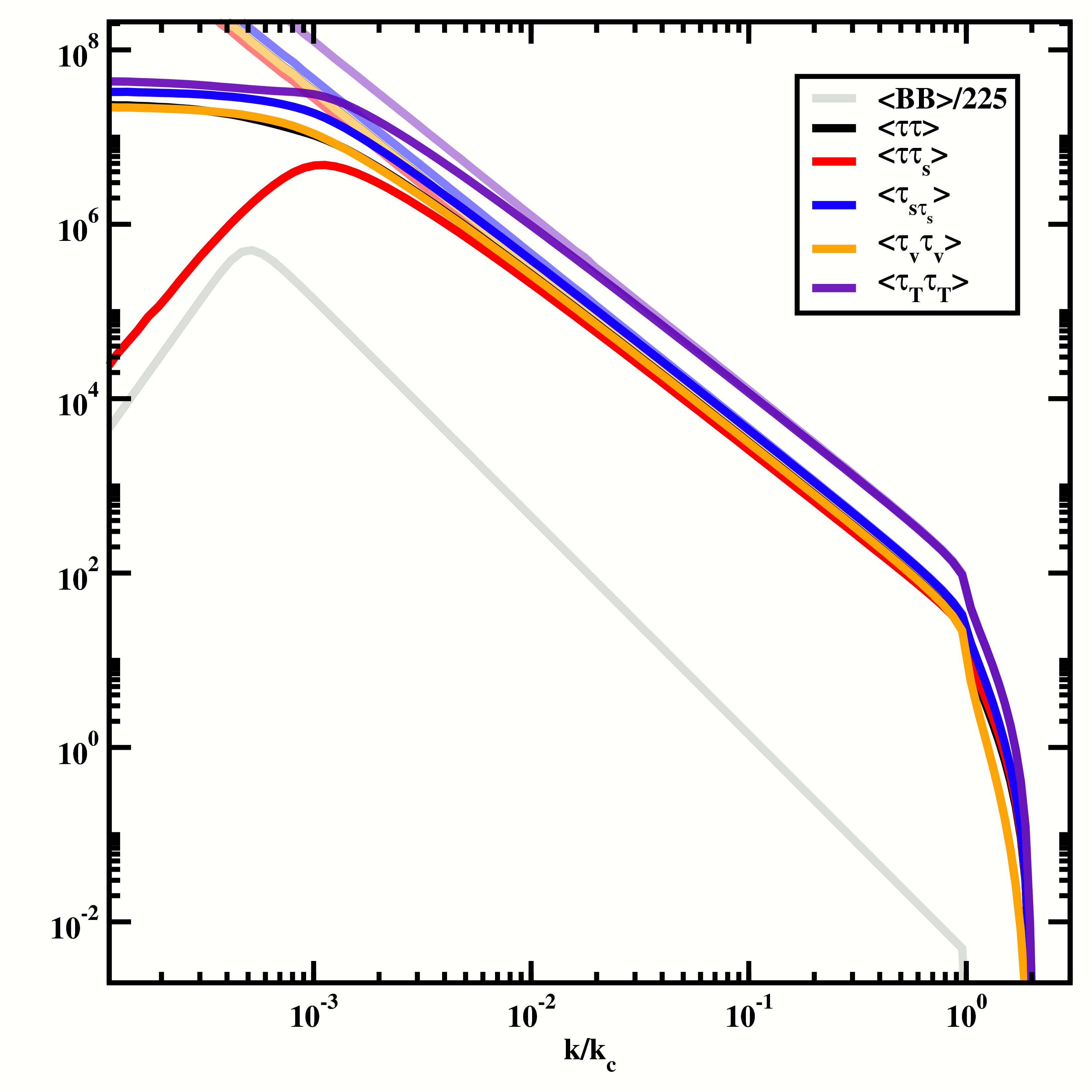}
\caption{Stress spectra for a causal magnetic field exponentially damped at $k\rightarrow k_c$ (left), and an inflationary field with $n_B=-5/2$ damped for $k\rightarrow 0$ (right). In the inflationary case numerical integrations are overlaid on the results for an undamped $n_B=-5/2$ field.}
\label{Figure-SpectraAlternatives}
\end{figure}

\begin{figure}
\centering
\includegraphics[width=0.47\textwidth]{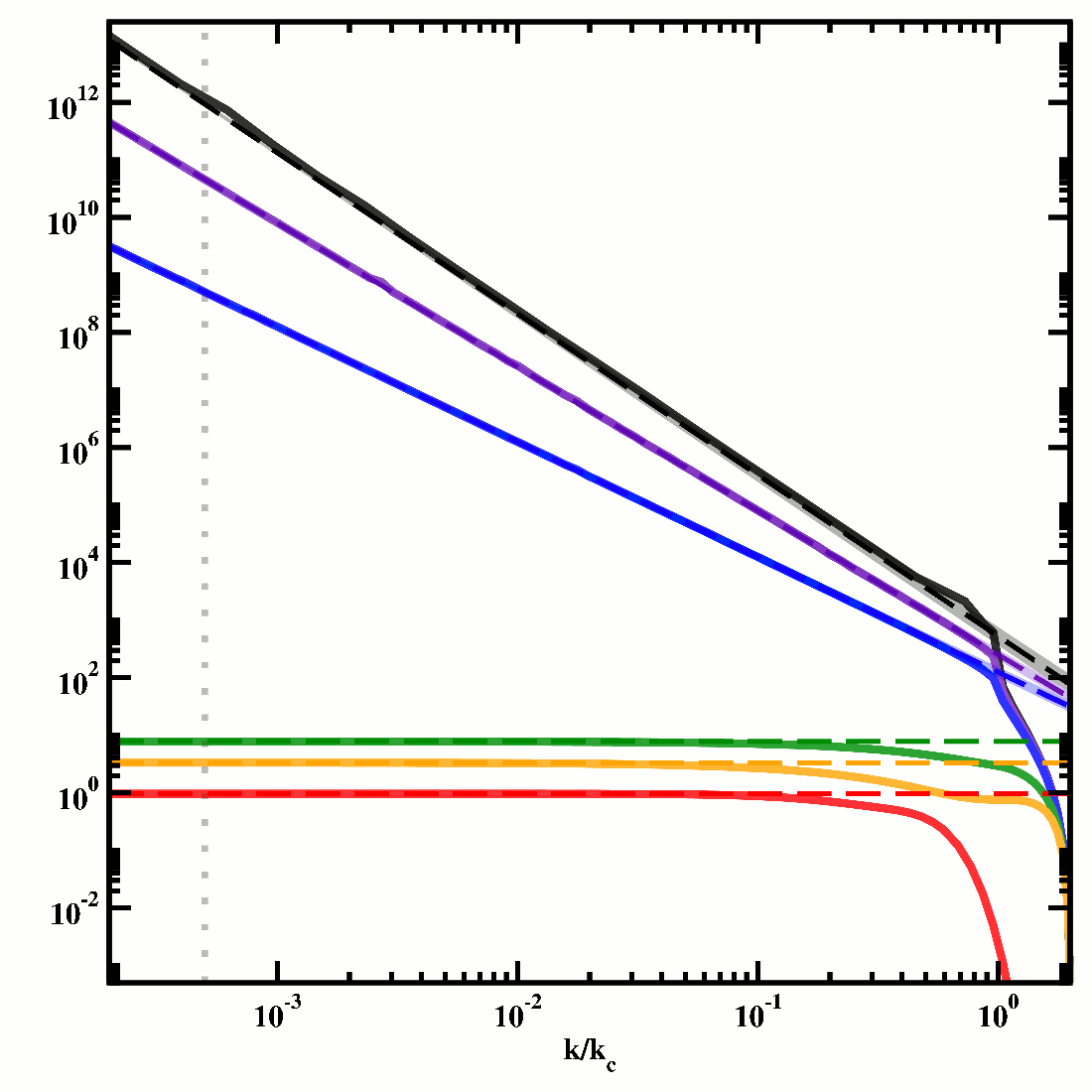}\;\;\includegraphics[width=0.47\textwidth]{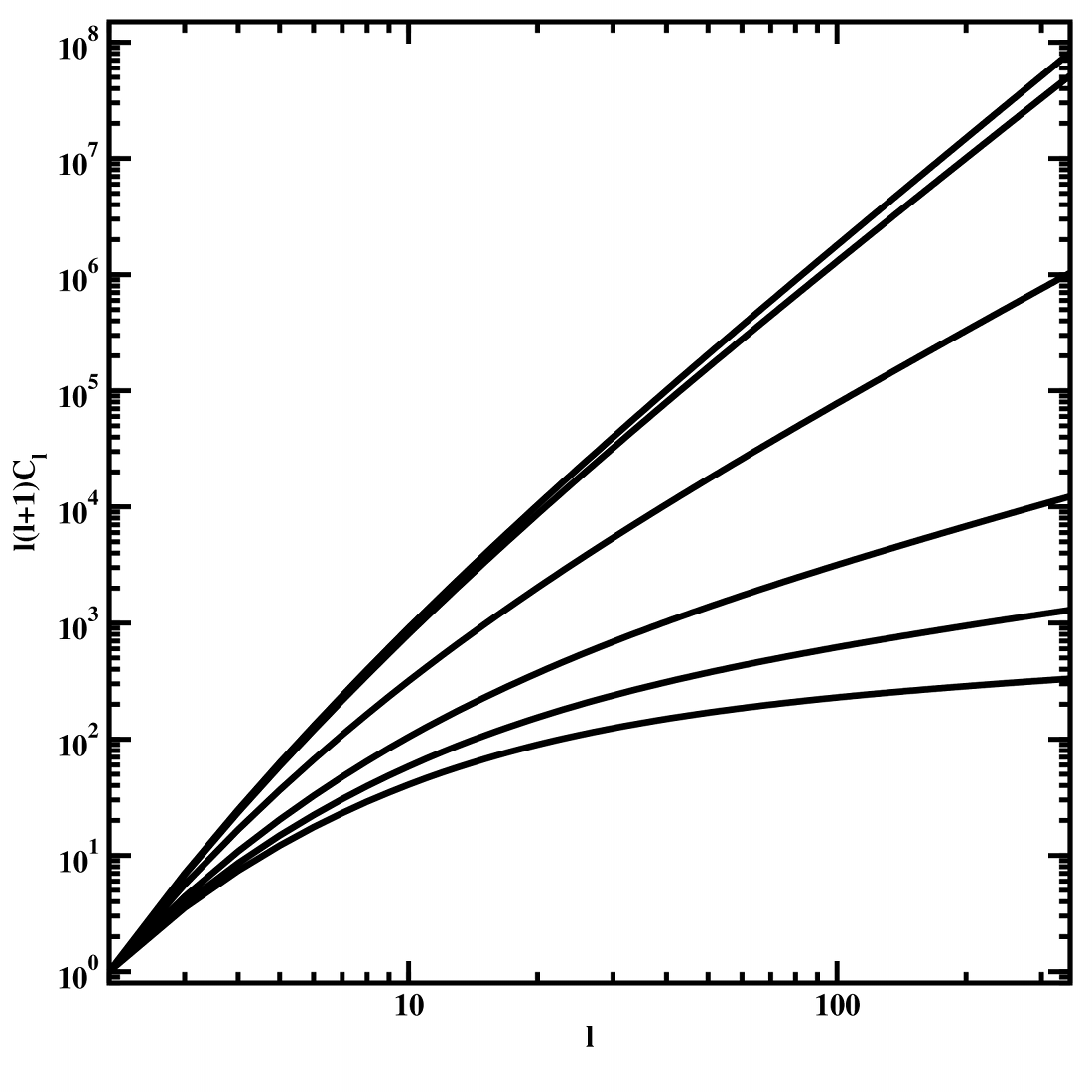}
\caption{Left: Reconstructed power spectra for (top to bottom) $n_B=-2.9$, $n_B=-2.75$, $n_B=-5/2$, $n_B=0$, $n_B=2$ and the damped causal magnetic fields. Shaded regions denote the $1$-$\sigma$ errors. Right: Large-Scale CMB angular power spectra from magnetised tensor perturbations, normalised to $\left.l(l+1)C_l\right|_{l=2}=1$, for (top to bottom) $n_B>-3/2$, $n_B=-3/2$, $n_B=-2$, $n_B=-5/2$, $n_B=-2.75$, $n_B=-2.9$. Signals from the damped causal and IR-controlled inflationary fields are indistinguishable on these scales from those for $n_B=2$ and $n_B=-5/2$ respectively.}
\label{Figure-Reconstructions}
\label{Figure-CMB}
\end{figure}

\label{Appendix-Yamazaki}

\end{document}